\documentclass[12pt]{article} 
\usepackage{amsmath} 
\usepackage{amssymb,amsfonts,euscript} 
\usepackage{hyperref} 
\renewcommand{\baselinestretch}{1.2} 
\newcommand\nn{\nonumber}

\newcommand{\D}{{\cal D}}

\newcommand{\cL}{\cal{L}}

\newcommand{\tg}{\tilde g} 
\newcommand{\Zint}{\mathbb{Z}} 
\newcommand{\Real}{\mathbb{R}} 
\newcommand{\bJ}{\bar{J}} 
\newcommand{\bz}{\bar{z}} 
\newcommand{\bw}{\bar{w}}

\def\l{\lambda} 
\def\m{\mu} 
\def\n{\nu} 
\def\r{\rho} 

\def\de{\delta} 
\def\k{\kappa}

\def\hB{\hat{B}}

\def\hg{\hat{g}}

\def\hb{\hat{B}}

\def\part{\partial}

\def\thickone{{\rm 1\mskip-4.5mu l}}

\newcommand{\s}{\sigma} 
\newcommand{\srange}{\sigma=0,...,N_c-1}

\newcommand{\ws}{\omega (h_\s)}

\newcommand{\hc}{$\hat{J}_{\gst}$} 
 
\newcommand{\tp}{{2\pi i}}

\newcommand{\Ord}{{\cal{O}}}

\newcommand{\sgb}{{\mbox{\scriptsize{\gb}}}}

\newcommand{\gfraks}{{\mbox{\scriptsize{\mbox{${\mathfrak g}$}}}}} 
\def\gb            {\mbox{$\hat{\mathfrak g}$}} 
 
\def\sm#1      {\mbox{\scriptsize $#1$}}

\def\d             {\mbox{$\mathbb D$}}

\def\srac#1#2{\smal{\frac{#1}{#2}}} 
 
\def\foot#1{\mbox{\footnotesize $#1$}} 
\def\scrs#1{\mbox{\scriptsize $#1$}} 
 
\def\smal#1{\mbox{\small $#1$}} 
\def\big#1{\mbox{\large $#1$}} 
\def\Big#1{\mbox{\Large $#1$}} 
\def\BIG#1{\mbox{\Huge $#1$}}

\def\hjb{\hat{\bar{J}}}

\def\gfrakh{\hat{\mathfrak g}}

\def\hc{^\dagger} 
 
\def\one{{\mathchoice {\rm 1\mskip-4mu l} {\rm 1\mskip-4mu} {\rm 1\mskip-4.5mu l} 
{\rm 1\mskip-5mu l}}} 
\def\d{\delta}

\def\nnsrs{n+\srac{n(s)}{\r(\s)}} 
 
\def\nrm{{n(r)\m}}

\def\nrn{{n(r)\n}} 
\def\mnrn{{-n(r),\n}} 
\def\nsn{{n(s)\n}} 
\def\ntd{{n(t)\d}} 
 
\def\nue{{n(u)\epsilon}} 
\def\nvk{{n(v)\kappa}} 
\def\nwl{{n(w)\lambda}}
 
\def\mnnrnsrs{{m+n+\srac{n(r)+n(s)}{\r(\s)}}} 
 
\def\mnrrs{{m+\srac{n(r)}{\r(\s)}}}

\def\nrrs{{\srac{n(r)}{\r(\s)}}}
\def\nrrsf{{\frac{n(r)}{\r(\s)}}}
\def\nsrs{{\srac{n(s)}{\r(\s)}}}
\def\scf{{\cal F}} 
\def\sG{{\cal G}}

\def\gfrak{\mbox{$\mathfrak g$}} 
 
\def\hj{\hat{J}} 
\def\nrn{{n(r)\n}} 
\def\nsn{{n(s)\n}} 
 
\def\schisig{{\foot{\chi(\s)}}} 

\def\ntd{{n(t)\delta}} 
\def\hc{^\dagger} 
\def\st{{\cal T}} 
\def\0b{\ } 
\def\pl{\partial} 
 
\def\Nrm{{N(r)\m}}

\def\Nsn{{N(s)\n}} 
\def\Ntd{{N(t)\d}}

\def\srange{\s=0,\ldots,N_c-1} 
\def\sm{{\cal M}} 
 
\def\ho{{\hat{\Omega}}}

\def\he{{\hat{e}}}
\def\hei{{ \hat{e}^{-1}}}
\def\heb{{\hat{\bar{e}\hspace{.02in}}\hspace{-.02in}}} 
\def\hebi{{ \hat{\bar{e}}^{-1}}}
\def\hx{{\hat{x}}}

\def\hp{\hat{p}}
\def\sh{{\cal H}}
\def\hG{\hat{G}}
\def\hpl{\hat{\pl}}
\def\k{\kappa}
\def\l{\lambda}

\def\su{{ \mathfrak{su} }} 
\def\so{{ \mathfrak{so} }} 
 
\def\bigspc{{ \quad \quad \quad \quad}}
\def\gscfwt{{ \hat{\Delta}_0 (\s)}}
\def\ep{{ \epsilon}}
\def\pom{{\tilde{\omega}}}

\def\hPs{{\hat{\Psi}}}

\def\bnrrs{{ \frac{\bar{n}(r)}{\r(\s)}}}
\def\sbnrrs{{ \srac{\bar{n}(r)}{\r(\s)}}}
\def\Dg{{ \D_{\sgb(\s)}}}
\def\bkspc{{ \!\!\!\!}}
\def\lpl{{ \overleftarrow{\pl}}}
\def\bpl{{ \bar{\pl}}}
\def\nsrsf{{ \frac{n(s)}{\r(\s)} }}
\def\bT{{ \bar{T}}}
\def\nrmu{{\nrm u}}
\def\nsnv{{\nsn v}}

\def\bst{{\bar{\st}}}
\def\ugp{{ \mathfrak{u} }}

\makeatletter 
\renewcommand{\@makefnmark}{\mbox{$^{\ddagger\@thefnmark}$}} 
\renewcommand{\subsection}{\@startsection 
{subsection}{2}{0pt
}{-\baselineskip}{0.5\baselineskip} 
{\normalfont\normalsize\bf}} 
\renewcommand{\section}{\@startsection 
{section}{2}{0pt
}{-\baselineskip}{0.5\baselineskip} 
{\bf\large}} 

\makeatother 
\numberwithin{equation}{section} 
\numberwithin{table}{section}

\newcounter{myfigctr}
\def\myfig#1{\refstepcounter{myfigctr}%
 \label{#1}%
} 
 
\newcommand{\publititle}[8] 
{ 
  \vspace*{-3cm} 
  \begin{flushright} #1 \\ {\tt #2} \end{flushright} 
  \vfill 
  \begin{center}{\Large
    \bfseries #3}\end{center} 
  \vskip 8mm 
  \begin{center}{\large #4}\end{center} 
  \begin{center}{\normalsize #5}\end{center} 
  \vskip 8mm 
  \nopagebreak 
  \noindent #6 
  \vfill 
  \begin{flushleft} #7 
  \end{flushleft} 
  \hrule width 6.7cm \vskip.1mm 
  {\small #8} 
  \thispagestyle{empty} 
  \clearpage 
} 
 
\begin{document} 
 
\publititle{ ${}$ \\ UCB-PTH-04/15 \\ LBNL-55076 \\ hep-th/0406003}{}{The General Twisted Open WZW String
}{M.B.Halpern$^{1a}$ and C. Helfgott$^{1b}$} 
{$^1$ {\em Department of Physics, University of California and \\
Theoretical Physics Group,  Lawrence Berkeley National 
Laboratory \\ 
University of California, Berkeley, California 94720, USA}
\\[2mm]} {We recently studied two large but disjoint classes of twisted open WZW strings: the open-string sectors of the WZW orientation orbifolds and 
the so-called basic class of twisted open WZW strings. In this paper, we discuss {\it all T-dualizations} of the basic class to construct the {\it general} 
twisted open WZW string -- which includes the disjoint classes above as special cases. For the general case, we give the {\it branes} and {\it twisted 
non-commutative geometry} at the classical level and the {\it twisted open-string KZ equations} at the operator level. Many examples of the general 
construction are discussed, including in particular the simple case of twisted free-bosonic open strings. We also revisit the open-string sectors of the 
general WZW orientation orbifold in further detail. For completeness, we finally review the {\it general twisted boundary state equation} which provides a 
complementary description of the general twisted open WZW string.
} {$^a${\tt halpern@physics.berkeley.edu} \\ $^b${\tt helfgott@socrates.berkeley.edu} 
} 
 
\clearpage 
 
\renewcommand{\baselinestretch}{.4}\rm 
{\footnotesize 
\tableofcontents 
} 
\renewcommand{\baselinestretch}{1.2}\rm

\section{Introduction}

At the level of examples, twisted scalar fields and orbifold theory [1-7] are almost as old as string theory itself. It is only in the last few years 
however, that the orbifold program [8-18] has in large part completed the local description of the general {\it closed-string orbifold} $A(H)/H$. The program 
constructs all orbifolds at once, using the {\it principle of local isomorphisms} [8,10-13,15,18] to map the symmetric theory $A(H)$ into the twisted sectors
of $A(H)/H$. The orbifold results are expressed in terms of a set of {\it duality transformations} \cite{Dual,More,Big}, which are discrete Fourier transforms 
constructed from the eigendata of the {\it H-eigenvalue problem} \cite{Dual,More,Big} in the symmetric theory. For the reader interested in particular topics, 
we offer the following list:

\noindent $\bullet$ the twisted current algebras and stress tensors of
all sectors of the general current-algebraic orbifold [8-12],

\noindent $\bullet$ the twisted affine primary fields, twisted operator
algebras and {\it twisted KZ equations} of all WZW orbifolds $A_g(H)/H$ \cite{Big,Big',Perm,so2n},

\noindent $\bullet$ the action formulation of all WZW
and coset orbifolds $A_{g/h}(H)/H$, in terms of {\it group
orbifold elements} with definite monodromy [13-18],

\noindent $\bullet$ the action formulation and {\it twisted Einstein equations}
of a large class of sigma-model orbifolds $A_M(H)/H$ \cite{Geom},

\noindent $\bullet$ free-bosonic avatars of these constructions\footnote{An abelian twisted KZ equation for the inversion orbifold $x\rightarrow -x$ was 
given earlier in Ref.~\cite{Froh}.} on abelian $g$ and 
the explicit form of their twisted vertex operators \cite{Big', Perm,Geom}.

\noindent A pedagogical review of the program is included in Ref.~\cite{so2n}.  Recent progress
at the level of characters has been reported in Refs.~[20,8,21-23].

More recently, the orbifold program has also been applied to {\it twisted open strings}:

\noindent $\bullet$ the WZW orientation orbifolds $A_g(H_-)/H_-$ and their {\it twisted open-string KZ equations} \cite{Orient1,Orient2},

\noindent $\bullet$ the action formulation and twisted Einstein equations of WZW, coset and sigma-model orientation orbifolds $A_M(H_-)/H_-$ \cite{Orient2},

\noindent $\bullet$ the basic class of twisted open WZW strings $A_g^{open}(H)/H$ and their corresponding twisted open-string KZ equations \cite{TwGiu}.

\noindent The {\it orientation orbifolds} are obtained by twisting closed-string theories by a discrete automorphism group $H_-$ which contains {\it world-sheet
orientation-reversing automorphisms}. Like conventional orientifolds [27-30], the orientation orbifolds contain an equal number of 
open- and closed-string sectors, but in distinction to orientifolds, both the open- and closed-string sectors are generically characterized by fractional 
moding. The {\it basic class} of twisted open WZW strings involves a rather different (Hamiltonian) formulation, generalizing the method given in 
Ref.~\cite{Giusto} for untwisted open WZW strings $A_g^{open}$. This class of twisted open strings is apparently disjoint from the open-string sectors of the
WZW orientation orbifolds. 

The method of Refs.~\cite{Giusto,TwGiu} works in the {\it open-string picture} of open WZW strings, which uses only a single set of current modes and Virasoro 
generators. This is in distinction to the {\it closed-string picture} of open strings, which uses both left- and right-mover current modes to describe 
boundary states (see Sec.~5).

More specifically, the method of Ref.~\cite{TwGiu} constructs a twisted open WZW string from the left-mover current modes and Virasoro generators in sector 
$\s$ of each closed-string WZW orbifold $A_g(H)/H$. Each resulting twisted open string in this basic class can be considered as a sector of an {\it open-string
WZW orbifold} $A_g^{open}(H)/H$ (see Fig.~1), and indeed the open-string orbifolds inherit their orbifold structure directly from the underlying closed-string 
WZW orbifolds. Alternately, the open-string orbifolds can be viewed as orbifolds by $H$ of the untwisted $H$-symmetric open WZW string $A_g^{open}(H)$ 
constructed in Ref.~\cite{Giusto}.

\vskip 10pt
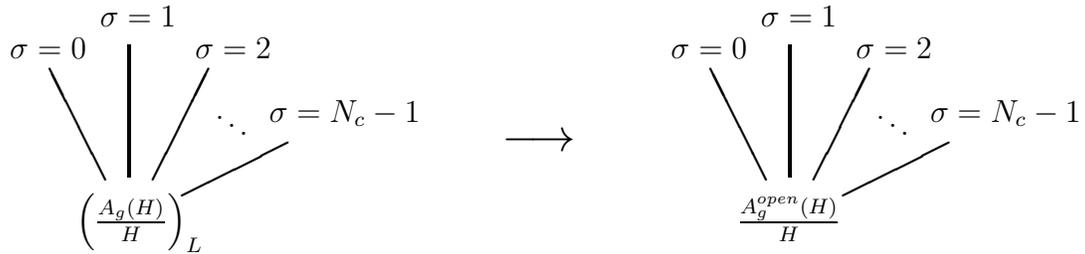
\begin{picture}(380,230)

\thicklines

\put(46,142){\line(-1,2){21}}
\put(55,143){\line(0,1){50}}
\put(64,142){\line(1,2){21}}
\put(75,136){\line(2,1){40}}
\put(35,123){$\left( \frac{A_g(H)}{H} \right)_L $}

\put(10,188){$\s=0$}
\put(44,200){$\s =1$}
\put(80,188){$\s =2$}
\put(87,158){$\ddots$}
\put(108,164){$\s =N_c -1$}

\put(196,153){$\Big{\longrightarrow}$}

\put(296,142){\line(-1,2){21}}
\put(305,143){\line(0,1){50}}
\put(314,142){\line(1,2){21}}
\put(325,136){\line(2,1){40}}
\put(285,123){$\frac{A_g^{open}(H)}{H} $}

\put(260,188){$\s=0$}
\put(294,200){$\s =1$}
\put(330,188){$\s =2$}
\put(337,158){$\ddots$}
\put(358,164){$\s =N_c -1$}

\put(15,83){$\left( \frac{A_g(H)}{H} \right)_L $: The left-mover data of any closed-string WZW orbifold.}
\put(15,56){$\frac{A_g^{open}(H)}{H} $: The corresponding open-string WZW orbifold in the basic class.}
\put(15,38){$N_c$: The number of conjugacy classes of symmetry group $H$ and the }
\put(36,25){number of sectors $\s$ of $A_g(H)/H$ or $A_g^{open}(H)/H$.} 

\put(3,2){Fig.\,\ref{fig:Giu-reln}: Construction of open-string WZW orbifold from closed-string WZW orbifold}

\end{picture}
\myfig{fig:Giu-reln}
\vskip 10pt

The present paper extends the Hamiltonian method of Refs.~\cite{Giusto,TwGiu} to construct the {\it general twisted open WZW string} by including {\it all 
T-dualizations} of the open strings in the basic class. This is accomplished for all twisted open WZW strings at once by introducing a mode-independent 
automorphism $\pom (\s)$ of the twisted current algebra in each sector $\s$ of each open-string orbifold (see Fig.~2). Then, each sector $\s$ has a unique, 
{\it $\pom$-independent Hamiltonian} which describes an entire {\it $\pom$-family of T-dual twisted open strings}: At the classical level, we insert the 
automorphism into the phase-space realization of the twisted currents, which induces $\pom$-dependent {\it non-commutative geometry} and {\it general}
$\pom$-dependent {\it WZW branes}. At the operator level, we find $\pom$-independent {\it spectra} and the {\it general} $\pom$-dependent {\it open-string 
KZ equations}. 

\vskip 10pt
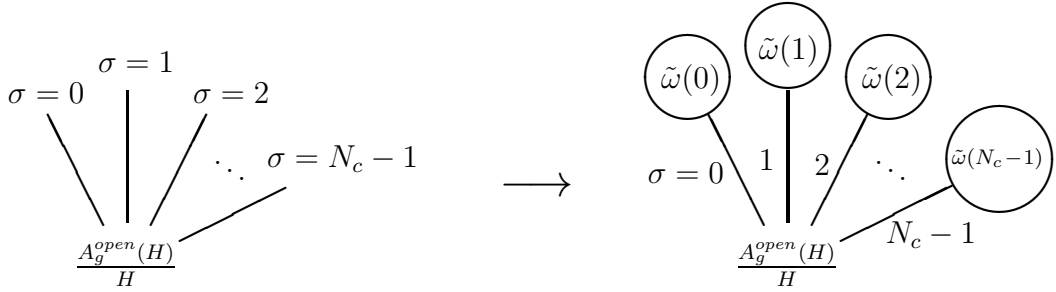
\begin{picture}(380,252)

\thicklines

\put(46,156){\line(-1,2){21}}
\put(55,157){\line(0,1){50}}
\put(64,156){\line(1,2){21}}
\put(75,150){\line(2,1){40}}
\put(35,137){$\frac{A_g^{open}(H)}{H}$}

\put(10,202){$\s=0$}
\put(44,214){$\s =1$}
\put(80,202){$\s =2$}
\put(87,172){$\ddots$}
\put(108,178){$\s =N_c -1$}

\put(196,167){$\Big{\longrightarrow}$}

\put(296,156){\line(-1,2){21}}
\put(305,157){\line(0,1){50}}
\put(314,156){\line(1,2){21}}
\put(325,150){\line(2,1){42}}
\put(285,137){$\frac{A_g^{open}(H)}{H}$}

\put(252,172){$\s =0$}
\put(294,177){$1$}
\put(315,175){$2$}
\put(342,149){$N_c-1$}

\put(267,212){\circle{30}}
\put(257,208){$\pom (0)$}
\put(305,224){\circle{30}}
\put(294,219){$\pom (1)$}
\put(343,212){\circle{30}}
\put(333,208){$\pom (2)$}
\put(337,172){$\ddots$}
\put(385,181){\circle{65}}
\put(367,179){$\scrs{\pom (N_c \!-\!1)}$}

\put(30,97){$\frac{A_g^{open}(H)}{H}$: The data of any open-string WZW orbifold.}
\put(30,72){$\pom(\s)$: An automorphism of the twisted current algebra in sector $\s$}
\put(61,59){of $A_g^{open}(H)/H$. For each sector $\s$, one obtains an $\pom$-family of}
\put(61,46){T-dual twisted open WZW strings.}

\put(12,22){Fig.\,\ref{fig:Giu-reln2}: From open-string WZW orbifold to T-dual twisted open WZW strings}

\end{picture}
\myfig{fig:Giu-reln2}
\vskip 10pt

Because our construction is immense, many explicit examples of our general results are discussed in the text, and for an overview, the reader may enjoy an 
early glance at Subsec.~$3.4$. Among these examples, special attention is given to: a) untwisted open WZW strings, b) twisted free-bosonic open strings and 
c) the open-string sectors of the WZW orientation orbifolds. In particular, we find that the open-string sectors of the WZW orientation orbifolds are T-dual 
to the sectors of the generalized open-string $\Zint_2$ permutation orbifolds (see Subsec.~$4.2$) in the basic class. Finally, Sec.~5 contains a review 
and examples of the {\it general twisted boundary state equation} \cite{Big,TwGiu}, which provides a complementary description of the general twisted open 
WZW string.

The present paper should be considered as a companion paper to Ref.~\cite{TwGiu}. In particular, we assume for brevity a familiarity with the various 
eigenvalue problems and duality transformations given there, and many of our results are derived by a so-called $\pom$-map (see Subsec.~$2.5$) from the 
analogous results in Ref.~\cite{TwGiu}.

\section{Classical Theory of the General Twisted Open WZW String}

\subsection{Twisted Open-String Current Algebras}

We begin our construction with the equal-time bracket\footnote{As in Refs.~\cite{Giusto,TwGiu}, our brackets are rescaled by an $i$, so that the
quantum correspondence is $\{ \hat{A},\hat{B} \} \rightarrow [ \hat{A} ,\hat{B}]$.} algebra of the twisted strip currents given in Ref.~\cite{TwGiu}:
\begin{subequations}
\label{EqT-Strip-CA}
\begin{gather}
\hj^{(\pm)}_\nrm (\xi,t,\s) \equiv \hj_\nrm (\pm \xi,t,\s) = \sum_{m\in \Zint} \hj_\nrm (m\!+\!\nrrs) e^{-i(m+ \nrrsf) (t\pm \xi)} ,\quad 0\leq \xi \leq \pi
\label{Eq 2.1a} \\
\hj^{(\pm )}_\nrm (-\xi,t,\s) =\hj^{(\mp )}_\nrm (\xi,t,\s) ,\quad \pl_\mp \hj^{(\pm)}_\nrm (\xi,t,\s) =0 ,\quad \pl_\pm \equiv \pl_t \pm \pl_\xi \\
\{ \hj_\nrm^{(+)} (\xi,t,\s) ,\hj_\nsn^{(+)} (\eta,t,\s) \} = 2\pi i \Big{(} \scf_{\nrm;\nsn}{}^{n(r)+n(s),\de} (\s)
\hj_{n(r)+n(s) ,\de}^{(+)} (\eta,t,\s) \quad \nn \\
\bigspc \bigspc \quad \quad +\de_{n(r)+n(s) ,0\, \text{mod } \r(\s)} \sG_{\nrm;\mnrn} (\s) \pl_\xi \Big{)} \de_{\nrrsf} (\xi -\eta) \\
\{ \hj_\nrm^{(+)} (\xi,t,\s) ,\hj_\nsn^{(-)} (\eta,t,\s) \} = 2\pi i \Big{(} \scf_{\nrm;\nsn}{}^{n(r)+n(s),\de} (\s)
\hj_{n(r)+n(s) ,\de}^{(-)} (\eta,t,\s) \quad \nn \\
\bigspc \bigspc \quad \quad +\de_{n(r)+n(s) ,0\, \text{mod } \r(\s)} \sG_{\nrm;\mnrn} (\s) \pl_\xi \Big{)} \de_\nrrsf (\xi +\eta) \label{J+J-} \\
\{ \hj_\nrm^{(-)} (\xi,t,\s) ,\hj_\nsn^{(-)} (\eta,t,\s) \} = 2\pi i \Big{(} \scf_{\nrm;\nsn}{}^{n(r)+n(s),\de} (\s)
\hj_{n(r)+n(s) ,\de}^{(-)} (\eta,t,\s) \quad \nn \\
\bigspc \bigspc \quad \quad -\de_{n(r)+n(s) ,0\, \text{mod } \r(\s)} \sG_{\nrm;\mnrn} (\s) \pl_\xi \Big{)} \de_{-\nrrsf} (\xi -\eta)  \label{J-J-} \\
\bigspc \s =0,\ldots ,N_c -1 \,.
\end{gather}
\end{subequations}
We emphasize that, as it must be for open strings, both the left- and the right-mover currents $\hj^{(\pm )}$ are constructed from the {\it same} set 
of twisted left-mover current modes. These current modes, which satisfy the twisted affine Lie algebra $\gfrakh (\s)$, are appropriated from sector $\s$ of 
any closed-string WZW orbifold $A_g(H)/H$. The number of sectors $N_c$ of $A_g(H)/H$ is the number of conjugacy classes of $H$. The fractional moding of 
the currents is described by the integers $\r(\s)$ and $n(r)$, where $\r(\s)$ is the order of the automorphism $h_\s \in H$ and $\{n(r) \}$ are the spectral 
indices of the $H$-eigenvalue problem \cite{Dual,More,Big} in $A_g(H)$. All quantities in the orbifold program are periodic $n(r) \rightarrow n(r)\pm \r(\s)$ 
in the spectral indices $n(r)$. The $H$-eigenvalue problem also leads to explicit formulae for the {\it duality transformations} \cite{Dual,More,Big} 
of orbifold theory, including the twisted tangent-space metric $\sG (\s)$ and twisted structure constants $\scf (\s)$ which appear in 
Eq.~\eqref{EqT-Strip-CA}. Other important quantities in Eq.~\eqref{EqT-Strip-CA} are the {\it phase-modified Dirac delta functions}
\begin{equation}
\de_\nrrsf (\xi\pm \eta) = \frac{1}{2\pi} \sum_{m\in \Zint} e^{-i (\mnrrs)(\xi \pm\eta)} = e^{-i \nrrs (\xi \pm\eta)} \de (\xi \pm\eta) =\de_{\frac{n(r)\pm 
  \r(\s)}{\r(\s)}} (\xi \pm \eta) \label{Eq2.2}
\end{equation}
which have been previously discussed in Refs.~\cite{Geom,TwGiu}. We note in particular that $\de_{n(r)/\r(\s)} (\xi \!+\!\eta)$ has support only at the {\it 
boundaries} of the strip $\xi \!=\!\eta \!=\!0,\pi$ -- and the appearance of this delta function reflects the interaction between a non-abelian charge at 
$\xi$ (or $\eta$) with a non-abelian {\it image charge} \cite{Giusto,TwGiu} at $-\eta$ (or $-\xi$). In distinction to closed-string orbifold theory, the 
phase modification of $\de_{n(r)/\r(\s)} (\xi-\eta)$ can be ignored on the strip
\begin{gather}
\de_{\nrrsf} (\xi -\eta) \equiv \de (\xi -\eta) \text{   when } 0\leq \xi ,\eta \leq \pi
\end{gather}
although for symmetry we often leave these phases in our results. Finally, we note the {\it boundary conditions} on the twisted currents
\begin{gather}
\hj_\nrm^{(+)} (0,t,\s) =\hj_\nrm^{(-)} (0,t,\s) ,\quad \hj_\nrm^{(+)} (\pi,t,\s) = e^{-\tp \nrrs} \hj_\nrm^{(-)} (\pi,t,\s) \label{hjBCs}
\end{gather}
which follow from their moding in Eq.~\eqref{Eq 2.1a}.

For detailed information on particular classes of closed-string WZW orbifolds, we direct the reader to the following references:

$\bullet$ the WZW permutation orbifolds [8,10,12-14,16]

$\bullet$ the inner-automorphic WZW orbifolds \cite{More,Big, Perm}

$\bullet$ the (outer-automorphic) charge conjugation orbifold on $\su (n\geq 3)$ \cite{Big'}

$\bullet$ the outer-automorphic WZW orbifolds on $\so (2n)$, including the triality orbifolds \linebreak
\indent $\quad$ on $\so(8)$ \cite{so2n}.

\noindent These references give the explicit forms of the twisted operator algebras (and twisted KZ equations), including the duality automorphisms
and current algebras $\gfrakh (\s)$ required in our construction.

\subsection{General Phase-Space Realization of the Twisted Currents}

Our next step is the key ingredient in generalizing the construction of Ref.~\cite{TwGiu}. In particular, for each twisted current algebra $\gfrakh (\s)$, we 
introduce an {\it $\pom$-family} of phase-space 
realizations of the twisted strip currents:
\begin{subequations}
\label{gen-canon-real2}
\begin{align}
\hj_\nrm^{(+)} (\xi) &\equiv 2\pi \he^{-1} (\hx(\xi))_\nrm {}^{\!\ntd} \hp_\ntd (\hb ,\xi) \nn \\
  & \bigspc \bigspc + \frac{1}{2} \pl_\xi \hx_\s^\ntd (\xi) \he (\hx(\xi))_\ntd {}^{\!\nsn} \sG_{\nsn;\nrm} (\s) \label{hj+} \\
\hj_\nrm^{(-)} (\xi) &\equiv \pom (n(r),\s)_\m{}^\k \Big{(} 2\pi \heb^{-1} (\hx(\xi))_{n(r),\k} {}^{\!\ntd} \hp_\ntd (\hb,\xi) \nn \\
& \bigspc \bigspc -\frac{1}{2} \pl_\xi \hx_\s^{\ntd} (\xi) \heb (\hx(\xi))_\ntd {}^{\!\nsn} \sG_{\nsn;n(r),\k} (\s) \Big{)} \label{hj-} \\
& \quad \quad \quad \pom \in Aut (\gfrakh (\s)) ,\quad  0\leq \xi \leq \pi ,\quad \srange \,.
\end{align}
\end{subequations}
These phase-space realizations differ from those of Ref.~\cite{TwGiu} primarily by the matrix $\pom$, which is any mode-number-independent 
{\it automorphism}\footnote{General automorphisms $\pom \in Aut(\gfrakh (\s))$ appeared earlier in the general twisted boundary-state equation 
of Ref.~\cite{TwGiu}, and are well-known in this context for untwisted open strings (see e.~g.~Ref.~\cite{FS}) -- where $\pom =\omega \in Aut(g)$ is an 
automorphism of the untwisted current algebra. The description of T-duality via the insertion of an automorphism in the phase-space realization of the currents
was first proposed in Ref.~\cite{Giusto}. The simplest example of T-duality is the case of DD and NN strings (see Subsec.~$2.8$), which correspond to 
$\omega =1$ and $-1$ respectively in the untwisted abelian limit of Eq.~\eqref{gen-canon-real2}.} {\it of the twisted current algebra}:
\begin{subequations}
\begin{gather}
\pom (n(r),\s)_\m{}^\k \pom (n(s),\s)_\n{}^\l \sG_{n(r),\k ;n(s),\l}(\s) =\sG_{\nrm;\nsn}(\s) \label{wwG=G} \\
\pom (n(r),\s)_\m{}^\k \pom (n(s),\s)_\n{}^\l \scf_{n(r),\k ;n(s),\l}{}^\ntd (\s) =\scf_{\nrm;\nsn}{}^{n(t),\ep} (\s) \pom (n(t),\s)_\ep{}^\de \,.\label{wwF=Fw}
\end{gather}
\end{subequations}
The special case $\pom =1$ describes the {\it basic class} studied in Ref.~\cite{TwGiu}, and, at any point in the construction below, the results of that 
reference can be obtained by setting $\pom =1$. In what follows, we shall see that all the realizations in the $\pom$-family are in fact related to each 
other by T-duality, and we therefore refer to $\pom$ as the {\it T-duality automorphism}.

Aside from the T-duality automorphism $\pom$, the geometric quantities which appear in the phase-space realization \eqref{gen-canon-real2} are defined as 
follows:
\begin{subequations}
\label{Field-Defns}
\begin{gather}
\hg (\st,\xi,t,\s) = e^{i\hx_\s^\nrm (\xi,t) \st_\nrm (T,\s)} \label{hg=eixt} \\
\widehat{Tr} \left( \sm (\st,\s) \st_\nrm (T,\s) \st_\nsn (T,\s) \right) =\sG_{\nrm;\nsn}(\s) \bigspc \quad \quad \quad \nn \\
  \bigspc \quad \quad =\de_{n(r)+n(s),0\,\text{mod }\r(\s)} \sG_{\nrm;\mnrn}(\s) \label{Eq 2.3a} \\
[\st_\nrm(T,\s) ,\st_\nsn(T,\s)] =i\scf_{\nrm;\nsn}{}^{n(r)+n(s),\de}(\s) \st_{n(r)+n(s),\de}(T,\s) \label{Eq 2.3b} 
\end{gather}
\begin{gather}
\he_\nrm (\st) \equiv -i \hg^{-1} (\st) \hpl_\nrm \hg (\st) \equiv \he_\nrm {}^\nsn \st_\nsn ,\quad \hpl_\nrm (\xi) \equiv
   \srac{\pl}{\pl \hx^\nrm (\xi)} \label{l-i-viel} \\
\heb_\nrm (\st) \equiv -i\hg (\st) \hpl_\nrm \hg^{-1} (\st) \equiv \heb_\nrm {}^\nsn \st_\nsn \label{r-i-viel} \\
\he_\nrm{}^\ntd \he^{-1}_\ntd {}^\nsn = \heb_\nrm{}^\ntd \heb^{-1}_\ntd {}^\nsn = \de_\nrm{}^\nsn = \de_\m^\n \de_{n(r)-n(s),0\, \text{mod } \r(\s)} \\
\hat{H}_{\nrm;\nsn;\ntd} (\hx) \equiv \hpl_\nrm \hb_{\nsn;\ntd} (\hx) \!+\! \hpl_\nsn \hb_{\ntd;\nrm} (\hx) \!+\! \hpl_\ntd \hb_{\nrm;\nsn} (\hx) \nn \\
= -i \widehat{Tr} \left( \sm (\st,\s) \he_\nrm (\st,\hx) [\he_\nsn (\st,\hx) ,\he_\ntd (\st,\hx)] \right) \\
\quad \quad \quad= \he (\hx)_\nrm{}^{n(r')\m'} \he(\hx)_\nsn {}^{n(s')\n'} \he(\hx)_\ntd {}^{n(t')\de'} \scf_{n(r')\m';n(s')\n' ;n(t')\de'} (\s) \label{H-Form3}
\end{gather}
\begin{gather}
\hp_\nrm (\hb ,\xi) \equiv \hp^\s_\nrm + \frac{1}{4\pi} \hb_{\nrm;\nsn} (\hx(\xi)) \pl_\xi \hx_\s^\nsn (\xi) \,.
\end{gather}
\end{subequations}
Here $\hg$ is the {\it group orbifold element}, $\he,\heb$ are the left- and right-invariant twisted vielbeins (with $\he (0)=1$), and $\hat{H}$ is the 
twisted torsion field. The twisted coordinates $\hx_\s$ and momenta $\hp^\s$ are expected to be canonical in the bulk $0<\xi <\pi$. The twisted 
representation matrices $\st$ satisfy the orbifold Lie algebra \eqref{Eq 2.3b} of sector $\s$, and the explicit forms of $\st$ and the twisted 
data matrix $\sm$ are given in Ref.~\cite{Big}. The totally antisymmetric twisted structure constants $\scf_{\bullet} (\s)$ in \eqref{H-Form3} are constructed 
as usual by lowering the last index of $\scf(\s)$ using the twisted metric $\sG_{\bullet} (\s)$. All these geometric quantities are familiar in closed-string
orbifold theory \cite{Geom}, as well as the orientation orbifolds \cite{Orient2} and open-string orbifold theory \cite{TwGiu}.

It is important to note however that the twisted current algebra $\gfrakh (\s)$ and hence the twisted currents $\hj^{(\pm)}$ themselves are $\pom$-independent,
which implies that all the other twisted fields above are {\it implicitly $\pom$-dependent}:
\begin{subequations}
\label{2.7}
\begin{gather}
\hg \equiv \hg(\pom) ,\quad \he \equiv \he (\pom) ,\quad \heb \equiv \heb (\pom) ,\quad \hx \equiv \hx (\pom) ,\quad \hp \equiv \hp (\pom) \\
\hat{H} \equiv \hat{H} (\pom) ,\quad \hB \equiv \hB (\pom) \,.
\end{gather}
\end{subequations}
In what follows, we will construct\footnote{The general phase-space realizations above can in principle be applied to the classical description of {\it any} 
twisted current-algebraic open string (related to orbifolds $A(H)/H$ of general affine-Virasoro constructions [33-39,10,12]). In this more general setting we 
expect that, for $\pom$ to describe a T-duality, it must also be a symmetry of the general twisted inverse inertia tensor ${\cL} \pom \pom = {\cL}$, as shown 
in Eq.~\eqref{Lww=L} for WZW orbifolds.} the general twisted open WZW string for all T-duality automorphisms $\pom$, leaving for future work the classification
of $\pom$ by equivalence classes. Many explicit examples of $\pom$ are found e.~g.~in Subsecs.~$2.8, \,3.4$ and $4.1$.

At any point in the following construction, one may consider the $\s=0$ sector -- which describes the general {\it untwisted} open WZW string:
\begin{subequations}
\label{s=0}
\begin{gather}
\s =0 : \quad n(r) =0 \\
\hj_\nrm^{(\pm)} \,\rightarrow \,J_a^{(\pm)} ,\quad \he_\nrm{}^\nsn ,\heb_\nrm{}^\nsn \,\rightarrow \, e_i{}^a ,\bar{e}_i{}^a ,\quad \hx_\s^\nrm \,
   \rightarrow \, x^i ,\quad \hp^\s_\nrm \,\rightarrow \,p_i \\
\sG \,\rightarrow \, G_{ab} ,\quad \scf \,\rightarrow f_{ab}{}^c  ,\quad \st_\nrm \,\rightarrow \,T_a \\
\pom (n(r),\s)_\m{}^\n \,\rightarrow \omega_a{}^b \in Aut(g) \\
a,i =1\ldots \text{dim }g \,.
\end{gather}
\end{subequations}
In this case, the modes $J_a(m)$ of the currents $J_a^{(\pm)}$ satisfy the affine Lie algebra [40-42,37] associated to $g$ and $T$ is any matrix 
rep of $g$. The special case of untwisted open WZW strings at $\omega =1$ was worked out in Ref.~\cite{Giusto}.

\subsection{The General Classical Hamiltonian and T-Duality}

As emphasized in Ref.~\cite{Giusto}, open-string theories are constructed from a single set of Virasoro generators $\{ L_\s (m)\}$. For the general twisted 
open WZW string, we begin with the following classical open-string stress tensors \cite{TwGiu}
\begin{subequations}
\begin{gather}
\hat{T}_\s^{(\pm )} (\xi,t) =\srac{1}{4\pi} \sG^{\nrm;\mnrn}(\s) \hj_\nrm^{(\pm )} (\xi,t) \hj_\mnrn^{(\pm )} (\xi,t) ,\quad 0\leq \xi \leq \pi \\
\hat{T}_\s^{(\pm )} (-\xi,t) =\hat{T}_\s^{(\mp )} (\xi,t) ,\quad \pl_\mp \hat{T}_\s^{(\pm )} (\xi,t) =0 \\
\hat{T}_\s^{(\pm )} (\xi,t) = \srac{1}{2\pi} \sum_{m\in \Zint} L_\s (m) e^{-im(t \pm \xi)}
\end{gather}
\end{subequations}
which, like the currents $\hj^{(\pm)}$, are appropriated from left-mover sector $\s$ of the underlying WZW orbifold $A_g(H)/H$. 
In terms of these stress tensors, the classical Hamiltonian of the twisted open WZW string is:
\begin{subequations}
\label{cl-Hamil}
\begin{gather}
\hat{H}_\s = L_\s (0) =\int_0^\pi \!\!d\xi (\hat{T}_\s^{(+)} (\xi,t) +\hat{T}_\s^{(-)} (\xi,t)) \bigspc \bigspc \bigspc \bigspc \\
=\srac{1}{4\pi} \!\int_0^\pi \!\!d\xi \sG^{\nrm;\mnrn}(\s) \left( \hj^{(+)}_\nrm (\xi,t) \hj^{(+)}_\mnrn (\xi,t) + \hj^{(-)}_\nrm (\xi,t) 
   \hj^{(-)}_\mnrn (\xi,t) \right)  \\
\pl_t \hat{A} = i\{ \hat{H}_\s ,\hat{A} \} ,\quad \srange \,.
\end{gather}
\end{subequations}
Like the currents, the stress tensors and hence the Hamiltonian are explicitly $\pom$-independent. Since each member of the $\pom$-family 
\eqref{gen-canon-real2} of phase-space realizations has the same Hamiltonian \eqref{cl-Hamil}, it follows that each realization is T-dual to all the others 
in its $\pom$-family.

With the phase-space realization \eqref{gen-canon-real2}, the general classical Hamiltonian can be reexpressed in terms of the phase-space variables as follows
\begin{subequations}
\label{canon-Hamil}
\begin{gather}
\hat{H}_\s =\int_0^\pi \!\!d\xi \hat{\sh}_\s (\hx (\xi) ,\hp (\xi)) \\
\hat{\sh}_\s (\hx(\xi),\hp(\xi)) \!=\! \hat{T}_\s^{(+)} (\xi,t) +\hat{T}_\s^{(-)} (\xi,t) \bigspc \bigspc \bigspc \bigspc  \nn \\
\quad = 2\pi \hG^{\nrm;\nsn} (\hx) \hp_\nrm (\hb) \hp_\nsn (\hb) \!+\! \frac{1}{8\pi} \pl_\xi \hx_\s^\nrm \pl_\xi \hx_\s^\nsn \hG_{\nrm;\nsn} (\hx) \label{Hamil-dens} 
\end{gather}
\begin{gather}
\hG_{\nrm;\nsn} (\hx) \!\equiv \!\he (\hx)_\nrm{}^{\!\!\ntd} \he (\hx)_\nsn{}^{\!\!\nue} \sG_{\ntd;\nue} (\s) \bigspc \nn \\
\quad \quad =\! \heb (\hx)_\nrm{}^{\!\!\ntd} \heb (\hx)_\nsn{}^{\!\!\nue} \sG_{\ntd;\nue} (\s) \\
\hG^{\nrm;\nsn} (\hx) \equiv \sG^{\ntd;\nue} (\s) \hei_\ntd {}^{\!\nrm} \hei_\nue {}^{\!\nsn} \\
 \hG_{\nrm;\ntd} (\hx) \hG^{\ntd;\nsn} (\hx) = \de_\nrm {}^\nsn  \label{hG^}
\end{gather}
\end{subequations}
where $\hG_{\bullet}$ and $\hG^{\bullet}$ are respectively the twisted Einstein metric of sector $\s$ and its inverse. To obtain this result, we have used the 
automorphic invariance \eqref{wwG=G} of the twisted tangent-space metric $\sG(\s)$ to cancel all the explicit $\pom$-dependence, and we note that the 
Hamiltonian density $\hat{\sh}_\s$ has the same phase-space form as that found in closed-string WZW orbifold theory \cite{Geom}. We remind the reader however
that all the phase-space variables which appear here are implicitly $\pom$-dependent. 

For the discussion below, it will be convenient to define the automorphic transform $\heb_\pom$ of the right-invariant twisted vielbein
\begin{subequations}
\begin{gather}
\heb_\pom (\hx)_\nrm{}^\nsn \equiv \heb (\hx)_\nrm{}^{n(s),\l} \pom^{-1} (n(s),\s)_\l{}^\n \\
\heb_\pom^{-1} (\hx)_\nrm{}^\nsn = \pom (n(r),\s)_\m{}^\k \heb^{-1} (\hx)_{n(r),\k}{}^\nsn \\
\hG_{\nrm;\nsn}(\hx) = \heb_\pom (\hx)_\nrm{}^\ntd \heb_\pom (\hx)_\nsn{}^\nue \sG_{\ntd;\nue}(\s) \label{Eq 2.7e}
\end{gather}
\end{subequations}
where we have used the automorphic invariance \eqref{wwG=G} to obtain the alternate expression \eqref{Eq 2.7e} for the twisted Einstein metric.
In terms of $\heb_\pom$, the phase-space realization of $\hj^{(-)}$ has the alternate form
\begin{align}
\hj_\nrm^{(-)} (\xi) &\equiv 2\pi \heb_\pom^{-1} (\hx(\xi))_\nrm {}^{\!\ntd} \hp_\ntd (\hb ,\xi) \nn \\
& \bigspc \bigspc - \frac{1}{2} \pl_\xi \hx_\s^\ntd (\xi) \heb_\pom (\hx(\xi))_\ntd {}^{\!\nsn} \sG_{\nsn;\nrm} (\s) \label{hj-_2}
\end{align}
which follows by Eq.~\eqref{wwG=G} from Eq.~\eqref{hj-}. As another example, the twisted-current boundary conditions in Eq.~\eqref{hjBCs} now take the 
$\heb_\pom$-form
\begin{subequations}
\label{Canon-BCs}
\begin{gather}
4\pi \Big{(} \heb_\pom^{-1}(\hx(\xi))_\nrm{}^\nsn - \he^{-1} (\hx(\xi))_\nrm {}^\nsn \Big{)} \hp_\nsn (\hb ,\xi) = \bigspc \bigspc \bigspc \nn \\
\bigspc =\pl_\xi \hx_\s^{\nsn} \Big{(} \heb_\pom (\hx(\xi))_\nsn {}^\ntd + \he (\hx(\xi))_\nsn {}^\ntd \Big{)} \sG_{\ntd;\nrm} (\s) \text{  at } \xi=0 \\
4\pi \Big{(} e^{-\tp\nrrs} \heb_\pom (\hx(\xi))^{-1}_\nrm {}^\nsn - \he (\hx(\xi))^{-1}_\nrm {}^\nsn \Big{)} \hp_\nsn (\hb ,\xi) =\bigspc \bigspc \bigspc \nn \\
=\pl_\xi \hx_\s^\nsn \Big{(} e^{-\tp\nrrs} \heb_\pom (\hx(\xi))_\nsn{}^\ntd + \he (\hx(\xi))_\nsn{}^\ntd \Big{)} \sG_{\ntd;\nrm} (\s) \text{  at } \xi=\pi
\end{gather}
\end{subequations}
where we have used Eqs.~\eqref{hj+} and \eqref{hj-_2}. For reference below, we observe that Eqs.~\eqref{hj-_2} and \eqref{Canon-BCs} differ from the
corresponding results in Ref.~\cite{TwGiu} only by the substitution $\heb \rightarrow \heb_\pom$.

\subsection{The Brackets of the Currents with the Coordinates}

In the method of Refs.~\cite{Giusto,TwGiu}, phase-space brackets are obtained by solving linear inhomogeneous partial differential equations derived from 
the current algebra and the phase-space realizations of the currents.

In this approach, the following tools are needed: We first give the so-called {\it inverse relations}
\begin{subequations}
\label{Inv-Relns}
\begin{gather}
\pl_\xi \hx_\s^\nrm (\xi) = \hj_\nsn^{(+)} (\xi) \sG^{\nsn;\ntd} (\s) \he^{-1} (\xi)_\ntd{}^\nrm \bigspc \bigspc \nn \\
\bigspc \bigspc -\hj_\nsn^{(-)} (\xi) \sG^{\nsn;\ntd} (\s) \heb_\pom^{-1} (\xi)_\ntd{}^\nrm  \label{x=fn(J)} \\
\hp_\nrm (\hb,\xi) = \frac{1}{4\pi} \left( \he(\xi)_\nrm{}^\nsn \hj_\nsn^{(+)} (\xi) + \heb_\pom (\xi)_\nrm {}^\nsn \hj_\nsn^{(-)} (\xi) \right) \label{hp=he-hJ}
\end{gather}
\end{subequations}
which follow for all $0\leq \xi \leq \pi$ from the phase-space realizations in Eqs.~\eqref{gen-canon-real2} and \eqref{hj-_2}. We will also need 
the spatial derivative of the group orbifold element $\hg$
\begin{subequations}
\label{plxi-hg}
\begin{gather}
\pl_\xi \hg(\st,\xi) = \pl_\xi \hx_\s^\nrm \hpl_\nrm \hg(\st,\xi) = i \left( \hg(\st,\xi) \hj^{(+)}(\st,\xi) +\hj^{(-)}(\st^\pom ,\xi) \hg(\st,\xi) \right) \label{pl_xi-hg} \\
\hj^{(+)} (\st,\xi) \equiv \hj_\nrm^{(+)} (\xi) \sG^{\nrm;\nsn} (\s) \st_\nsn \label{Mat-hj-Defn} \\
\hj^{(-)} (\st^\pom ,\xi) \equiv \hj_\nrm^{(-)} (\xi) \sG^{\nrm;\nsn}(\s) \st_\nsn^\pom \label{Mat-hj-Defn2} \\
\st^\pom_\nrm \equiv \pom (n(r),\s)_\m{}^\k \st_{n(r),\k} \label{st-pom-Defn} 
\end{gather}
\begin{gather}
\hj_\nrm^{(+)} (\xi) = \widehat{Tr} \left( \sm(\st,\s) \hj^{(+)} (\st,\xi) \st_\nrm \right) \\
\hj_\nrm^{(-)} (\xi) = \widehat{Tr} \left( \sm(\st,\s) \hj^{(-)} (\st^\pom,\xi) \st^\pom_\nrm \right) 
\end{gather}
\end{subequations}
which is obtained via the chain rule from Eqs.~\eqref{Inv-Relns} and the definitions of the vielbeins in Eq.~\eqref{Field-Defns}. This result is expressed in
terms of the matrix currents $\hj^{(+)}(\st)$ and $\hj^{(-)} (\st^\pom)$, where the automorphic transform $\st^\pom$ of the twisted representation matrix
is defined in Eq.~\eqref{st-pom-Defn}. This automorphic transform satisfies the same orbifold Lie algebra \eqref{Eq 2.3b} as the original $\st$
\begin{subequations}
\begin{gather}
[\st_\nrm^\pom ,\st_\nsn^\pom ] =i\scf_{\nrm;\nsn}{}^\ntd (\s) \st_\ntd^\pom \\
\widehat{Tr} (\sm \st^\pom_\nrm \st^\pom_\nsn ) = \sG_{\nrm;\nsn}(\s) 
\end{gather}
\end{subequations}
and $\st^\pom$ also appears when we reexpress the right-invariant matrix vielbein
\begin{gather}
\heb_\nrm (\st) =-i\hg (\st) \hpl_\nrm \hg^{-1} (\st) = (\heb_\pom )_\nrm {}^\nsn \st^\pom_\nsn
\end{gather}
in terms of the automorphic transform of the right-invariant vielbein.

Using the inverse relations in Eq.~\eqref{Inv-Relns} and the equal-time twisted current algebra \eqref{EqT-Strip-CA}, we have followed the steps of 
Ref.~\cite{TwGiu} to find partial differential equations for the bracket of $\hj^{(+)}$ with $\hx$
\begin{align}
&\pl_\eta \{ \hj_\nrm^{(+)} (\xi,t,\s), \hx_\s^\nsn (\eta,t) \} = \bigspc \bigspc  \bigspc \quad \quad \nn \\
&\quad =\smal{ \!\tp \left( \scf_{\nrm;\ntd}{}^{\!\!\nue} (\s)
\hj_\nue^{(+)} (\eta) \!+\! \sG_{\nrm;\ntd}(\s) \pl_\xi \right) \times }
\nn \\
& \bigspc \bigspc \bigspc \smal{ \times \de_\nrrsf (\xi \!-\!\eta)
\sG^{\ntd;n(v)\kappa}(\s) \he^{-1} (\eta)_{n(v)\kappa}{}^{\!\!\nsn} } \nn \\
&\quad \quad \smal{ -\tp \left( \scf_{\nrm;\ntd}{}^{\!\!\nue}(\s)
\hj_\nue^{(-)} (\eta) \!+\! \sG_{\nrm;\ntd}(\s) \pl_\xi \right) \times }
\nn \\
& \bigspc \bigspc \bigspc \smal{ \times \de_\nrrsf (\xi \!+\!\eta)
\sG^{\ntd;n(v)\kappa}(\s) \heb^{-1}_\pom (\eta)_{n(v)\kappa}{}^{\!\!\nsn} } \nn
\\
&\quad \quad \smal{ +\{ \hj_\nrm^{(+)} (\xi) ,\hx^\ntd_\s (\eta) \} } \big{(}
\smal{\hpl_\ntd \he^{-1} (\eta)_\nue{}^\nsn \sG^{\nue;n(v) \kappa}(\s)
\hj_{n(v)\kappa}^{(+)} (\eta) } \nn \\
&\bigspc \bigspc \bigspc  \smal{ -\hpl_\ntd \heb^{-1}_\pom (\eta)_\nue{}^\nsn
\sG^{\nue;n(v)\kappa}(\s) \hj_{n(v)\kappa}^{(-)} (\eta) } \big{)}  \label{Eq3.11}
\end{align}
and similarly for $\{ \hj^{(-)} ,\hx \}$ by $\xi \rightarrow -\xi$. We note again that Eq.~\eqref{Eq3.11} differs from Eq.~($3.11$) of Ref.~\cite{TwGiu} 
only by the substitution $\heb \rightarrow \heb_\pom$. The natural solution to this set of PDEs is the $\heb \rightarrow \heb_\pom$ form of the 
corresponding solution in Ref.~\cite{TwGiu}:
\begin{subequations}
\label{hJ-hx-Soln}
\begin{gather}
\{ \hj_\nrm^{(+)} (\xi,t,\s) ,\hx_\s^\nsn (\eta,t)\} =-\tp \Big{(} \hei (\eta)_\nrm{}^\nsn
  \de_{\nrrsf} (\xi -\eta) \bigspc \bigspc \quad \nn \\
\bigspc \bigspc \bigspc \quad \quad + \heb_\pom^{-1} (\eta)_\nrm{}^\nsn \de_{\nrrsf} (\xi+\eta) \Big{)} \\
\{ \hj_\nrm^{(-)} (\xi,t,\s) ,\hx_\s^\nsn (\eta,t)\} =-\tp \Big{(} \heb_\pom^{-1} (\eta)_\nrm{}^\nsn
  \de_{-\nrrsf} (\xi -\eta) \bigspc \bigspc \quad \nn \\
\bigspc \bigspc \bigspc \bigspc + \hei (\eta)_\nrm{}^\nsn \de_{-\nrrsf} (\xi+\eta) \Big{)} \\
0\leq \xi ,\eta \leq \pi ,\quad \srange \,.
\end{gather}
\end{subequations}
It is not difficult to check that these brackets satisfy the twisted current boundary conditions in Eq.~\eqref{hjBCs}, but considerable algebra is 
necessary to verify explicitly that \eqref{hJ-hx-Soln} solves the differential equations. In particular, one needs the phase-space 
realizations \eqref{gen-canon-real2}, \eqref{hj-_2} of the twisted currents, as well as the following geometric identities:
\begin{subequations}
\label{Misc-ho-Ids}
\begin{gather}
\heb_\pom (\xi)_\nrm{}^\nsn = -\he (\xi)_\nrm{}^\ntd \ho_\pom (\xi)_\ntd{}^\nsn \label{ho-pom-Defn} \\
\hei (\xi)_\nrm{}^\nue \hpl_\nue \ho_\pom (\xi)_\nsn{}^\ntd = -\scf_{\nrm;\nsn}{}^\nue (\s) \ho_\pom (\xi)_\nue{}^\ntd \\
\heb_\pom^{-1}(\xi)_\nrm{}^\nue \hpl_\nue \ho_\pom (\xi)_\nsn{}^\ntd = -\ho_\pom(\xi)_\nsn{}^\nue \scf_{\nue;\nrm}{}^\ntd (\s) 
\end{gather}
\begin{gather}
\ho_\pom (\xi)_\nrm{}^\ntd \ho_\pom (\xi)_\nsn{}^\nue \sG_{\ntd;\nue} (\s) = \sG_{\nrm;\nsn} (\s) \\
\ho_\pom (\xi)_\nrm{}^\nue \ho_\pom (\xi)_\nsn{}^\nvk \scf_{\nue;\nvk}{}^\ntd (\s) = \scf_{\nrm;\nsn}{}^\nue (\s) \ho_\pom (\xi)_\nue{}^\ntd \\
\heb_\pom (\xi)_\nrm{}^\nue \heb_\pom (\xi)_\nsn{}^\nvk \scf_{\nue;\nvk}{}^\nwl (\s) \heb_\pom^{-1} (\xi)_\nwl{}^\ntd \bigspc \bigspc \nn \\
  \bigspc =-\he_\nrm{}^\nue \he_\nsn{}^\nvk \scf_{\nue;\nvk}{}^\nwl (\s) \hei_\nwl{}^\ntd 
\end{gather}
\begin{gather}
\hpl_\nrm \he_\nsn{}^\ntd - \hpl_\nsn \he_\nrm{}^\ntd = \he_\nrm{}^\nue \he_\nsn{}^\nvk \scf_{\nue;\nvk}{}^\ntd (\s) \label{CM} \\
\hei_\nrm{}^\ntd \hpl_\ntd \hei_\nsn{}^\nue - \hei_\nsn{}^\ntd \hpl_\ntd \hei_\nrm{}^\nue = \scf_{\nsn;\nrm}{}^\ntd (\s) \hei_\ntd{}^\nue \,. \label{Inv-CM}
\end{gather}
\end{subequations}
Here we have introduced the automorphic transform $\ho_\pom$ of the twisted adjoint action $\ho$: 
\begin{subequations}
\begin{gather}
\hg (\st,\xi,t,\s) \st_\nrm \hg^{-1} (\st,\xi,t,\s) \equiv \ho (\xi)_\nrm{}^\nsn \st_\nsn = \ho_\pom (\xi)_\nrm{}^\nsn \st_\nsn^\pom \\
\ho_\pom (\xi)_\nrm{}^\nsn \equiv \ho (\xi)_\nrm{}^{n(s),\l} \pom^{-1} (n(s),\s)_\l{}^\n \,.
\end{gather}
\end{subequations}
The relations in \eqref{CM}, \eqref{Inv-CM} are respectively the twisted Cartan-Maurer and inverse twisted Cartan-Maurer identities (the CM identities 
also hold for $\he \rightarrow \heb_\pom$). Geometric identities analogous to those above are well-known \cite{Geom} in the geometry of ordinary WZW orbifolds. 

Finally, we give the $\hj \hg$ brackets 
\begin{subequations}
\label{Eq2.18}
\begin{align}
\{ \hj_\nrm^{(+)} (\xi,t,\s) ,\hg(\st,\eta,t,\s) \} =& \, 2\pi \Big{(} \hg(\st,\eta) \st_\nrm \de_\nrrsf (\xi \!-\!\eta) \bigspc \nn \\
   &\bigspc  - \st_\nrm^\pom \hg(\st,\eta) \de_\nrrsf (\xi \!+\!\eta) \Big{)} \\
\{ \hj_\nrm^{(-)} (\xi,t,\s) ,\hg(\st,\eta,t,\s) \} =& \,2\pi \Big{(} -\st_\nrm^\pom \hg(\st,\eta) \de_{-\nrrsf} (\xi \!-\!\eta) \bigspc  \nn \\
   &\bigspc  +\hg(\st,\eta) \st_\nrm \de_{-\nrrsf} (\xi \!+\!\eta) \Big{)} \\
\bigspc \bigspc \quad \quad \quad &\srange 
\end{align}
\end{subequations}
which follow from Eq.~\eqref{hJ-hx-Soln} and the chain rule. Being linear in $\hg$, these brackets are easily quantized and will play a central role in
the operator theory of Sec.~3.

As in the twisted current algebra \eqref{EqT-Strip-CA}, the boundary terms proportional to $\de_{\pm n(r)/\r(\s)} (\xi+\eta)$ in Eqs.~\eqref{hJ-hx-Soln} and 
\eqref{Eq2.18} can be interpreted \cite{Giusto,Orient1,TwGiu} as the interaction of a non-abelian charge at $\xi$ (or $\eta$) with a non-abelian image charge 
at $-\eta$ (or $-\xi$).

\subsection{The $\pom$-Map}

Although it was not emphasized in Refs.~\cite{Giusto} and \cite{TwGiu}, all the classical open-string results of these two papers are determined by their 
Hamiltonian formulations, as given above.

This is an important point because all the results above (including in particular the $\pom$-family of phase-space realizations \eqref{gen-canon-real2}
and the geometric identities \eqref{Misc-ho-Ids}) can be obtained from the corresponding results of the basic class \cite{TwGiu} at $\pom =1$ by the following 
simple ``$\pom$-map"
\begin{subequations}
\label{Eq2.17}
\begin{gather}
\hj_\nrm^{(+)} \rightarrow \hj_\nrm^{(+)} ,\quad \hj^{(+)}(\st) \rightarrow \hj^{(+)} (\st) ,\quad \he \rightarrow \he ,\quad \hg \st \rightarrow \hg \st \quad
\end{gather}
\begin{gather} 
\quad \hj_\nrm^{(-)} \rightarrow \hj_\nrm^{(-)} ,\quad \hj^{(-)}(\st) \rightarrow \hj^{(-)} (\st^{\pom}) ,\quad \heb \rightarrow \heb_\pom 
    ,\quad \st \hg \rightarrow \st^\pom \hg \\
 \ho \rightarrow \ho_\pom 
\end{gather}
\end{subequations}
where $\pom \in Aut(\gfrakh (\s))$ is the T-duality automorphism. The $\pom$-map may therefore be applied to the remaining classical results of 
Ref.~\cite{TwGiu} as a {\it shortcut} to complete the classical theory of the general twisted open WZW string. Note that one replaces the twisted 
representation matrix $\st$ by its automorphic transform $\st^\pom$ (see Eq.~\eqref{st-pom-Defn}) only when $\st$ appears in the right-mover matrix current 
and when it acts on the left of $\hg$.

For brevity, the results below are generally obtained from the $\pom$-map, omitting details of the corresponding direct but lengthy computations.

\subsection{Coordinate Space and Twisted Non-Commutative Geometry}

The first step in the coordinate-space formulation of our theory is to solve for the twisted momenta in terms of the time derivatives 
of the twisted coordinates:
\begin{subequations}
\label{Coord-Spc}
\begin{align}
\!\!\!\pl_t \hx_\s^\nrm (\xi,t) &= i\{ \hat{H}_\s ,\hx_\s^\nrm (\xi,t) \} \bigspc \bigspc \bigspc \label{[H,x]} \\
&= \sG^{\nsn;\ntd} (\s) \Big{(} \hei (\xi)_\nsn{}^{\!\!\nrm} \hj_\ntd^{(+)} (\xi) + \!\heb^{-1}_\pom (\xi)_\nsn{}^{\!\!\nrm} \hj_\ntd^{(-)} (\xi) \Big{)} \quad \label{plt-hx1} \\
&= 4\pi \hG^{\nrm;\nsn} (\hx(\xi,t)) \hp_\nsn (\hB ,\xi,t)  \label{plt-hx} \\
\!\!\!\hp_\nrm (\hB,\xi,t) &= \frac{1}{4\pi} \hG_{\nrm;\nsn} (\hx(\xi,t)) \pl_t \hx_\s^\nsn (\xi,t) \,. \label{p=plt-hx-G}
\end{align}
\end{subequations}
This result can be obtained by evaluating the bracket in Eq.~\eqref{[H,x]} or via the $\pom$-map from the $\pom =1$ result in Eq.~$(3.18)$ of
Ref.~\cite{TwGiu}. Similarly, by chain rule from Eq.~\eqref{plt-hx1}, one obtains the classical open-string equations of motion
\begin{subequations}
\label{cl-EOM}
\begin{gather}
\pl_+ \hg(\st,\xi,t,\s) =2i\hg (\st,\xi,t,\s) \hj^{(+)} (\st,\xi,t,\s) \bigspc \\
\pl_- \hg(\st,\xi,t,\s) =-2i \hj^{(-)} (\st^\pom ,\xi,t,\s) \hg(\st ,\xi,t,\s) ,\quad \pl_\pm \equiv \pl_t \pm \pl_\xi
\end{gather}
\begin{gather}
\hj^{(+)} (\st,\xi,t,\s) = -\srac{i}{2} \hg^{-1} (\st,\xi,t,\s) \pl_+ \hg(\st,\xi,t,\s) \\
\hj^{(-)} (\st^\pom ,\xi,t,\s) = -\srac{i}{2} \hg(\st,\xi,t,\s) \pl_- \hg^{-1} (\st,\xi,t,\s) \\
\pl_- \left( \hg^{-1} (\st) \pl_+ \hg(\st) \right) = \pl_+ \left( \hg(\st) \pl_- \hg^{-1} (\st) \right) =0
\end{gather}
\end{subequations}
in terms of the group orbifold elements. Furthermore, we may use Eq.~\eqref{p=plt-hx-G} to eliminate the momenta and obtain a variety of other
coordinate-space results.

As a first example, we give the coordinate-space form of the twisted currents
\begin{subequations}
\label{Eq2.23}
\begin{gather}
\hj_\nrm^{(+)} (\xi,t,\s) =\srac{1}{2} \pl_+ \hx_\s^\nsn (\xi,t) \he (\xi,t)_\nsn{}^\ntd \sG_{\ntd;\nrm}(\s) \\
\hj_\nrm^{(-)} (\xi,t,\s) =\srac{1}{2} \pl_- \hx_\s^\nsn (\xi,t) \heb_\pom (\xi,t)_\nsn{}^\ntd \sG_{\ntd;\nrm}(\s) 
\end{gather}
\end{subequations}
which follows from the phase-space realizations. The relation between the matrix currents in Eq.~\eqref{cl-EOM} and the index currents in Eq.~\eqref{Eq2.23}
is given in Eq.~\eqref{plxi-hg}.

As a second example, the {\it bulk Lagrange density} of the general twisted open WZW string
\begin{subequations}
\label{LDens}
\begin{align}
& \hat{{\cL}}_\s \equiv \pl_t \hx_\s^\nrm \hp_\nrm - \hat{\sh}_\s \nn \\ 
& \bigspc = \frac{1}{8\pi} (\hG_{\nrm;\nsn} \!+\! \hB_{\nrm;\nsn}) \pl_+ \hx^\nrm_\s \pl_- \hx^\nsn_\s ,\quad 0< \xi <\pi \\
& \frac{1}{8\pi} \hG_{\nrm;\nsn} \pl_+ \hx^\nrm_\s \pl_- \hx^\nsn_\s = -\frac{1}{8\pi} \widehat{Tr}
  \left( \sm (\st,\s) \hg^{-1} (\st) \pl_+ \hg(\st) \hg^{-1} (\st) \pl_- \hg(\st) \right) 
\end{align}
\end{subequations}
is obtained via the $\pom$-map from the corresponding result in the basic class, or directly from Eqs.~\eqref{Hamil-dens} and \eqref{Coord-Spc}. 
As expected \cite{AS,Giusto,TwGiu}, this bulk density has the same form as (the sigma-model form of) the Lagrange density in the corresponding closed-string 
WZW orbifold sector \cite{Big,Geom}. We remind the reader however that all the fields in this action are implicitly $\pom$-dependent, as shown in 
Eq.~\eqref{2.7}.

A complete action for untwisted open WZW strings in terms of group elements and auxiliary boundary variables was given in Ref.~\cite{GT-TNB}, but the complete 
action formulation for the general twisted open WZW string is an open problem. We mention however that the complete action for the open-string sectors of 
the WZW orientation orbifolds was given in Ref.~\cite{Orient2}, in terms of nothing but group orbifold elements on the {\it solid half cylinder}. This may
be an interesting direction to consider in the general case.

Finally, we give the {\it twisted non-commutative geometry} of the general twisted open WZW string:
\begin{subequations}
\label{Tw-NC-Geom}
\begin{gather}
 \{ \hx_\s^\nrm (\xi,t) ,\hx_\s^\nsn (\eta,t) \} =\left\{ \begin{array}{ll} -i\pi\hPs_{\s,\pom}^{\nrm;\nsn} (0,0,t) &\text{if } \xi=\eta=0, \\
   i\pi \hPs_{\s,\pom}^{\nrm;\nsn} (\pi,\pi,t) & \text{if } \xi=\eta=\pi, \\ 0 & \text{otherwise} \end{array} \right. 
\end{gather}
\begin{align}
\!\!\!&\!\hPs_{\s,\pom}^{\nrm;\nsn}(\xi,\xi,t) = -\hPs_{\s,\pom}^{\nsn;\nrm} (\xi,\xi,t) \\
&\, =\!\sG^{\ntd;\nue}(\s) e^{-2i\frac{n(t)}{\r(\s)}\xi} \!\left( \hebi_\pom (\xi)_\ntd{}^{\!\!\!\nrm} \hei(\xi)_\nue{}^{\!\!\!\nsn} 
   \!-\!\hei(\xi)_\nue{}^{\!\!\!\nrm} \hebi_\pom (\xi)_\ntd{}^{\!\!\!\nsn} \right) \quad \quad \nn \\
&\,\, =\!\hei (\xi)_\nue{}^{\!\!\!\nrm} \sG^{\nue;\ntd} (\s) \times \nn \\
&\bigspc \bigspc \times \Big{(} e^{-2i \frac{n(t)}{\r(\s)}\xi} \ho_\pom^{-1}(\xi)_\ntd{}^{\!\!\!\nvk} 
   \!-\!\ho_\pom (\xi)_\ntd{}^{\!\!\!\nvk} e^{2i\frac{n(v)}{\r(\s)}\xi} \Big{)} \hei(\xi)_\nvk{}^{\!\!\!\nsn}  \\
& \bigspc \quad \he (\xi) \equiv \he (\hx (\xi,t)) ,\quad \heb_\pom (\xi) \equiv \heb (\hx(\xi,t)) \pom^{-1} ,\quad \ho_\pom (\xi) 
   \equiv \ho (\hx(\xi,t)) \pom^{-1} \\
& \bigspc \bigspc \bigspc \pom \in Aut(\gfrakh (\s)) ,\quad \srange \,.
\end{align}
\end{subequations}
The direct derivation of this result involves solving a set of PDEs for the $\{ \hx ,\hx \}$ brackets, but the result \eqref{Tw-NC-Geom} is more quickly 
obtained by the $\pom$-map from Eq.~$(3.38)$ of Ref.~\cite{TwGiu}. Conversely, the twisted non-commutative geometry of the open-string orbifolds 
$A_g^{open}(H)/H$ (i.~e.~the basic class) can be obtained from Eq.~\eqref{Tw-NC-Geom} by setting $\pom =\thickone$. We emphasize that, as expected, the 
coordinate brackets \eqref{Tw-NC-Geom} vanish in the bulk $0<\xi <\pi$, in parallel with the corresponding coordinate brackets of closed-string WZW orbifold 
theory \cite{Geom}. The general twisted non-commutative geometry in Eq.~\eqref{Tw-NC-Geom} is a central result of this paper.

A simple case of the general result \eqref{Tw-NC-Geom} is the non-commutative geometry of the general {\it untwisted} open WZW string
\begin{subequations}
\label{Untw-NC-Geom}
\begin{gather}
\{ x^i (\xi,t), x^j(\eta,t) \} = \left\{ \begin{array}{ll} -i\pi \Psi_{\omega}^{ij} (0,0,t) & \text{if } \xi=\eta=0, \\
   i\pi \Psi_{\omega}^{ij} (\pi,\pi,t) & \text{if } \xi=\eta =\pi , \\ 0 & \text{otherwise} \end{array} \right. \\
\Psi_{\omega}^{ij}(\xi,\xi,t) =-\Psi_{\omega}^{ji} (\xi,\xi,t) = G^{ab} \Big{(} \bar{e}^{-1}_{\omega} (\xi)_a{}^i e^{-1}(\xi)_b{}^j - e^{-1}(\xi)_b{}^i 
   \bar{e}_{\omega}^{-1}(\xi)_a{}^j \Big{)} \nn \\
\bigspc = e^{-1}(\xi)_a{}^i G^{ab} \Big{(} \Omega_{\omega}^{-1} (\xi) -\Omega_{\omega} (\xi) \Big{)}_{b}^{\,\,c} e^{-1}(\xi)_{c}{}^j \\
e(\xi) \equiv e(x(\xi,t)) ,\quad \bar{e}_\omega (\xi) \equiv \bar{e}(x(\xi,t)) \omega^{-1} ,\quad \Omega_\omega (\xi)
    \equiv \Omega (x(\xi,t)) \omega^{-1} \\
a,b,c ,i,j = 1,\ldots ,\text{ dim }g ,\quad \omega \in Aut(g)
\end{gather}
\end{subequations}
which is obtained via Eq.~\eqref{s=0} in the $\s=0$ sector. Here $e$ and $\bar{e}$ are the untwisted left- and right-invariant vielbeins on $g$ and $G^{ab}$
is the untwisted tangent-space metric. The result \eqref{Untw-NC-Geom} generalizes the $\omega =1$ non-commutative geometry of Ref.~\cite{Giusto}. 

Further examples of these general coordinate-space results are found in Subsecs.~$2.8$, $4.2$ and $4.3$. As in Refs.~\cite{Giusto,TwGiu}, one may use the 
known brackets above and the inverse relations \eqref{Inv-Relns} to straightforwardly compute the remaining phase-space brackets -- without solving any
additional PDEs.

\subsection{The General WZW Brane}

As our next coordinate-space topic, we discuss the {\it general WZW branes} which live at the ends of the general twisted open WZW string.

We begin with the boundary conditions on the twisted currents and coordinates
\begin{gather}
\hj_\nrm^{(+)} (0,t) = \hj_\nrm^{(-)} (0,t) ,\quad \hj_\nrm^{(+)} (\pi,t) = e^{-\tp \nrrs} \hj_\nrm^{(-)} (\pi,t) \label{hj-BCs}
\end{gather}
\begin{subequations}
\label{hx-BCs}
\begin{align}
&\pl_t \hx_\s^{\nrm}(\xi,t) \Big{(} \heb_\pom (\xi,t)_\nrm{}^{\!\nsn} - \he(\xi,t)_\nrm{}^{\!\nsn} \Big{)} \nn \\
&   \bigspc  =\pl_\xi \hx^\nrm (\xi,t) \Big{(} \heb_\pom (\xi,t)_\nrm{}^{\!\nsn} +\he (\xi,t)_\nrm{}^{\!\nsn} \Big{)} \,\,\, \text{at } \xi=0\\
&\pl_t \hx_\s^{\nrm}(\xi,t) \Big{(} e^{\tp \srac{n(s)}{\r(\s)}} \heb_\pom (\xi,t)_\nrm{}^{\!\nsn} - \he(\xi,t)_\nrm{}^{\!\nsn} \Big{)} \nn \\
&   \quad \quad \quad =\pl_\xi \hx^\nrm (\xi,t) \Big{(} e^{\tp \srac{n(s)}{\r(\s)}} \heb_\pom (\xi,t)_\nrm{}^{\!\nsn} +\he (\xi,t)_\nrm{}^{\!\nsn} \Big{)} \,\,\,\text{at } \xi=\pi
\end{align}
\end{subequations}
where Eq.~\eqref{hj-BCs} was given earlier, and Eq.~\eqref{hx-BCs} is then obtained from the coordinate-space form of the currents in Eq.~\eqref{Eq2.23}. The
current boundary conditions can also be rewritten in terms of the matrix currents
\begin{subequations}
\label{MCBCs}
\begin{gather}
\hj^{(+)} (\st,0,t,\s) \!= \! \hj^{(-)} (\st,0,t,\s) ,\,\,\,\, \hj^{(+)}(\st,\pi,t,\s) \!= \!E(T,\s) \hj^{(-)}(\st ,\pi,t,\s) E(T,\s)^\ast \label{MCBCa} \\
\hj^{(-)} (\st,\xi,t,\s) \equiv \hj_\nrm^{(-)} (\xi,t,\s) \sG^{\nrm;\nsn}(\s) \st_\nsn \label{hj_-(st)}
\end{gather}
\end{subequations}
and we note that the right-mover matrix current $\hj^{(-)} (\st)$ is {\it not} the same as the current $\hj^{(-)} (\st^\pom)$ defined in 
Eq.~\eqref{Mat-hj-Defn2}. To obtain this result, we used the selection rule of the twisted representation matrices \cite{Big}
\begin{subequations}
\label{E(T)}
\begin{gather}
e^{\tp \nrrsf} \st_\nrm (T,\s) =E(T,\s) \st_\nrm (T,\s) E^\ast (T,\s)\\
E(T,\s)_\Nrm{}^\Nsn = \de_\m{}^\n \de_{N(r)-N(s),0\, \text{mod }R(\s)} e^{-\tp \frac{N(r)-N(s)}{R(\s)}}
\end{gather}
\end{subequations}
where the diagonal matrix $E(T,\s)$ is the eigenvalue matrix of the extended $H$-eigenvalue problem \cite{Big,TwGiu} in sector $\s$ of the underlying 
closed-string orbifold $A_g(H)/H$.

We would like to express the matrix-current boundary conditions \eqref{MCBCs} in terms of the group orbifold elements themselves. For this, one needs
to define a unitary matrix $\tilde{W}$ by the following {\it twisted linkage relation}:
\begin{gather}
\st_\nrm^\pom = \pom (n(r),\s)_\m{}^\n \st_{n(r)\n} = \tilde{W}^{\hc} (\st,\s) \st_\nrm \tilde{W} (\st,\s) \,. \label{Tw-Link}
\end{gather}
According to this definition, the matrix $\tilde{W} \in Aut(\gfrakh (\s))$ is the action of the T-duality automorphism $\pom$ in twisted rep $\st$. 
The twisted linkage relation is a close analogue of the ordinary linkage relation of orbifold theory \cite{Big}
\begin{gather}
T_a^{\omega} \equiv \ws_a{}^b T_b =W^{\hc} (h_\s;T) T_a W(h_\s;T)  \label{Link}
\end{gather}
where $W\in Aut(g)$ is the action of the automorphism $\omega$ in untwisted rep $T$. In particular, Eq.~\eqref{Tw-Link} reduces to the ordinary linkage 
relation in untwisted open-string sector $\s=0$. The existence of solutions $\tilde{W}$ to Eq.~\eqref{Tw-Link} follows for the same reasons 
\cite{Big,Big',so2n} that solutions $W$ exist for the ordinary linkage relation. For inner automorphisms $\pom$ of the twisted current algebra, the matrix 
$\tilde{W}$ exists because $\st^\pom$ is by definition unitarily equivalent to $\st$ in that case. When $\pom$ is an outer automorphism and $\st^\pom \not 
\simeq \st$, one must consider $\hg(\st)$ and $\st$ as multiplets
\begin{gather}
\hg(\st)\rightarrow \left( \begin{array}{cccc} \hg(\st) &&& \\ &\hg(\st^\pom) && \\ &&\hg(\st^{\pom^2}) & \\ &&& \ddots \end{array} \right) ,\quad
\st \rightarrow \left( \begin{array}{cccc} \st &&& \\ &\st^\pom && \\ &&\st^{\pom^2} & \\ &&& \ddots \end{array} \right) \label{233}
\end{gather}
which collect the automorphism cycle of $\pom$. Then following Refs.~\cite{Big,Big',so2n}, the twisted linkage relation is easily solved for $\tilde{W}$, 
which is an off-diagonal matrix of size shown in Eq.~\eqref{233}.

Using the definitions \eqref{Mat-hj-Defn2} and \eqref{hj_-(st)} of the two right-mover matrix currents $\hj^{(-)}$, as well as the twisted linkage relation
\eqref{Tw-Link}, we find the following relation between the two:
\begin{gather}
\hj^{(-)} (\st,\xi,t,\s) = \tilde{W}(\st,\s) \hj^{(-)}(\st^\pom ,\xi,t,\s) \tilde{W}^{\hc} (\st,\s) \,.
\end{gather}
This identity gives an alternate form of the matrix-current boundary conditions \eqref{MCBCs}
\begin{subequations}
\label{Mat-J-BC-W}
\begin{gather}
\hj^{(+)} (\st,0,t,\s) =\tilde{W}(\st,\s) \hj^{(-)}(\st^\pom ,0,t,\s) \tilde{W}^{\hc} (\st,\s) \quad \\
\hj^{(+)}(\st,\pi,t,\s) = E(T,\s) \tilde{W} (\st,\s) \hj^{(-)}(\st^\pom ,\pi,t,\s) \tilde{W}^{\hc} (\st,\s) E(T,\s)^\ast 
\end{gather}
\end{subequations}
which explicitly involves the matrix $\tilde{W}$.

With these boundary conditions and Eqs.~($2.27$c,d), we finally obtain the desired form of the boundary conditions in terms of group orbifold elements:
\begin{gather}
\hg^{-1} (\st,\xi,t,\s) \pl_+ \hg(\st,\xi,t,\s) =\bigspc \bigspc \bigspc \bigspc \bigspc \nn \\
=\left\{ \begin{array}{cc} \tilde{W}(\st,\s) \hg(\st,\xi,t,\s) \pl_- \hg^{-1} (\st,\xi,t,\s) \tilde{W}^{\hc} (\st,\s) & \text{at } \xi=0 \\
   E(T,\s) \tilde{W} (\st,\s) \hg(\st,\xi,t,\s) \pl_- \hg^{-1} (\st,\xi,t,\s) \tilde{W}^{\hc} (\st,\s) E(T,\s)^\ast &\text{at } \xi=\pi \,.
   \end{array} \right.  \label{GOE-BCs}
\end{gather}
This is the most natural description of the {\it general WZW branes} at the ends of the general twisted open WZW string. The quantities $E(T,\s)$ and 
$\tilde{W}(\st,\s)$ are defined in Eqs.~\eqref{E(T)} and \eqref{Tw-Link} respectively. Following the discussion above, we emphasize that, in the case of 
outer-automorphic $\pom$ and $\st^\pom \not \simeq \st$, the boundary conditions \eqref{GOE-BCs} relate different fields $\hg (\st^{\pom^n})$ in the 
multiplet \eqref{233}.

The general description \eqref{GOE-BCs} of {\it branes on group orbifolds} is another central result of this paper. Many explicit examples of this result are 
found below.

\subsection{Example: Twisted Free-Bosonic Open Strings}

We turn next to a large class of twisted free-bosonic open strings which is easily obtained in the formal {\it abelian limit} of our general construction. 

As background for the free-bosonic case, we begin with the untwisted left-mover sector of a closed-string CFT on abelian $g$
\begin{subequations}
\label{eq235}
\begin{gather}
\{J_a(m) ,J_b (n)\} =mG_{ab} \de_{m+n,0} ,\quad f_{ab}{}^c =0 ,\quad [T_a ,T_b] =0 \\
e_i{}^a = - \bar{e}_i{}^a = \de_i{}^a ,\quad B_{ij} =H_{ijk} =0 ,\quad i,j,a,b =1,\ldots ,\text{dim }g \label{eq235a} 
\end{gather}
\end{subequations}
where $i,j$ and $a,b$ label the bosons and the $T$'s are the abelian momenta. Then, twisting a general automorphism of the abelian current algebra
\begin{gather}
\ws \in H \subset Aut(g) ,\quad \s= 0,\ldots ,N_c -1 \label{AbAuto}
\end{gather}
one obtains the left-mover sector $\s$ of a free-bosonic orbifold. Moreover, following the development of this paper, we find the abelian data for the 
corresponding $\pom$-family of twisted free-bosonic open strings:
\begin{subequations}
\label{Ab-Data}
\begin{gather}
\scf_{\nrm;\nsn}{}^\ntd =0 ,\quad [\st_\nrm ,\st_\nsn] =[\st_\nrm^\pom ,\st_\nsn^\pom ]=0 \\
\{ \hj_\nrm (m\!+\!\nrrs) ,\hj_\nsn (n\!+\!\nsrs) \} =(m\!+\!\nrrs) \sG_{\nrm;\mnrn}(\s) \de_{m+n+\frac{n(r)+n(s)}{\r(\s)},0} \label{Ab-TCA} \\
\{ \hj_\nrm (m\!+\!\nrrs) ,\hg(\st,\xi,t) \} = \hg(\st,\xi,t) \st_\nrm e^{i(m+\nrrsf)(t+\xi)} \bigspc \quad \quad  \nn \\
  \bigspc \bigspc \bigspc -\st_\nrm^\pom \hg(\st,\xi,t) e^{i(m+\nrrsf)(t-\xi)} \\
\he_\nrm{}^{\!\!\nsn} = -\heb_\nrm{}^{\!\!\nsn} = \de_\m^\n \de_{n(r)-n(s) ,0\, \text{mod } \r(\s)} \\
(\heb_\pom )_\nrm{}^\nsn = -\pom^{-1} (n(r),\s)_\m{}^\n \de_{n(r)-n(s) ,0\, \text{mod } \r(\s)} \\
\hG_{\nrm;\nsn} =\sG_{\nrm;\nsn}(\s) =\schisig_\nrm \schisig_\nsn U(\s)_\nrm{}^a U(\s)_\nsn{}^b G_{ab} \label{Ab-hG} \\
\hat{B}_{\nrm;\nsn} =\hat{H}_{\nrm;\nsn;\ntd} =0 ,\quad \s=0,\ldots ,N_c -1  \label{Ab-hB} \\
\hat{{\cL}}_\s = \srac{1}{8\pi} \sG_{\nrm;\mnrn} (\s) \pl_+ \hx^\nrm_\s \pl_- \hx^\mnrn_\s ,\quad 0< \xi <\pi
\end{gather}
\end{subequations}
Here $\pom$, which satisfies Eq.~\eqref{wwG=G}, is any automorphism\footnote{Although abelian current algebras (such as Eq.~\eqref{Ab-TCA}) can be considered 
as formal limits of non-abelian current algebras (such as Eq.~\eqref{EqT-Strip-CA}), the automorphism group of the abelian current algebra is of course larger
than the automorphism group of the non-abelian starting point.} of the twisted abelian current algebra \eqref{Ab-TCA}, and the unitary matrices $U(\s)$ in 
Eq.~\eqref{Ab-hG} solve the H-eigenvalue problem \cite{Dual,More,Big} for the automorphism $\ws$ in Eq.~\eqref{AbAuto}. It is clear from Eqs.~\eqref{eq235a} 
and \eqref{Ab-hB} that we are not treating the most general twisted free-bosonic open string, which can also involve a class of twisted $\hB$ fields, 
$\hB \neq 0$. Similarly, we will not discuss any particular target-space compactification of the twisted coordinates.
 
One may now obtain the classical description of these twisted free-bosonic open strings by substitution of the abelian data \eqref{Ab-Data} into 
the general results above. For example, one obtains the $\pom$-family of phase-space realizations of the twisted abelian currents
\begin{subequations}
\begin{gather}
\hj_\nrm^{(\pm )} (\xi,t) =\sum_{m\in \Zint} \hj_\nrm (m\!+\!\nrrs ) e^{-i (m+\nrrsf )(t \pm \xi)} \\
\hj_\nrm^{(+)} (\xi,t) =2\pi \hat{p}_\nrm^\s (\xi,t) +\srac{1}{2} \pl_\xi \hx_\s^\nsn (\xi,t) \sG_{\nsn;\nrm}(\s) \\
\hj_\nrm^{(-)} (\xi,t) =\pom(n(r),\s)_\m{}^\n \Big{(} \!-2\pi \hat{p}_{n(r)\n}^\s (\xi,t) +\srac{1}{2} \pl_\xi \hx_\s^{n(s)\k} (\xi,t) \sG_{n(s)\k;n(r)\n}(\s) \Big{)} 
\end{gather}
\end{subequations}
from the general realization \eqref{gen-canon-real2}, and the corresponding $\pom$-family of branes in twisted free-bosonic sector $\s$
\begin{subequations}
\label{Ab-hx-BCs}
\begin{gather}
\pl_t \hx_\s^\nrn (\xi,t) \Big{(} \pom^{-1} (n(r),\s) \!+\!\one \Big{)}_\n^{\,\,\m} = \pl_\xi \hx_\s^\nrn (\xi,t) \Big{(}\pom^{-1} (n(r),\s) \!-\! \one \Big{)}_\n^{\,\,\,\m} 
  \,\,\,\text{  at } \xi=0 \\
\pl_t \hx_\s^\nrn (\xi,t) \Big{(} e^{\tp \nrrsf} \pom^{-1} (n(r),\s) \!+\!\one \Big{)}_\n^{\,\,\m} \bigspc \bigspc \bigspc \bigspc \nn \\
 \bigspc = \pl_\xi \hx_\s^\nrn (\xi,t) \Big{(} e^{\tp \nrrsf} \pom^{-1} (n(r),\s) \!-\! \one \Big{)}_\n^{\,\,\m} \,\,\,\text{  at } \xi =\pi 
\end{gather}
\end{subequations}
follow from the general boundary conditions \eqref{hx-BCs}. In these results, the basic class \cite{TwGiu} of twisted free-bosonic open strings is found at 
$\pom =\thickone$. 

The simplest case\footnote{Twisted scalar fields and orbifold sectors were first considered in Ref.~\cite{2faces2}, and this reference led directly to the 
study of DN (ND) strings in Ref.~\cite{WS}.}  of twisted free-bosonic open strings is realized when we appropriate the initial left-mover data from a simple 
inversion orbifold $x \rightarrow -x$. In this case, the resulting twisted open strings have $\r(\s) =2$, $\bnrrs =\srac{1}{2}$ and $\pom =\pm 1$:
\begin{gather}
DN: \,\, \pom =1 \quad ; \quad ND: \,\, \pom =-1 \,. \label{DN-ND}
\end{gather}
For this pair of half-integer-moded strings, T-duality is nothing but reversal of the branes. This example is very simple, but we remind the reader that each 
twisted sector $\s$ comes equipped (via the $H$-eigenvalue problem \cite{Dual,More,Big}) with a {\it set} $\{ n(r) \}$ of spectral indices and hence a 
corresponding {\it set} of mixed boundary conditions \eqref{Ab-hx-BCs} on the twisted coordinates $\hx_\s^\nrm$. Mixed boundary conditions with $\hat{B}=0$ 
have also been observed in the open-string sectors of the free-bosonic orientation orbifolds \cite{Orient1,Orient2} and in free-bosonic open-string 
orbifolds \cite{TwGiu}.

Even simpler free-bosonic open strings are obtained in the untwisted case, where each end of the string lies on the same type of brane:
\begin{subequations}
\begin{gather}
\pl_t x^i (\xi,t) \de_i{}^a (\omega^{-1} +\one )_a{}^b = \pl_\xi x^i(\xi,t) \de_i{}^a (\omega^{-1} -\one )_a{}^b \,\,\,\text{ at } \xi =0,\pi \\
\omega \in Aut(g) \,.
\end{gather}
\end{subequations}
This result is obtained by using Eqs.~\eqref{s=0} and \eqref{eq235} in the $\s =0$ sector of Eq.~\eqref{Ab-hx-BCs}. For a single coordinate (with 
tangent-space and Einstein metric $G=1$), one encounters the cases (see e.~g.~Ref.~\cite{Giusto})
\begin{subequations}
\label{DD-NN}
\begin{gather}
S = \srac{1}{8\pi} \int \!\!dt \int_0^\pi \!\!d\xi \pl_+ x \pl_- x \\
\!\!\{ J^{(+)}(\xi,t) ,J^{(\pm)}(\eta,t) \} \!=\!\tp \pl_\xi \de (\xi \mp \eta) ,\,\,\, \{ J^{(-)}(\xi,t) ,J^{(\pm)}(\eta,t) \} \!=\!-\tp \pl_\xi \de (\xi \pm \eta) 
\end{gather}
\begin{gather}
DD: \,\,\omega = 1 ,\,\quad J^{(+)} (\xi) = 2\pi p(\xi) + \srac{1}{2} \pl_\xi x(\xi) \bigspc \nn \\
\bigspc  J^{(-)} (\xi) = -2\pi p(\xi) + \srac{1}{2} \pl_\xi x(\xi) \\
NN: \,\, \omega =-1 ,\quad J^{(+)} (\xi) = 2\pi p(\xi) + \srac{1}{2} \pl_\xi x(\xi) \bigspc \nn \\
\bigspc  J^{(-)} (\xi) = 2\pi p(\xi) - \srac{1}{2} \pl_\xi x(\xi)
\end{gather}
\end{subequations}
which describe the classic example of T-duality.

Returning to the twisted case, we give the non-commutative geometry of the twisted free-bosonic open strings 
\begin{subequations}
\label{247}
\begin{gather}
\hat{\Psi}^{\nrm;\nsn}_{\s,\pom} (\xi,\xi) =\sG^{n(r)\de ;\nsn}(\s) \left( e^{2i\nrrsf \xi} \pom^{-1} (n(r),\s) -e^{-2i \nrrsf \xi} \pom (n(r),\s)
  \right)_\de^{\,\,\,\m} \\
\{ \hx_\s^\nrm (\xi,t) ,\hx^\nsn_\s (\eta,t)\} = i\pi \de_{n(r)+n(s),0\,\text{mod }\r(\s)} \sG^{n(r),\de ;\mnrn}(\s) \times \bigspc \quad \quad \quad \nn \\
  \bigspc \times \left\{ \begin{array}{cc} (\pom (n(r),\s) -\pom^{-1} (n(r),\s) )_\de{}^\m & \text{if } \xi=\eta =0, \\
     (e^{\tp \nrrsf} \pom^{-1} (n(r),\s) -e^{-\tp \nrrsf} \pom (n(r),\s) )_\de{}^\m & \text{if } \xi =\eta =\pi ,\\
 0& \text{otherwise} \end{array} \right.  \label{Ab-xx-Comm}
\end{gather}
\end{subequations}
which is obtained by inserting the abelian data \eqref{Ab-Data} into the general non-commutative geometry \eqref{Tw-NC-Geom}. This result reduces to the untwisted non-commutative geometry
\begin{subequations}
\begin{gather}
\{ x^i (\xi,t) ,x^j (\eta,t) \} = \left\{ \begin{array}{cc} i\pi G^{ab} \de_b{}^j (\omega -\omega^{-1} )_a{}^c \de_c{}^i & \text{ if } \xi=\eta =0,\pi \\
   0 & \text{ otherwise } \end{array} \right. \label{248} \\
a,i =1\ldots \text{dim } g ,\quad g \,\text{ abelian}
\end{gather}
\end{subequations}
in sector $\s=0$ (see Eqs.~\eqref{s=0} and \eqref{eq235}). The open-string geometries \eqref{247} and \eqref{248} are commutative for DN, ND, DD and NN 
coordinates\footnote{Non-commutative geometry [45-49] is also found for NN coordinates when $B\neq 0$.}. Simple non-commutative examples are 
easily constructed with two or more coordinates, using e.~g.~$\pom =1, \,\r(\s)=3$ in the twisted geometry \eqref{Ab-xx-Comm}
or by using 
\begin{gather}
\omega = \left( \begin{array}{cc} \cos (\srac{2\pi}{3}) & \sin (\srac{2\pi}{3}) \\ -\sin (\srac{2\pi}{3}) & \cos (\srac{2\pi}{3}) \end{array} \right)
\end{gather}
in the untwisted geometry \eqref{248}.
 
For use below, we also give the equations of motion of the twisted coordinates
\begin{subequations}
\label{Ab-EOMs}
\begin{gather}
\pl_t \hx_\s^\nrm (\xi,t) =\sG^{\nrm ;n(s)\ep}(\s) \Big{(} \de_\ep{}^\n \hj_\nsn^{(+)} (\xi,t) -\pom^{-1} (n(s),\s)_\ep{}^\n \hj_\nsn^{(-)} (\xi,t) \Big{)} \\
\pl_\xi \hx_\s^\nrm (\xi,t) =\sG^{\nrm ;n(s)\ep}(\s) \Big{(} \de_\ep{}^\n \hj_\nsn^{(+)} (\xi,t) +\pom^{-1}(n(s),\s)_\ep{}^\n \hj_\nsn^{(-)} (\xi,t) \Big{)} 
\end{gather}
\end{subequations}
and the following forms of the twisted momenta:
\begin{subequations}
\begin{gather}
\hp_\nrm^\s (\xi,t) =\srac{1}{4\pi} \left( \hj_\nrm^{(+)} (\xi,t,\s) - \pom^{-1} (n(r),\s)_\m{}^\k \hj^{(-)}_{n(r)\k}(\xi,t,\s) \right) \label{Ab-hp} \\
 =\srac{1}{4\pi} \sG_{\nrm;\nsn}(\s) \pl_t \hx_\s^\nsn (\xi,t) \,.
\end{gather}
\end{subequations}
Using the automorphic invariance \eqref{wwG=G} of $\sG$, these results are obtained in the abelian limits of Eqs.~\eqref{Inv-Relns} and \eqref{plt-hx1}. 
Note that the boundary conditions \eqref{Ab-hx-BCs} also follow from these equations of motion.

As a final topic in this subsection, we will obtain the complete equal-time algebras of the twisted free-bosonic open strings. Following Ref.~\cite{TwGiu}, 
we begin with the solution of the equations of motion \eqref{Ab-EOMs}
\begin{subequations}
\label{Ab-hx-Mode}
\begin{gather}
\hx_\s^{0\m} (\xi,t) =\hat{q}^{0\m} +\sG^{0\m ;0\n}(\s) \BIG{(} \Big{(} \one -\pom (0,\s) \Big{)}_\n^{\,\,\de} \hj_{0\de}(0) t + \Big{(}\one +\pom (0,\s) 
  \Big{)}_\n^{\,\,\de} \hj_{0\de}(0) \xi \nn \\
\bigspc + i\sum_{m\neq 0} e^{-imt} \Big{(} e^{-im\xi} \one -e^{im\xi} \pom(0,\s) \Big{)}_\n^{\,\,\de} \frac{\hj_{0\de}(m)}{m} \BIG{)} 
\end{gather}
\begin{gather}
\bar{n}(r)\! \neq \!0\!: \,\,\,\hx_\s^\nrm (\xi,t)=\hat{q}^\nrm +i\sG^{n(r)\de;\nsn}(\s) \sum_{m\in \Zint} e^{-i(m+\nsrsf)t} \times \bigspc \quad \quad \nn \\
\bigspc \times \Big{(} e^{-i(m+\nsrsf )\xi} \one -e^{i(m+\nsrsf )\xi} \pom(n(r),\s) \Big)_\de^{\,\,\,\m} \frac{\hj_\nsn (m+\nsrsf)}{m+\nsrsf} 
\end{gather}
\end{subequations}
where the $\hat{q}$'s are the zero modes of the twisted coordinates. The brackets of the zero modes with the twisted current modes are obtained 
as follows:
\begin{subequations}
\begin{gather}
\{ \hj_\nrm (\mnrrs) ,\hx^\nsn_\s (\xi,t) \} = \bigspc \bigspc \bigspc \nn \\
  \bigspc -i\de_{n(r)}^{\,\,\,n(s)} \Big{(} e^{i(m+\nrrsf)(t+\xi)} \one -\pom(n(r),\s) e^{i(m+\nrrsf)(t-\xi)} \Big{)}_\m^{\,\,\n} \label{Ab-Jx-Comm} \\
\Rightarrow \quad \{ \hj_\nrm (\mnrrs) ,\hat{q}^\nsn \} = -i (\one -\pom(0,\s))_\m{}^\n \de_{m+\nrrsf ,0} \de_{n(s),0 \,\text{mod }\r(\s)} \,.\label{Jq-Comm}
\end{gather}
\end{subequations}
Here Eq.~\eqref{Ab-Jx-Comm} is equivalent to the abelian limit of Eq.~\eqref{hJ-hx-Soln}, and Eq.~\eqref{Jq-Comm} follows by insertion of the mode expansions 
\eqref{Ab-hx-Mode} into Eq.~\eqref{Ab-Jx-Comm}.

Using the twisted momenta \eqref{Ab-hp}, the mode expansions \eqref{Ab-hx-Mode} and the brackets \eqref{Jq-Comm}, we may now compute the complete equal-time 
{\it quasi-canonical algebra} of the twisted free-boson system:
\begin{subequations}
\label{Ab-Alg}
\begin{gather}
\{ \hp_\nrm^\s (\xi,t) ,\hp_\nsn^\s (\eta,t)\} = \frac{i}{8\pi} \de_{n(r)+n(s),0\,\text{mod }\r(\s)} \times \bigspc \bigspc \bigspc \bigspc \bigspc \nn \\
 \quad \quad  \times \pl_\xi \Big{(} \pom^{-1} (n(r),\s)_\m{}^\de \de_{-\nrrsf}(\xi +\eta)
  -\pom(n(r),\s )_\m{}^\de \de_{\nrrsf} (\xi+\eta) \Big{)} \sG_{n(r)\de ;\mnrn}(\s) \\
\{ \hx_\s^\nrm (\xi,t) ,\hp_\nsn^\s (\eta,t)\} = i\de_\nsn{}^\nrm \de (\xi-\eta) \bigspc \bigspc \bigspc \bigspc \quad \quad \nn \\
  \quad - \srac{i}{2} \de_{n(r)-n(s),0\,\text{mod }\r(\s)} \Big{(} \pom(n(r),\s)_\n{}^\m \de_{\nrrsf}(\xi+\eta) +\pom^{-1} (n(r),\s)_\n{}^\m \de_{-\nrrsf} 
  (\xi+\eta) \Big{)} 
\end{gather}
\begin{gather}
\{ \hx_\s^\nrm (\xi,t) ,\hx^\nsn_\s (\eta,t) \}= i\pi \de_{n(r)+n(s),0\,\text{mod }\r(\s)} \sG^{n(r)\de ;\mnrn}(\s) \times \bigspc \quad \quad \quad \nn \\
  \quad \quad \times \left\{ \begin{array}{cc} (\pom (n(r),\s) -\pom^{-1} (n(r),\s) )_\de{}^\m & \text{if } \xi=\eta =0, \\
     (e^{\tp \nrrsf} \pom^{-1} (n(r),\s) -e^{-\tp \nrrsf} \pom (n(r),\s) )_\de{}^\m & \text{if } \xi =\eta =\pi ,\\
 0& \text{otherwise} \end{array} \right.  \label{xx-Mode} \\
0\leq \xi, \eta \leq \pi \,.
\end{gather}
\end{subequations}
To obtain these results, we used summation identities given in App.~A of Ref.~\cite{Orient1}, and we also required that the coordinate brackets 
$\{ \hx ,\hx \}$ vanish in the bulk. This requirement uniquely fixes the $\{ \hat{q} ,\hat{q} \}$ algebra
\begin{subequations}
\label{qq-Comm}
\begin{gather}
\{ \hat{q}^{0\m} ,\hat{q}^{\nrn} \}= \pi i \sG^{0\de ;\nrn}(\s) \Big{(} \pom (0,\s) -\pom^{-1}(0,\s) \Big{)}_\de^{\,\,\m} \\
\bar{n}(r) \!\neq \!0\!: \,\, \{\hat{q}^\nrm ,\hat{q}^\nsn \}=\pi \sG^{n(r)\de ;\nsn}(\s) \times \bigspc \bigspc \bigspc \bigspc \quad \nn \\
  \times \Big{\{} \!\cot (\pi \nrrs ) \Big{(} 2\one -\pom (n(r),\s) +\pom^{-1}(n(r),\s) \Big{)}_\de^{\,\,\m} + \nn \\\
  \bigspc \bigspc \bigspc \quad \quad +\,i  \Big{(} \pom(n(r),\s) -\pom^{-1}(n(r),\s) \Big{)}_\de^{\,\,\m} \Big{\}} 
\end{gather}
\end{subequations}
and we have checked that all of the Jacobi identities among the $\hj$'s and $\hat{q}$'s are satisfied trivially.

We note in particular that: 1) the general quasi-canonical algebra \eqref{Ab-Alg} is canonical in the bulk, and 2) the coordinate brackets $\{ \hx,\hx\}$ 
in Eq.~\eqref{xx-Mode} agree exactly with the result \eqref{Ab-xx-Comm} given earlier. This agreement (which is based solely on the weaker assumption that 
$\{ \hx ,\hx \}$ vanishes in the bulk) is an important consistency check for the twisted free-bosonic open strings.
 
As simple examples, we list the following cases:
\begin{subequations}
\begin{gather}
DN: \quad \{ \hx^1 (\xi,t) ,\hp_1 (\eta,t) \} =i (\de (\xi-\eta) - \de_{1/2} (\xi +\eta) ) \quad \quad\\
ND: \quad \{ \hx^1 (\xi,t) ,\hp_1 (\eta,t) \} =i (\de (\xi-\eta) + \de_{1/2} (\xi +\eta) ) \quad \quad \\
DD: \quad \{ x(\xi,t) ,p(\eta,t) \} = i (\de (\xi-\eta) - \de (\xi +\eta) ) \bigspc \\
NN: \quad \{ x(\xi,t) ,p(\eta,t) \} = i (\de (\xi-\eta) + \de (\xi +\eta) ) \bigspc \,.
\end{gather}
\end{subequations}
All other phase-space brackets in the respective quasi-canonical algebras vanish identically. The results for DD and NN agree with those of Ref.~\cite{Giusto} 
(whose $q$ is related to ours by $\hat{q}(\s =0) =2q$ for both DD and NN).

\section{Operator Theory of the General Twisted Open WZW String}

\subsection{The Twisted Open-String Operator Algebra}

We begin this part of our discussion with the twisted affine-Sugawara constructions of the general open-string operator theory \cite{TwGiu}:
\begin{subequations}
\label{Eq3.1}
\begin{gather}
\hat{T}_\s^{(\pm )} (\xi,t) =\srac{1}{2\pi} {\cL}_{\sgb(\s)}^{\nrm;\mnrn}(\s) :\!\hj^{(\pm )}_\nrm (\xi,t) \hj^{(\pm )}_\mnrn (\xi,t) \!: 
   =\srac{1}{2\pi} \sum_{m \in \Zint} L_\s (m) e^{-im(t\pm \xi)} \label{StrTens} \\
\hj_\nrm^{(\pm )} (\xi,t) = \sum_{m\in \Zint} \hj_\nrm (m\!+\!\nrrs ) e^{-i(m+\nrrsf) (t\pm \xi)} \\
\!\!\hat{T}_\s^{(\pm )} (-\xi,t) \!=\!\hat{T}_\s^{(\mp )} (\xi,t) ,\,\,\, \hj_\nrm^{(\pm)} (-\xi,t) \!=\!\hj_\nrm^{(\mp)} (\xi,t) ,\,\,\, \pl_\mp \hat{T}_\s^{(\pm)} (\xi,t)
   \!=\! \pl_\mp \hj_\nrm^{(\pm)} (\xi,t) \!=\!0
\end{gather}
\begin{align}
&L_\s(m) = {\cL}_{\sgb(\s)}^{\nrm;\mnrn}(\s)  \sum_{p\in \Zint} :\!\hj_\nrm (p\!+\!\nrrs) \hj_{\mnrn} (m-p\!-\!\nrrs) \!: \nn \\
& \quad = {\cL}_{\sgb(\s)}^{\nrm;\mnrn}(\s) \Big{(} \sum_{p\in \Zint} :\!\hj_\nrm (p\!+\!\nrrs) \hj_{\mnrn} (m-p\!-\!\nrrs) \!:_M  \nn \\
&\quad \bigspc \bigspc \bigspc  -i\srac{\bar{n}(r)}{\r(\s)} \scf_{\nrm;\mnrn}{}^{0,\de}(\s) \hj_{0,\de} (m) \Big{)} +\gscfwt \de_{m,0} \label{Eq 3.1a} 
\end{align}
\begin{align}
\!\!\!&\!\!\!:\!\hj_\nrm (m\!+\!\nrrs) \hj_\nsn (n\!+\!\nsrs)\! :_M \,\equiv \theta (m\!+\!\nrrs \geq 0) \hj_\nsn (n\!+\!\nsrs) \hj_\nrm (m\!+\!\nrrs) \quad \nn \\
& \quad \quad \bigspc \bigspc + \theta (m\!+\!\nrrs <0) \hj_\nrm (m\!+\!\nrrs) \hj_\nsn (n\!+\!\nsrs) 
\end{align}
\begin{gather}
\gscfwt \equiv \sum_{r,\m ,\n} {\cL}_{\sgb(\s)}^{\nrm;\mnrn}(\s) \srac{\bar{n}(r)}{2\r(\s)} (1-\srac{\bar{n}(r)}{\r(\s)}) \sG_{\nrm;\mnrn}(\s) \label{gscfwt}
\end{gather}
\begin{align}
\bkspc&\bkspc [\hj_\nrm(m\!+\!\srac{n(r)}{\r(\s)}), \hj_\nsn(n\!+\!\srac{n(s)}{\r(\s)}) ] \!=\! i\scf_{\nrm;\nsn}{}^{\!\!\!\!\!n(r)+
n(s),\d}(\s) \hj_{n(r)+n(s),\d} (m\!+\!n\!+\!\srac{n(r)+n(s)}{\r(\s)}) \quad \nn\\
 &\bigspc \bigspc \bigspc \quad \quad + (m\!+\!\srac{n(r)}{\r(\s)}) \d_{m+n+\frac{n(r)+n(s)}{\r(\s)},\,0} \sG_{\nrm;-\nrn}(\s) \label{tw-ALA}
\end{align}
\begin{gather}
[L_\s (m), \hj_\nrm (n\!+\!\nrrs)] = -(n\!+\!\nrrs) \hj_\nrm (m\!+\!n\!+\!\nrrs) \\
[L_\s (m) ,L_\s (n)] =(m\!-\!n) L_\s (m\!+\!n) +\de_{m+n,0} \frac{\hat{c}}{12} m(m^2 -1) \\
\hat{c} =2{\cL}_{\sgb(\s)}^{\nrm;\mnrn}(\s) \sG_{\nrm;\mnrn}(\s) =2L_g^{ab} G_{ab} =c_g \label{Eq 3.2g} 
\end{gather}
\begin{align}
 \hat{H}_\s &=L_\s(0)= \!\int_0^\pi \!\!\!d\xi \,(\hat{T}_\s^{(+)} (\xi,t) +\hat{T}_\s^{(-)} (\xi,t) ) \nn \\
& =\!\frac{1}{2\pi} \!\int_0^{\pi} \!\!\!d\xi \,{\cL}_{\sgb(\s)}^{\nrm;\mnrn}(\s) \!:\!\hj_\nrm^{(+)}(\xi,t) \hj_\mnrn^{(+)} (\xi,t) 
   \!+\!\hj_\nrm^{(-)}(\xi,t) \hj_\mnrn^{(-)}(\xi,t)\!: \label{Eq 3.2i}
\end{align}
\begin{gather}
\pl_t \hat{A}(\xi,t) = i[\hat{H}_\s ,\hat{A}(\xi,t)] \,.
\end{gather}
\end{subequations}
Here $[\,,\,]$ is commutator and the twisted affine Lie algebra $\gfrakh(\s)$ is given explicitly in Eq.~\eqref{tw-ALA}. The symbol $:\, \cdot \,:_M$ denotes 
mode normal ordering \cite{Dual,More,Big} and ${\cL}_{\sgb(\s)}(\s)$ is the twisted inverse inertia tensor which, like $\sG(\s)$, is invariant \cite{TwGiu} 
under the T-duality automorphism $\pom \in Aut (\gfrakh (\s))$:
\begin{gather}
{\cL}_{\sgb(\s)}^{n(r),\k ;n(s),\l} (\s) \pom (n(r),\s)_\k{}^\m \pom(n(s),\s)_\l{}^\n ={\cL}_{\sgb(\s)}^{\nrm;\nsn} (\s) \,. \label{Lww=L} 
\end{gather}
The explicit form of ${\cL}_{\sgb(\s)}(\s)$ for any given WZW orbifold sector appears in Refs.~\cite{Dual,More,Big}. The central charge $\hat{c} =c_g$ in 
Eq.~\eqref{Eq 3.2g} is the same as that of the affine-Sugawara construction [42,50-53,37] on $g =\oplus_I \gfrak^I$ in the 
underlying untwisted theory. In the classical (high-level) limit
\begin{gather}
{\cL}_{\sgb(\s)}^{\bullet} (\s) \, \rightarrow \, \srac{1}{2} \sG^{\bullet}(\s) ,\quad [ \,, \, ] \, \rightarrow \, \{ \,, \, \} 
\end{gather}
the operator results \eqref{Eq3.1} reduce to the corresponding classical results of Subsecs.~$2.1$ and $2.3$. We remind that $\bar{n}(r)$ in 
Eq.~\eqref{gscfwt} is the pullback of the spectral index $n(r)$ to its fundamental range $0\leq \!\bar{n} \!<\!\r(\s)$. 

We note in particular that, as in the classical theory, the operator currents and the operator stress tensors are independent of the T-duality automorphism 
$\pom$. It follows that, for each twisted current algebra $\gfrakh (\s)$, the general open-string operator Hamiltonian \eqref{Eq 3.2i} and the bulk 
momentum\footnote{The twisted current-current form of the (non-conserved) bulk momentum is discussed in Ref.~\cite{TwGiu}.} are also $\pom$-independent.
The entire $\pom$-family of twisted open WZW strings are therefore related to each other by operator T-duality\footnote{It is an interesting open question 
to identify the appropriate $\pom$ which describes coset$\leftrightarrow$WZW T-dualities such as $\su(2) \! \oplus \! \ugp (1) / \ugp (1)' \simeq \su(2)$.}.

As far as open-string states are concerned, we first consider the {\it scalar twist-field state} $|0\rangle_\s$, which satisfies:
\begin{equation}
\label{ScTFSt}
\hj_{\nrm} (\mnrrs \geq 0) |0\rangle_\s = {}_\s \langle 0|\hj_\nrm (\mnrrs \leq 0) =0 \,.
\end{equation}
From the defining relations in Eq.~\eqref{ScTFSt} and the mode-ordered forms \eqref{Eq 3.1a} of the Virasoro generators we find the $\pom$-independent 
conformal weight $\gscfwt$ of the scalar twist-field \cite{TwGiu}:
\begin{subequations}
\label{ScTF2}
\begin{gather}
L_\s (m)\hc =L_\s (-m) \\
(L_\s (m\geq 0) -\de_{m,0} \gscfwt )|0\rangle_\s = {}_\s \langle 0|(L_\s (m\leq 0) -\de_{m,0} \gscfwt )=0 \\
(\hat{H}_\s -\gscfwt )|0\rangle_\s = {}_\s \langle 0|(\hat{H}_\s -\gscfwt )=0 \\
\gscfwt = \frac{x_\gfraks /2}{x_\gfraks +\tilde{h}_\gfraks} \sum_r \srac{\bar{n}(r)}{2\r(\s)} (1-\srac{\bar{n}(r)}{\r(\s)})
   \text{ dim} [\bar{n}(r)] \,. \label{gscfwt2}
\end{gather}
\end{subequations}
The general formula for $\gscfwt$ is given in Eq.~\eqref{gscfwt}, while the simplified form given in \eqref{gscfwt2} holds when the underlying untwisted
theory is permutation-invariant
\begin{gather}
g=\oplus_I \gfrak^I ,\quad \gfrak^I \simeq \text{simple }\gfrak ,\quad k_I =k ,\quad x_\gfraks = \frac{2k}{\psi^2_\gfraks} \label{perminv}
\end{gather}
which includes all the basic types of WZW orbifolds. In Eq.~\eqref{gscfwt2}, the quantities $\tilde{h}_\gfraks$ and $x_\gfraks$ are respectively the dual
Coxeter number of $\gfrak$ and the invariant level of affine $\gfrak$, while $\dim [\bar{n}(r)]$ is the degeneracy of the eigenvalue $E_{n(r)}(\s)$ in the
$H$-eigenvalue problem \cite{Dual,More,Big}. These results for the scalar twist-field state are the same as those obtained in closed-string orbifold theory,
and further evaluation of $\gscfwt$ is given for specific cases in Refs.~\cite{Big,Big',Perm,so2n}.

The scalar twist-field state is the twisted affine primary state with $\st =0$. More generally, {\it twisted affine primary fields} $\hg(\st)$ and their
corresponding {\it twisted affine primary states} $|\st \rangle_\s$ are discussed in Refs.~\cite{Chr,Big,Perm,TwGiu}. As in these references, the group 
orbifold elements of the previous section are the classical (high-level) limit of the twisted affine primary fields. To study these topics in further detail, 
we will use the operator form of the $\pom$-map \eqref{Eq2.17}:
\begin{subequations}
\label{Eq3.0}
\begin{gather}
\hj_\nrm^{(\pm )} \rightarrow \hj_\nrm^{(\pm )} ,\quad \hg (\st) \st \rightarrow \hg (\st,\pom) \st ,\quad \st \hg (\st) \rightarrow \st^\pom \hg (\st,\pom) \\
| \st \rangle_\s \st \rightarrow | \st,\pom \rangle_\s \st ,\quad \st | \st \rangle_\s \rightarrow \st^\pom | \st,\pom \rangle_\s \,.
\end{gather}
\end{subequations}
Here $\pom \in Aut(\gfrakh (\s))$ is the T-duality automorphism and the automorphic transform $\st^\pom$ of $\st$ is defined in Eq.~\eqref{st-pom-Defn}. 

The $\pom$-map \eqref{Eq3.0} and the results of Refs.~\cite{Perm,TwGiu} then allow us to write down the action of the twisted currents on the twisted affine 
primary fields and states:
\begin{subequations}
\label{Eq3.2}
\begin{gather}
[\hj_\nrm (m\!+\!\nrrs) ,\hg(\st,\pom,\xi,t)] =\hg(\st,\pom,\xi,t) \st_\nrm e^{i(m+\nrrs)(t+\xi)} \bigspc \bigspc \nn \\
\bigspc \bigspc \bigspc \bigspc -\!\st_\nrm^\pom \hg(\st,\pom,\xi,t) e^{i(m+\nrrs)(t-\xi)} \label{Eq 3.2h} \\
\hj_\nrm (m\!+\!\nrrs \geq 0) | \st,\pom \rangle_\s = \de_{m+\nrrsf ,0} \Big{(} |\st ,\pom \rangle_\s \st_\nrm -\st_\nrm^\pom |\st ,\pom \rangle_\s \Big{)} \label{AfPrSt} \\
\hj_\nrm (m\!+\!\nrrs \geq 0) (\tilde{W} (\st,\s) | \st,\pom \rangle_\s )= \de_{m+\nrrsf ,0} \Big{(} ( \tilde{W}(\st,\s) |\st ,\pom \rangle_\s ) \st_\nrm \bigspc \quad \nn \\
  \bigspc \bigspc \bigspc \bigspc \bigspc \quad \quad -\st_\nrm (\tilde{W}(\st,\s) |\st ,\pom \rangle_\s ) \Big{)} \,. \label{W-AfPrSt}
\end{gather}
\end{subequations}
We note first that Eq.~\eqref{Eq 3.2h} is the operator analogue of the classical brackets \eqref{Eq2.18}. The action \eqref{AfPrSt} of the twisted 
currents on the twisted affine primary states follows from the commutator \eqref{Eq 3.2h} when the states are created by the twisted affine primary fields 
on the scalar twist-field state \cite{Perm}. In Eq.~\eqref{W-AfPrSt}, $\tilde{W} \in Aut (\gfrakh (\s))$ is the action of the T-duality automorphism $\pom$ in 
twisted rep $\st$, which satisfies the twisted linkage relation \eqref{Tw-Link}. This last result shows that the action of the twisted currents on the 
particular linear combinations $\tilde{W}(\st) |\st ,\pom \rangle$ is in fact $\pom$-independent\footnote{Note however that the $\tilde{W}$-transformed 
states in Eq.~\eqref{W-AfPrSt} are not themselves $\pom$-independent.}. Then since the Virasoro operators are $\pom$-independent quadratics in the twisted 
currents, we see that, as expected, the entire $\pom$-family of T-dual twisted open WZW strings has the same spectrum (conformal weights):
\begin{gather}
\hat{\Delta} (\st,\pom) = \hat{\Delta} (\st) ,\quad \pom \in Aut (\gfrakh (\s)) \,. \label{Spectrum}
\end{gather}
The explicit form of $\hat{\Delta}(\st)$ is given in the following subsection.
In what follows we return to the simpler notation $\hg (\st,\pom) =\hg(\st)$ for the twisted affine primary field.

We emphasize that the results ($3.8$b,c) hold in the non-abelian case, where the primary fields create the primary states. In the abelian (free-boson) case, 
on the other hand, the primary states are created by the primary fields only when there is a zero mode $\hat{q}$ canonically conjugate to the current zero 
modes $\hj (0)$. Thus e.~g. Eq.~\eqref{AfPrSt} holds in the case of a single NN coordinate 
\begin{subequations}
\begin{gather}
\omega =-1 : \quad T^\omega = \omega T = -T \\
J (m \geq 0) |T ,\omega \rangle = \de_{m,0} ( |T,\omega \rangle T + T |T ,\omega \rangle ) = \de_{m,0} 2T |T,\omega \rangle \label{NNAfPrSt} \\
L (m\geq 0) |T,\omega \rangle = \de_{m,0} \Delta (T) |T,\omega \rangle ,\quad \Delta (T) = 2T^2
\end{gather}
\end{subequations}
where $T$ is the abelian momentum. For the corresponding DD case, however, $\hat{q}$ commutes with $\hj(0)$ and the abelian limit of Eq.~\eqref{AfPrSt} 
does not hold. Eigenstates of $J(0)$ which satisfy Eq.~\eqref{NNAfPrSt} exist in any case, but for DD they are not created by the primary fields. Because 
the DD and NN Virasoro generators are identical ($L(0) = \srac{1}{2} J(0)^2 + R$) the DD spectrum on these eigenstates is the same as the NN spectrum. (In 
conventional language, this spectral equality corresponds to compactification of DD and NN on dual lattices.) More generally, the $\pom$-independence of the 
Virasoro generators guarantees that the spectral equality \eqref{Spectrum} can always be maintained for $\pom$-families of twisted free-bosonic open strings.

Returning to the general non-abelian case, we close this subsection with the commutators of the Virasoro generators with the twisted affine primary fields
\begin{subequations}
\label{310}
\begin{gather}
[L_\s (m) ,\hg(\st,\xi,t)] =\hg (\st,\xi,t) \Big{(} -\!\srac{i}{2} \overleftarrow{\pl_+} +m \Dg (\st) \Big{)} e^{im(t+\xi)} \bigspc \bigspc \bigspc \nn \\
 \bigspc \bigspc \quad \quad +e^{im(t-\xi)} \Big{(} -\!\srac{i}{2} \pl_- +m\Dg (\st^\pom ) \Big{)} \hg (\st,\xi,t) \label{Eq 3.3a} \\
 \quad \quad \quad = e^{imt} \left( -i\cos (m\xi) \pl_t +\sin(m\xi) \pl_\xi \right) \hg(\st,\xi,t) \bigspc \quad \quad \nn \\
 \bigspc \bigspc +me^{imt} \left( e^{im\xi} \hg(\st,\xi,t) \Dg (\st) +e^{-im\xi} \Dg (\st^\pom) \hg(\st,\xi,t) \right) \label{Eq 3.3b} \\
\Dg (\st) \equiv {\cL}_{\sgb(\s)}^{\nrm;\mnrn} (\s) \st_\nrm \st_\mnrn = U(T,\s) D_g (T) U\hc (T,\s) \\
D_g (T) =L_g^{ab} T_a T_b \\
\Dg (\st^\pom)= {\cL}_{\sgb(\s)}^{\nrm;\mnrn} (\s) \st_\nrm^\pom \st_\mnrn^\pom =\Dg (\st) 
\end{gather}
\end{subequations}
which follow via the $\pom$-map \eqref{Eq3.0} from the corresponding $\pom =\thickone$ results of Ref.~\cite{TwGiu}. Here $D_g (T)$ is the conformal weight
matrix of rep $T$ of $g$ under the affine-Sugawara construction [42,50-53,37] and $\Dg (\st)$ is the corresponding twisted 
conformal weight matrix \cite{Big} of the twisted open WZW string.

\subsection{The Twisted Vertex Operator Equations and the One-Sided Notation}

As another application of the $\pom$-map \eqref{Eq3.0}, one can obtain the two-sided and then the one-sided forms \cite{Giusto,Orient1,TwGiu} of the twisted
vertex operator equations from the corresponding results of Ref.~\cite{TwGiu}. For example, the following two-sided form of the {\it twisted vertex operator 
equations}\footnote{By solving the twisted vertex operator equations of closed-string orbifold theory \cite{Big} in the abelian limit, the twisted vertex 
operators of closed-string free-bosonic orbifolds were obtained in Ref.~\cite{Big'}. It would be interesting to explicitly solve the corresponding 
(see Eq.~\eqref{Ab-Data}) free-bosonic form of Eq.~\eqref{TwoS-TVOE} for the twisted vertex operators of the general twisted free-bosonic open string. 
An example of this is discussed for the free-bosonic orientation orbifolds in Subsec.~$4.5$.}
\begin{subequations}
\label{TwoS-TVOE}
\begin{gather}
\srac{1}{2} \pl_+ \hg(\st,\xi,t) =2i{\cL}_{\sgb(\s)}^{\nrm ;\mnrn}(\s) :\!\hj_\nrm^{(+)} (\xi,t) \hg(\st,\xi,t) \st_\mnrn \!:_M \bigspc \nn \\
  \quad \quad +i\hg(\st,\xi,t) \D_{\sgb(\s)} (\st) -2i{\cL}_{\sgb(\s)}^{\nrm ;\mnrn}(\s) \sbnrrs \hg(\st,\xi,t) \st_\nrm \st_\mnrn \nn \\
  \quad \quad -2i{\cL}_{\sgb(\s)}^{\nrm ;\mnrn}(\s) \frac{e^{-2i\bnrrs \xi}}{1-e^{-2i\xi}} \st_\nrm^\pom \hg(\st,\xi,t) \st_\mnrn \bigspc 
\end{gather}
\begin{gather}
\srac{1}{2} \pl_- \hg(\st,\xi,t) =-2i{\cL}_{\sgb(\s)}^{\nrm ;\mnrn}(\s) :\!\st_\nrm^\pom \hj_\mnrn^{(-)} (\xi,t) \hg(\st,\xi,t)\!:_M \bigspc \nn \\
  \quad \quad +i\D_{\sgb(\s)}(\st) \hg(\st,\xi,t) -2i{\cL}_{\sgb(\s)}^{\nrm ;\mnrn}(\s) \srac{\overline{-n(r)}}{\r(\s)} \st_\nrm^\pom \st_\mnrn^\pom 
  \hg(\st,\xi,t) \nn \\
  \quad \quad -2i{\cL}_{\sgb(\s)}^{\nrm;\mnrn}(\s) \frac{e^{2i\frac{\overline{-n(r)}}{\r(\s)} \xi}}{1-e^{2i\xi}} \st_\nrm^\pom \hg(\st,\xi,t) \st_\mnrn \bigspc 
\end{gather}
\begin{gather}
:\! \hj_\nrm (m\!+\! \nrrs ) \hg(\st,\xi,t) \!:_M \equiv \theta (m\!+\!\nrrs <0) \hj_\nrm (m\!+\!\nrrs ) \hg(\st,\xi,t) + \bigspc \bigspc \nn \\
\bigspc \bigspc \bigspc +\theta (m\!+\! \nrrs \geq 0) \hg(\st,\xi,t) \hj_\nrm (m\!+\!\nrrs ) \label{Jg_M}
\end{gather}
\end{subequations}
are obtained with $\pl_\pm =\pl_t \pm \pl_\xi$ by the $\pom$-map from Eq.~$(4.31)$ of Ref.~\cite{TwGiu}. With Refs.~\cite{Giusto,TwGiu}, we emphasize that 
these vertex operator equations are {\it generically singular} at $\xi =0,\pi$ -- and hence so are the twisted affine primary fields themselves. These 
singularities result as a charge and its image approach each other at the strip boundary. The formal mechanism which generates these singularities in 
open-string WZW theory is reviewed in App.~A, where we discuss the properties of the constituent affine primary fields $\hg =\hg_- \hg_+$. In special cases, 
such as the open-string sectors of the WZW orientation orbifolds (see Sec.~4), the singularity at $\xi =0$ is suppressed.

For brevity, we will present most of our results only in the simpler {\it one-sided notation} \cite{Giusto, TwGiu}, which is defined as follows
\begin{subequations}
\label{Eq3.36}
\begin{gather}
\tilde{\hg}(\st,\bz,z,\s)^{\Nsn ;\Nrm} \equiv \hg(\st,\bz,z,\s)_{\Nrm}{}^{\Nsn} \label{Eq 3.36a} 
\end{gather}
\begin{gather}
(A \hg(\st,\bz,z,\s) B )_{\Nrm}{}^{\Nsn} = A_{\Nrm}{}^{\Ntd} \hg_{\Ntd}{}^{N(u)\ep} B_{N(u)\ep}{}^{\Nsn} \bigspc \bigspc \nn \\
\bigspc =\!-\tilde{\hg}^{N(u)\ep;\Ntd} B_{N(u)\ep}{}^{\!\!\!\Nsn} (\bar{A})_{\Ntd}{}^{\!\!\!\Nrm} \label{Eq 3.36b} \\
=\!-(\tilde{\hg}(\st,\bz,z,\s) B \otimes \bar{A} )^{\Nsn;\Nrm} \quad \, \label{Eq 3.36c} \\
\bar{A}\equiv -A^t ,\quad (A^t )_{\Nrm}{}^{\Nsn} \equiv A_{\Nsn}{}^{\Nrm} 
\end{gather}
\end{subequations}
where superscript $t$ is matrix transpose. In our application, we will need in particular the automorphically-transformed matrices $\bar{\st}^\pom$ 
\begin{subequations}
\label{Eq3.37}
\begin{gather}
\bar{\st}^\pom_\nrm (T,\s) \equiv -\st_\nrm^\pom (T,\s)^t =-\pom (n(r),\s)_\m{}^\n \st_\nrn (T,\s)^t \bigspc \label{337a} \\
  \bigspc =\pom (n(r),\s)_\m{}^\n \st_\nrn (\bT,\s) =\st_\nrm^\pom (\bar{T},\s) ,\quad \bT =-T^t \label{Eq 3.37a} \\
[\bar{\st}^\pom_\nrm ,\bar{\st}^\pom_\nrm ] =i\scf_{\nrm;\nsn}{}^{n(r)+n(s),\de}(\s) \bar{\st}^\pom_{n(r)+n(s),\de} \label{Eq 3.4b}
\end{gather}
\end{subequations}
which are the images on the right of the matrices $\st^\pom$. The first form of $\bar{\st}^\pom$ in \eqref{Eq 3.37a} was obtained using a result
of Ref.~\cite{Orient1}. Moreover, as shown in Eqs.~\eqref{Eq 3.36b}, \eqref{Eq 3.36c}, $\bar{\st}^\pom$ always acts on the right indices of $\tilde{\hg}$, 
while $\st$ acts on the left indices. This $B\!\otimes \!\bar{A}$ bookkeeping should be born in mind even though we sometimes neglect the ordering in the
tensor product
\begin{gather}
\bar{\st}^\pom \!\otimes \!\st  \simeq \st \!\otimes \!\bar{\st}^\pom \label{Eq3.38}
\end{gather}
for notational convenience. The matrices $\bar{\st}^\pom$ satisfy the same orbifold Lie algebra \eqref{Eq 2.3b} satisfied by the twisted representation 
matrices $\st ,\st^\pom$. 

In the one-sided notation, the commutators of the twisted fields take the form:
\begin{subequations}
\begin{gather}
[\hj_\nrm (m\!+\!\nrrs) ,\tilde{\hg}(\st,\xi,t)] =\tilde{\hg}(\st,\xi,t) \Big{(} \st_\nrm e^{i(m+\nrrs)(t+\xi)} +\bigspc \bigspc \nn \\
\bigspc \bigspc \bigspc \bigspc +\!\bar{\st}_\nrm^\pom e^{i(m+\nrrs)(t-\xi)} \big{)} \label{311} 
\end{gather}
\begin{gather}
[L_\s (m) ,\tilde{\hg}(\st,\xi,t)] = e^{imt} \left( -i\cos (m\xi) \pl_t +\sin(m\xi) \pl_\xi \right) \tilde{\hg}(\st,\xi,t) \bigspc \nn \\
 \bigspc \bigspc +me^{imt} \tilde{\hg}(\st,\xi,t) \left( e^{im\xi} \Dg (\st) +e^{-im\xi} \Dg (\bar{\st}^\pom) \right) \\
\Dg (\bar{\st}^\pom) ={\cL}_{\sgb(\s)}^{\nrm;\mnrn}(\s) \bar{\st}_\nrm^\pom \bar{\st}_\mnrn^\pom ={\cL}_{\sgb(\s)}^{\nrm;\mnrn}(\s) \bar{\st}_\nrm
   \bar{\st}_\mnrn \nn \\
  =\Dg (\st)^t =\Dg (\st) \,. \label{Eq 3.8c}
\end{gather}
\end{subequations}
We remind the reader that, in spite of the identity \eqref{Eq 3.8c}, the twisted conformal weight matrix $\Dg (\bar{\st}^\pom)$ acts on the right indices
of $\tilde{\hg}$. Similarly, the one-sided forms of the general twisted open-string vertex-operator equations
\begin{subequations}
\label{One-S-TVOE}
\begin{gather}
\srac{1}{2} \pl_+ \tilde{\hg}(\st,\xi,t) =2i{\cL}_{\sgb(\s)}^{\nrm ;\mnrn}(\s) \BIG{(} :\!\hj_\nrm^{(+)}(\xi,t) \tilde{\hg}(\st,\xi,t)\!:_M \bigspc 
  \bigspc \quad \quad \nn \\
  \quad \quad -\sbnrrs \tilde{\hg}(\st,\xi,t) \st_\nrm +\frac{e^{-2i\bnrrs \xi}}{1-e^{-2i\xi}} \tilde{\hg}(\st,\xi,t) \bar{\st}_\nrm^\pom 
  \BIG{)} \st_\mnrn +i\tilde{\hg}(\st,\xi,t) \Dg (\st) \\
\srac{1}{2} \pl_- \tilde{\hg}(\st,\xi,t) =2i{\cL}_{\sgb(\s)}^{\nrm ;\mnrn}(\s) \BIG{(} :\!\hj_\nrm^{(-)} (\xi,t) \tilde{\hg}(\st,\xi,t)\!:_M \bigspc 
  \bigspc \quad \quad \nn \\
  \quad \quad -\sbnrrs \tilde{\hg}(\st,\xi,t) \bar{\st}_\nrm^\pom +\frac{e^{2i\bnrrs \xi}}{1-e^{2i\xi}} \tilde{\hg}(\st,\xi,t) \st_\nrm \BIG{)} 
  \bar{\st}_\mnrn^\pom +i\tilde{\hg}(\st,\xi,t) \Dg(\bar{\st}^\pom)
\end{gather}
\begin{gather}
[\pl_+ ,\pl_-] \tilde{\hg}(\st,\xi,t) =0 
\end{gather}
\end{subequations}
follow from the two-sided forms in Eq.~\eqref{TwoS-TVOE}. 

As another application of the one-sided notation, we present the explicit form of the $\pom$-independent conformal weights of the twisted affine primary states
\begin{subequations}
\begin{gather}
L_\s (m\geq 0) \Big{(} |\st ,\pom \rangle_\s \one \otimes \tilde{W}(\st)^t \Big{)} = \de_{m,0} \Big{(} |\st ,\pom \rangle_\s \one \otimes 
   \tilde{W}(\st)^t \Big{)} \hat{\Delta} (\st) \\
\hat{\Delta} (\st) \equiv {\cL}_{\sgb(\s)}^{0\m ;0\n} (\s) ( \st_{0\m} - \bst_{0\m}) (\st_{0\n} -\bst_{0\n} ) \nn \\
\bigspc -i {\cL}_{\sgb(\s)}^{\nrm;\mnrn} \sbnrrs \scf_{\nrm;\mnrn}{}^{0\delta} (\st_{0\de} - \bst_{0\de} ) + \gscfwt  
\end{gather}
\end{subequations}
which was promised in Subsec.~$3.1$. This result for $\hat{\Delta}(\st)$ (which in general needs further diagonalization \cite{Perm}) follows directly from 
Eq.~\eqref{Eq3.1} and the one-sided form of Eq.~\eqref{W-AfPrSt}.

\subsection{The General Twisted Open-String KZ System}

We now present the third central result of this paper, namely the one-sided form of the {\it general twisted open-string KZ system}:
\begin{subequations}
\label{Eq3.39}
\begin{gather}
\tilde{\hg}_R (\st,\bz,z) \equiv \tilde{\hg} (\st,\xi,t) z^{-\Dg (\st)} \otimes \bz^{-\Dg (\bar{\st}^\pom)} \label{hg-Red} \\
\tilde{\hat{F}}_\s (\st,\bz,z) \equiv {}_\s \langle 0| \tilde{\hg}_R (\st^{(1)},\bz_1 ,z_1) \ldots \tilde{\hg}_R (\st^{(n)} ,\bz_n ,z_n) |0 \rangle_\s \label{Fred} \\
\bkspc \!\! \pl_i \tilde{\hat{F}}_{\!\s} (\st,\bz,z) \!=\! \tilde{\hat{F}}_{\!\s} (\st,\bz,z) \hat{W}_i (\st,\bz,z,\s) ,\,\,\,\,\bpl_i \tilde{\hat{F}}_{\!\s}
  (\st,\bz,z) \!=\! \tilde{\hat{F}}_{\!\s} (\st,\bz,z) \hat{\bar{W}}_{\!i} (\st,\bz,z,\s) 
\end{gather}
\begin{gather}
\hat{W}_i (\st,\bz,z,\s) =2{\cL}_{\sgb(\s)}^{\nrm;\mnrn}(\s) \Big{(} \sum_{j\neq i} \left( \frac{z_j}{z_i} \right)^\bnrrs \frac{\st_\nrm^{(j)} \otimes
  \st_\mnrn^{(i)}}{z_i -z_j} \bigspc \bigspc \quad \quad \nn \\
  +\sum_j \left( \frac{\bz_j}{z_i} \right)^\bnrrs \frac{\bar{\st}_\nrm^{(j),\pom} \otimes \st_\mnrn^{(i)}}{z_i -\bz_j} -\sbnrrs \frac{1}{z_i} 
  \st_\nrm^{(i)} \st_\mnrn^{(i)} \Big{)} \\
\hat{\bar{W}}_{\!i} (\st,\bz,z,\s) =2{\cL}_{\sgb(\s)}^{\nrm;\mnrn}(\s) \Big{(} \sum_{j\neq i} \left( \frac{\bz_j}{\bz_i} \right)^\bnrrs 
  \frac{\bar{\st}_\nrm^{(j),\pom} \otimes \bar{\st}_\mnrn^{(i),\pom}}{\bz_i -\bz_j} \bigspc \bigspc \quad \quad  \nn \\
  +\sum_j \left( \frac{z_j}{\bz_i} \right)^\bnrrs \frac{\st_\nrm^{(j)} \otimes \bar{\st}_\mnrn^{(i),\pom}}{\bz_i -z_j} -\sbnrrs \frac{1}{\bz_i} 
  \bar{\st}_\nrm^{(i),\pom} \,\bar{\st}_\mnrn^{(i),\pom} \Big{)} \label{bWi} \\
\tilde{\hat{F}}_\s (\st,\bz,z) \sum_{i=1}^n (\st_{0,\m}^{(i)} \otimes \one + \one \otimes \bar{\st}_{0,\m}^{(i),\pom} )=0 \label{GWI} 
\end{gather}
\begin{gather}
z_i \equiv e^{i(t_i +\xi_i)} ,\quad \bz_i \equiv e^{i(t_i -\xi_i)} \\
\pl_i \equiv \srac{\pl}{\pl z_i} =\srac{-i}{2z_i} \pl_{+i} ,\quad \bpl_i \equiv \srac{\pl}{\pl \bz_i}= \srac{-i}{2\bz_i} \pl_{-i} ,\quad \pl_{\pm i} 
  \equiv \pl_{t_i} \pm \pl_{\xi_i} \\
\left( \frac{z_j}{z_i} \right)^\bnrrs =e^{i\bnrrs (t_j+\xi_j -t_i -\xi_i)} ,\quad \left( \frac{\bz_j}{z_i} \right)^\bnrrs 
   =e^{i\bnrrs (t_j -\xi_j -t_i -\xi_i)} \\
\left( \frac{z_j}{\bz_i} \right)^\bnrrs =e^{i\bnrrs (t_j+\xi_j -t_i +\xi_i)} ,\quad \left( \frac{\bz_j}{\bz_i} \right)^\bnrrs 
   =e^{i\bnrrs (t_j -\xi_j -t_i +\xi_i)} \\
\st^{(i)} \equiv \st (T^{(i)},\s) ,\quad \bar{\st}^{(i),\pom} \equiv -\st^\pom (T^{(i)},\s)^t = \pom(\s) \st (\bT^{(i)},\s) ,\quad i=1\ldots n \\
\pom \in Aut( \gfrakh (\s)) ,\quad \,\,\, \srange \,.
\end{gather}
\end{subequations}
This system can be obtained directly from Eq.~$(4.41)$ of Ref.~\cite{TwGiu}, using the $\pom$-map \eqref{Eq3.0}. Here $|0\rangle_\s$ is the scalar twist field 
state defined in Eq.~\eqref{ScTFSt}, and following Ref.~\cite{TwGiu}, we have removed the standard cylinder factors from the reduced affine primary fields 
$\tilde{\hg}_R$. The relation in Eq.~\eqref{GWI} is the global Ward identity associated to the residual symmetry algebra (zero modes) of the twisted current 
algebra $\gfrakh(\s)$. We remind that $\st ,\bar{\st}^\pom$ are twisted representation matrices satisfying the orbifold Lie algebra \eqref{Eq 2.3b}, while the 
integers $\bar{n}(r) ,\r(\s)$ are defined in the $H$-eigenvalue problem \cite{Dual,More,Big} of the underlying untwisted theory. Because the twisted inverse 
inertia tensor ${\cL}$ is invariant under the T-duality automorphism $\pom$ (see Eq.~\eqref{Lww=L}), the $\pom$ factors can in fact be omitted from the terms 
in Eq.~\eqref{bWi} with two $\pom$'s.

From a physical point of view, the twisted open-string KZ system \eqref{Eq3.39} shows in the clearest possible terms that the only effect of the duality
automorphism $\pom$ is to globally ``rotate'' the non-abelian image charges $\bar{\st}^{(i)}$ at $\bz_i$.

The general twisted open-string KZ system \eqref{Eq3.39} has the form of a ``doubled" but standard {\it chiral} orbifold KZ system \cite{Big,Big',Perm,so2n} on 
$2n$ variables with a specific choice of twisted representation matrices:
\begin{subequations}
\label{Eq3.41}
\begin{gather}
\hat{F}_\s (\st ,\{ z\}) \equiv \tilde{\hat{F}}_\s (\st,\bz,z) ,\quad \pl_\kappa \hat{F}_\s (\st,\{z\}) =\hat{F}_\s (\st,\{z\}) \hat{W}_\kappa (\st,\{z\},\s) 
\end{gather}
\begin{align}
\hat{W}_\kappa (\st,\{z\},\s) &= 2{\cL}_{\sgb(\s)}^{\nrm;\mnrn}(\s) \BIG{[} \!\sum_{\r \neq \kappa} \left( \frac{z_\r}{z_\kappa} 
  \right)^{\!\nrrsf} \!\!\frac{1}{z_{\kappa \r}} \st_\nrm^{(\r)} \otimes \st_{\mnrn}^{(\kappa)} \bigspc \nn \\
& \bigspc \bigspc \bigspc \quad \quad -\frac{1}{z_\kappa} \nrrs \st_\nrm^{(\kappa)} \st_{\mnrn}^{(\kappa)} \BIG{]} 
\end{align}\vspace{-0.25in}
\begin{gather}
\hat{F}_\s (\st,\{z\}) \sum_{\kappa =1}^{2n} \st_{0,\m}^{(\kappa )} =0 ,\quad \forall \m \\
\pl_\kappa \equiv \frac{\pl}{\pl z_\kappa} ,\quad z_{\kappa \r} \equiv z_\kappa -z_\r ,\quad z_\kappa \equiv \left\{ \begin{array}{ll} z_\kappa , 
   &\kappa =1\ldots n, \\ \bz_{\kappa -n} ,&\kappa =n+1 \ldots 2n \end{array} \right. \\
\st_\nrm^{(\kappa)} \!\equiv \!\left\{ \begin{array}{ll} \st_\nrm (T^{(\kappa)}\!,\s) &\kappa =1\ldots n \\
   \st_\nrm^\pom (\bT^{(\kappa -n)}\!,\s) & \kappa =n+1\ldots 2n  \end{array} \right. \,.
\end{gather}
\end{subequations}
Then it follows as expected \cite{Cardy,Giusto,Orient1,TwGiu} that the $n$-point correlators of our twisted open WZW strings have the same general structure
as the $2n$-point correlators of ordinary (closed-string) orbifold theory. 

We finally remark that, if our initial data was taken from the $\s=0$ sector of any closed-string WZW orbifold (including the trivial orbifold with $H=1$), 
then the general twisted open-string operator algebra reduces to the following untwisted algebra:
\begin{subequations}
\label{G-Untw-OS}
\begin{gather}
[J_a(m) ,J_b(n)] =if_{ab}{}^c J_c(m+n) +G_{ab} m\de_{m+n,0} ,\quad a,b =1,\ldots ,\text{dim } g \label{ALA} \\
[J_a (m) ,\tg (T,\bz,z)] =\tg (T,\bz,z) (T_a z^{-m} +\bT_a^{\omega} \bz^{-m}) \\
L_g (m) = L_g^{ab} \sum_{p\in \Zint} :J_a (p) J_b (m-p): ,\quad L_g^{ab} = \oplus_I \frac{\eta_I^{a(I)b(I)}}{2k_I +Q_I} \\
[L_g(m) ,\tilde{g}(T,\bz,z)] = \tilde{g}(T,\bz,z) \Big{(} (\lpl z + mD_g(T)) z^m + (\overleftarrow{\bpl} \bz + mD_g (\bT^{\omega})) \bz^m \Big{)} 
\end{gather}
\begin{gather}
D_g(T) = L_g^{ab} T_a T_b ,\quad D_g(\bT^\omega ) = L_g^{ab} \bT_a^\omega \bT_b^\omega = D_g(T) ,\quad \bT_a^\omega \equiv \omega_a{}^b \bT_b =-\omega_a{}^b T_b^t \\
[T_a ,T_b] =if_{ab}{}^c T_c ,\quad [\bT_a^\omega ,\bT_b^\omega ]=if_{ab}{}^c \bT_c^\omega ,\quad \omega \in Aut(g) \,.
\end{gather}
\end{subequations}
Here Eq.~\eqref{ALA} is the untwisted affine Lie algebra [40-42,37] on $g$, and $L_g(m)$ are the modes of the affine-Sugawara construction on $g$ 
[42,50-53,37]. The corresponding general untwisted open-string KZ system 
\begin{subequations}
\label{322}
\begin{gather}
\tilde{F}(T,\bz,z) \equiv \langle 0 | \tilde{g}(T^{(1)},\bz_1 ,z_1) \ldots \tilde{g}(T^{(n)},\bz_n ,z_n) |0\rangle \prod_{i=1}^n z_i^{-D_g(T^{(i)})} \otimes 
   \bz_i^{-D_g(\bT^{(i),\omega})} \\
\pl_i \tilde{F} =\tilde{F} 2L_g^{ab} \Big{(} \sum_{j\neq i} \frac{T_a^{(i)} \otimes T_b^{(j)}}{z_i -z_j} +\sum_j \frac{T_a^{(i)} \otimes 
   \bT_b^{(j),\omega }}{z_i -\bz_j} \Big{)} \\
\bpl_i \tilde{F} =\tilde{F} 2L_g^{ab} \Big{(} \sum_{j\neq i} \frac{\bT_a^{(i),\omega } \otimes \bT_b^{(j),\omega }}{\bz_i -\bz_j} +\sum_j 
   \frac{\bT_a^{(i),\omega} \otimes T_b^{(j)}}{\bz_i -z_j} \Big{)} \\
\tilde{F} \sum_{i=1}^n (T_a^{(i)} \!\otimes \!\one + \one \!\otimes \!\bT_a^{(i),\omega} )=0 
\end{gather}
\end{subequations} 
is similarly obtained from Eq.~\eqref{Eq3.39} in the $\s=0$ sector. In this case, the non-abelian image charges $\bT =-T^t$ are globally rotated by an 
automorphism of the untwisted affine algebra. These untwisted open-string KZ equations were stated previously (with $T^{'} \equiv T^\omega$) without 
derivation in Eq.~$(6.33)$ of Ref.~\cite{Orient1}. The special case $\omega =1$ was first derived in Ref.~\cite{Giusto}.

\subsection{Overview of Examples and Subexamples}

Among the twisted open WZW strings in our general construction, we focus here on those classes of examples which have to some extent 
been studied before:

\noindent $\bullet$ Twisted Free-Bosonic Open Strings

The classical description of twisted free-bosonic open strings was given in Subsec.~$2.8$, including their branes and quasi-canonical algebra. These
results were obtained from our general construction by choosing the underlying group manifold to be {\it abelian} (see Eq.~\eqref{eq235}) and substituting 
the twisted free-bosonic data in Eq.~\eqref{Ab-Data}. 

Owing to the linearity of free-boson theories, the corresponding operator results for this class of examples can be obtained by the global substitution 
of commutators for brackets
\begin{gather}
\{ \hat{A} ,\hat{B} \} \, \rightarrow \, [\hat{A} ,\hat{B}] \label{319}
\end{gather}
in all the results of Subsec.~$2.8$. This substitution and the data in Eq.~\eqref{Ab-Data} allow us to apply the operator results of Sec.~3 to the twisted 
free-bosonic open strings. We have already used this prescription implicitly in comparing NN and DD spectra in Subsec.~$3.1$. As further examples of this 
prescription, the open-string sectors of the free-bosonic orientation orbifolds \cite{Orient1, Orient2} are revisited in Subsecs.~$4.3$ and $4.5$.

\noindent $\bullet$ The Basic Class of Twisted Open WZW Strings

The {\it basic class} of twisted open WZW strings \cite{TwGiu} provided the framework for the general construction given here, and is obtained by setting the 
T-duality automorphism $\pom =\thickone$ in all the results of this paper. As an example, we note the classical description of the WZW branes for the
basic class \cite{TwGiu}
\begin{subequations}
\begin{gather}
\heb_\pom (\hx) =\heb (\hx) ,\quad \st^\pom =\st ,\quad \tilde{W} (\st,\s) =\one \\
\hg^{-1} (\st,\xi,t,\s) \pl_+ \hg(\st,\xi,t,\s) =\bigspc \bigspc \bigspc \bigspc \nn \\
=\left\{ \begin{array}{cc} \hg(\st,\xi,t,\s) \pl_- \hg^{-1} (\st,\xi,t,\s) & \text{ at }\xi =0 \\ E(T,\s) \hg(\st,\xi,t,\s) \pl_- \hg^{-1} (\st,\xi,t,\s)
   E^\ast (T,\s) & \text{ at }\xi =\pi \end{array} \right.
\end{gather}
\end{subequations}
as a special case of our general result in Eq.~\eqref{GOE-BCs}. We remind that $E(T,\s)$ in Eq.~\eqref{E(T)} is the eigenvalue matrix of the extended 
$H$-eigenvalue problem \cite{Big, TwGiu} in sector $\s$ of the underlying WZW orbifold $A_g(H)/H$.

This class of examples is special because it is a collection of {\it open-string orbifolds} $A_g^{open} (H)/H$ of open strings $A_g^{open}(H)$, with 
orbifold sectors $\srange$ (where $N_c$ is the number of conjugacy classes of $H$). Bearing in mind that orbifolds involve interactions among sectors, 
it is an open question whether any other complete orbifolds can be assembled from the twisted open strings of our general construction (see the related
comments on orientation orbifolds at the end of this section). Ref.~\cite{TwGiu} also discusses an explicit non-abelian example of the non-commutative 
geometry in the basic class, as well as the twisted free-bosonic constructions in the basic class.

\noindent $\bullet$ The General Untwisted Open WZW String

The general untwisted open WZW string is the special case of our general construction with $\s \!=\!0,\,\pom \!=\!\omega \!\in \!Aut (g)$. In this case, 
the WZW branes are the same at both ends of the string
\begin{subequations}
\label{326}
\begin{gather}
g^{-1} (T,\xi,t) \pl_+ g(T,\xi,t) = W(T) g(T,\xi,t) \pl_- g^{-1} (T,\xi,t) W^{\hc} (T) \,\,\,\text{at } \xi =0,\pi \label{318a} \\
T_a^{\omega} \equiv \omega_a{}^b T_b = W^{\hc} (T) T_a W(T) ,\quad W(T) \in Aut(g) \label{Link2}
\end{gather}
\end{subequations}
where $\omega \in Aut(g)$ is the (untwisted) T-duality automorphism, $W(T)$ is the action of $\omega$ in rep $T$ and Eq.~\eqref{Link2} is isomorphic to the 
ordinary linkage relation \eqref{Link} of orbifold theory. As discussed in Refs.~\cite{Big,Big', so2n} and Subsec.~$2.6$, rep $T$ can be reducible when $W$ 
is an outer automorphism of $g$. The non-commutative geometry for this case is given in Eq.~\eqref{Untw-NC-Geom} and the corresponding open-string KZ system 
appears in Eq.~\eqref{322}.

The subexample with $\omega =W=\thickone$ is the set of basic untwisted open WZW strings constructed in Ref.~\cite{Giusto} -- and this case is also a subset of 
the basic class above. An explicit non-abelian example of the non-commutative geometry for this case was given in an appendix of Ref.~\cite{TwGiu} 
(see also Ref.~\cite{ARS}). The description \eqref{326} of the WZW branes for $\omega =W =\thickone$ was first given in Ref.~\cite{AS}, and additional 
discussion of branes on group manifolds appears in Refs.~[60-69].

\noindent $\bullet$ The Permutation-Twisted Open WZW Strings

The open-string permutation orbifolds
\begin{gather}
\frac{A_g^{open}(H)}{H} ,\quad g =\oplus_{I=1}^K \gfrak^I ,\quad \gfrak^I \simeq \text{ simple } \gfrak ,\quad k_I =k ,\quad H \subseteq S_K
\end{gather}
(i.~e.~the basic class of permutation-twisted open WZW strings at $\pom=1$) were briefly discussed in Ref.~\cite{TwGiu}. More generally, permutation-twisted 
open WZW strings with arbitrary $\pom$ are obtained in our construction when we appropriate the initial data from the left-mover sectors of any 
closed-string WZW permutation orbifold \cite{Big,Big',Perm} on semisimple $g$. This data includes e.~g.~the twisted current algebra $\gfrakh (\s)$
\begin{subequations}
\begin{gather}
[\hj_{\hat{j}aj} (m\!+\!\srac{\hat{j}}{f_j(\s)}) , \hj_{\hat{l}bl} (n\!+\!\srac{\hat{l}}{f_l(\s)})] =\de_{jl} \Big{(} if_{ab}{}^c 
   \hj_{\hat{j} +\hat{l},cj} (m\!+\!n \!+\!\srac{\hat{j}+\hat{l}}{f_j(\s)}) + \bigspc \bigspc \nn \\
\bigspc \bigspc \bigspc \bigspc +\eta_{ab} kf_j(\s) (m\!+\!\srac{\hat{j}}{f_j(\s)}) \de_{m+n+\frac{\hat{j}+\hat{l}}{f_j(\s)} ,0} \Big{)} \\
\nrm \rightarrow \hat{j}aj ,\quad \nrrs \rightarrow \srac{\hat{j}}{f_j(\s)} ,\quad a= 1,\ldots ,\text{dim } \gfrak ,\quad \bar{\hat{j}} =0,\ldots ,f_j(\s) -1 \label{PermLabels}
\end{gather}
\end{subequations}
which in this case is called a {\it general orbifold affine algebra} \cite{Chr,Big,Big',Perm}. This result is given in the cycle notation 
[14-16] for permutations, where $f_j(\s)$ is the length of cycle $j$, and $\hat{j}$ labels the position in each cycle, for example:
\begin{subequations} 
\begin{align} 
& \Zint_\l :  f_j(\s)\! = \!\rho(\s), \,\, \bar{\hat{j}} = 0,\ldots, \rho(\s)\! -\!1, \,\, j = 0,\ldots, 
   \srac{\lambda}{\rho(\s)}\! -\!1, \,\,\s = 0,\ldots, \rho(\s)\!-\!1 \label{62a} \\ 
& \Zint_\l, \,\,\l = \text{prime}: \quad \r(\s) = \l , \quad  \bar{\hat{j}}\! =\! 0,\dots, \l \!-\!1,  
\quad  j\!=\!0, \quad \s =1, \ldots, \l -1 \, \\  
\!\!& \!\!S_N :\,\, f_j(\s) \!=\! \s_j, \,\, \s_{j+1} \leq \s_j, \,\, \bar{\hat{j}} =0,\dots, \s_j \!-\!1, \,\, j \!=\! 0, \dots, n(\vec{\s})-1, \,\, 
    \sum_{j=0}^{n(\vec{\s})-1} \!\!\s_j = \!N \,. \label{62c} 
\end{align} 
\end{subequations} 
The permutation-twisted open strings also contain a {\it general orbifold Virasoro algebra} (with fractional-moded Virasoro
generators), which is similarly inherited from the permutation orbifolds \cite{Chr,DV2,Perm}. We will return to this subject in Sec.~5.

A large subset of these open strings are described by the following twisted representation matrices 
\begin{subequations}
\label{st-Perms}
\begin{gather}
\st_{\hat{j}aj} (T,\s) = T_a t_{\hat{j}j}(\s) ,\quad t_{\hat{j}j}(\s)_{\hat{l}l}{}^{\hat{m}m} = \de_{jl} \de_l{}^m \de_{\hat{j} +\hat{l}-\hat{m},0\, 
   \text{mod } f_j(\s)} \\
\pom (\srac{\hat{j}}{f_j(\s)} ,\s)_j{}^l = e^{\tp \frac{\n_j \hat{j}}{f_j(\s)}} (\pom_P )_j{}^l  \label{321b} \\
\st^\pom_{\hat{j}aj} (T,\s) = T_a e^{\tp \frac{\n_j \hat{j}}{f_j(\s)}} (\pom_P )_j{}^l t_{\hat{j},l}(\s) ,\quad  
   \bar{\st}^\pom_{\hat{j}aj} (T,\s) = \bT_a e^{\tp \frac{\n_j \hat{j}}{f_j(\s)}} (\pom_P )_j{}^l t_{-\hat{j},l}(\s) \\
[T_a ,T_b] =if_{ab}{}^c T_c ,\quad t_{\hat{j}j} (\s) t_{\hat{l}l}(\s) = \de_{jl} t_{\hat{j} +\hat{l},j} (\s)
\end{gather}
\end{subequations}
where $T$ is any irrep of simple $\gfrak$ and $\bT =-T^t$. The matrix $\pom_P$ in Eq.~\eqref{321b} is a permutation which can exchange cycles of the same 
length and $\n_j$ is a cycle-dependent integer. This class of T-duality automorphisms $\pom$ can be further generalized to include Lie automorphisms acting on 
the index $a$, but we will not explore this possibility here. Together with the labelling \eqref{PermLabels} and the twisted inverse inertia tensor 
\cite{Big,Big',Perm}
\begin{gather}
{\cL}_{\sgb(\s)}^{\hat{j}aj ;\hat{l}bl}(\s) = \frac{\de_{jl}}{f_j(\s)} \frac{\eta^{ab}}{2k +Q_\gfraks} \de_{\hat{j}+\hat{l} ,0\,\text{mod }f_j(\s)}
\end{gather}
the twisted representation matrices in Eq.~\eqref{st-Perms} provide the explicit form of the twisted KZ system \eqref{Eq3.39} for the permutation-twisted
open WZW strings.

The branes of these strings are described by the following special case of Eq.~\eqref{GOE-BCs}:
\begin{subequations}
\label{325}
\begin{gather}
\hg^{-1} (\st,\xi,t) \pl_+ \hg(\st,\xi,t) = \bigspc \bigspc \bigspc \bigspc \nn \\
\bigspc \quad \quad =\left\{ \begin{array}{cc} \tilde{W}(\s) \hg(\st,\xi,t) \pl_- \hg^{-1}(\st,\xi,t) \tilde{W}^{\hc} (\s)
   & \text{at } \xi=0 \\   E(\s) \tilde{W}(\s) \hg(\st,\xi,t) \pl_- \hg^{-1}(\st,\xi,t) \tilde{W}^{\hc} (\s) E(\s)^\ast & \text{at } \xi=\pi 
   \end{array} \right. \label{Perm-Branes} \\
E(\s)_{\hat{j}j}{}^{\hat{l}l} = \de_{\hat{j}j}{}^{\hat{l}l} E_{\hat{j}j}(\s) = \de_{\hat{j}j}{}^{\hat{l}l} e^{-\tp \frac{\hat{j}}{f_j(\s)}} \\
\tilde{W}^{\hc} (\s) t_{\hat{j}j}(\s) \tilde{W}(\s) = e^{\tp \frac{\n_j \hat{j}}{f_j(\s)}} (\pom_P )_j{}^l t_{\hat{j},l}(\s) \,. \label{318c}
\end{gather}
\end{subequations}
Here we have used the simplification $E(T,\s) =E(\s)$ for the eigenvalue matrices, which is well-known in the theory of WZW permutation orbifolds 
\cite{Big,Big',Perm}. The solution of the twisted linkage relation $\tilde{W}(\st,\s) =\tilde{W}(\s)$ is similarly independent of twisted 
rep $\st$. The general solution of the reduced linkage relation \eqref{318c} is beyond the scope of this paper, but we remind that 
$\pom =\tilde{W} =\thickone$ for the special case of the basic permutation-twisted open strings \cite{TwGiu}.

In the same fashion, the known left-mover data \cite{Big,Big',Perm,so2n} of the various closed-string WZW orbifolds $A_g(H)/H$ on {\it simple} $g$ can be 
straightforwardly substituted in the general construction above to obtain the corresponding sets of twisted open WZW strings, including in particular 
their branes, twisted non-commutative geometry and twisted open-string KZ equations.

\noindent $\bullet$ The Open-String Orientation-Orbifold Sectors

Another class of twisted open strings contained in our general construction is the set of open-string sectors of the WZW {\it orientation orbifolds} 
\cite{Orient1,Orient2}
\begin{gather}
\frac{A_g (H_-)}{H_-} ,\quad H_- \subset Aut (g\oplus g) \label{327}
\end{gather}
where $H_-$ contains {\it world-sheet orientation-reversing automorphisms}. The closed-string sectors of these orbifolds arise by twisting the 
orientation-preserving subgroup of $H_-$ (and hence these sectors form complete WZW orbifolds by themselves), while the open-string sectors of these 
orbifolds are obtained by twisting the orientation-reversing automorphisms. It follows that, in contrast to the twisted open strings of the basic class, 
the open-string orientation-orbifold sectors do not by themselves form an (open-string) orbifold. Like conventional orientifolds [27-30], the orientation 
orbifolds have an equal number of closed- and open-string sectors, but in distinction to orientifolds, both the open- and closed-string sectors of the 
orientation orbifolds are generically characterized by {\it fractional moding}. Presumably, the open-string sectors of the orientation orbifolds are 
associated to their closed-string sectors by non-planar processes, but we will not study this issue here.

In the following section we revisit the open-string orientation-orbifold sectors as a special case of our general construction, which allows us to discuss
e.~g.~their phase-space realizations, non-commutative geometries and T-dualities for the first time.

\section{Example: Open-String WZW Orientation-Orbifold Sectors}

\subsection{Identification in the General Construction}

As a large example, we turn next to the identification of the open-string sectors of the {\it WZW orientation orbifolds} \cite{Orient1,Orient2} 
\begin{gather}
\frac{A_g(H_-)}{H_-} ,\quad H_- \subseteq \Zint_2 (\text{world-sheet}) \times H ,\quad H \subset Aut(g) \label{41}
\end{gather}
in our general construction. The notation in Eq.~\eqref{41} is a more explicit form of that given in Eq.~\eqref{327}. These twisted open WZW strings are 
characterized by an {\it orbifold Virasoro algebra} (including extra twisted Virasoro generators) and a doubled central charge $\hat{c} =2c_g$, both of
which directly reflect the twisting of the orientation-reversing automorphisms of closed-string WZW models.

We begin this section by reviewing some important background information about the open-string orientation-orbifold sectors. In particular, we may read
off from Ref.~\cite{Orient1} the twisted current algebra $\hat{\gfrak}_O (\s)$
\begin{gather}
[\hj_\nrmu (m\!+\!\nrrs \!+\!\srac{u}{2}) ,\hj_\nsnv (n\!+\!\nsrs \!+\!\srac{v}{2})] = \bigspc \bigspc \bigspc \bigspc \nn \\
  = i\scf_{\nrm;\nsn}{}^{n(r)+n(s),\de}(\s) \hj_{n(r)+n(s),\de ,u+v} (m\!+\!n\!+\! \srac{n(r)+n(s)}{\r(\s)} \!+\!\srac{u+v}{2}) \bigspc \bigspc \nn \\
\bigspc +2 (m\!+\!\nrrs \!+\!\srac{u}{2}) \de_{u+v,0\, \text{mod }2} \de_{m+n+\frac{n(r)+n(s)}{\r(\s)} +\frac{u+v}{2} ,0} \sG_{\nrm;\mnrn}(\s) \label{Eq 4.1a}
\end{gather}
of each open-string orientation-orbifold sector. The notation here follows Ref.~\cite{Orient1}, where $\sG_{\bullet}(\s)$ and $\scf(\s)$ are the ordinary 
twisted tangent-space metric and twisted structure constants of space-time orbifold theory \cite{Dual,More,Big}, while $u,v$ label the additional two-component
structure of these open strings. All quantities in this section are doubly-periodic 
\begin{gather}
n(r)\rightarrow n(r) \pm \r(\s) ,\quad u\rightarrow u \pm 2
\end{gather}
in the extended spectral indices $n(r),u$, where $\r(\s)$ is the order of the Lie part $h_\s$ of the orientation-reversing automorphism $\hat{h}_\s$ in 
the underlying untwisted closed string (see Eq.~\eqref{OR-Auto}). We note in particular that $-u$ is equivalent to $u$, and we denote the pullbacks 
of the spectral indices to their fundamental ranges by $\bar{n}(r) \in \{ 0\ldots \r(\s) -1 \}$ and $\bar{u} \in \{ 0,1 \}$.

Ref.~\cite{Orient1} also gives the fractional-moded {\it orbifold Virasoro algebra} \cite{Chr,DV2, Perm,Orient1,Orient2}
\begin{subequations}
\label{43}
\begin{gather}
\bkspc \hat{L}_u (m\!+\!\srac{u}{2}) =\bigspc \bigspc \bigspc \bigspc \bigspc \bigspc \bigspc \quad \quad \nn \\
  \quad =\!\srac{1}{2} {\cL}_{\sgb(\s)}^{\nrm;\mnrn}(\s) \sum_{v=0}^1 \sum_{p\in \Zint} :\hj_{\nrm v} (p\!+\!\nrrs \!+\!\srac{v}{2}) \hj_{\mnrn ,u-v} 
    (m\!-\!p\!-\!\nrrs \!+\!\srac{u-v}{2})\!: \\
[\hat{L}_u (m\!+\!\srac{u}{2}) , \hj_{\nrm v} (m\!+\!\nrrs \!+\!\srac{v}{2}) ] = -(n\!+\!\nrrs \!+\!\srac{u}{2}) 
   \hj_{\nrm, u+v} (m\!+\!n\!+\!\nrrs \!+\!\srac{u+v}{2}) \\
[\hat{L}_u (m\!+\!\srac{u}{2}) ,\hat{L}_v (n\!+\!\srac{v}{2})] =(m\!-\!n\!+\!\srac{u-v}{2}) \hat{L}_{u+v} (m\!+\!n\!+\!\srac{u+v}{2}) \bigspc \bigspc \nn \\
  \bigspc \bigspc \bigspc +\de_{m+n+\frac{u+v}{2},0} \frac{2c_g}{12} (m\!+\!\srac{u}{2}) ((m\!+\!\srac{u}{2})^2 -1) \label{OrbVirAlg} \\
L_\s (m) \equiv \hat{L}_0(m) \bigspc \bigspc \bigspc \bigspc \bigspc \bigspc \bigspc \nn \\
 \quad \quad =\srac{1}{2} {\cL}_{\sgb(\s)}^{\nrm;\mnrn}(\s) \sum_{u=0}^1 \sum_{p\in \Zint} :\!\hj_\nrmu (p\!+\!\nrrs \!+\!\srac{u}{2}) \hj_{\mnrn ,-u}
   (m\!-\!p\!-\!\nrrs \!-\!\srac{u}{2})\!: 
\end{gather}
\begin{gather}
[L_\s (m), \hj_\nrmu (n\!+\!\nrrs \!+\!\srac{u}{2}) ] = -(n\!+\!\nrrs \!+\!\srac{u}{2}) \hj_\nrmu (m\!+\!n\!+\!\nrrs \!+\!\srac{u}{2}) \\
[L_\s(m) ,L_\s(n)] =(m\!-\!n) L_\s (m\!+\!n) + \de_{m+n,0} \frac{2c_g}{12} m(m^2 -1) \label{Eq 4.1e} 
\end{gather}
\end{subequations}
in each open-string sector, where ${\cL}^{\bullet}_{\sgb(\s)}(\s)$ is the ordinary twisted inverse inertia tensor of space-time orbifold theory 
\cite{Dual,More,Big} and $\{ L_\s (m) \}$ are the ordinary Virasoro generators. The quantity $c_g$ is the central charge of the left- or right-mover
affine-Sugawara construction [42,50-53,37] in the underlying untwisted WZW theory. The orbifold Virasoro algebra has the following 
equal-time form
\begin{subequations}
\begin{gather}
\hat{\Theta}_u^{(\pm)} (\xi,t) \equiv \srac{1}{2\pi} \sum_{m\in \Zint} \hat{L}_u (m\!+\!\srac{u}{2}) e^{-i(m+\frac{u}{2})(t \pm \xi)} ,\quad 
   \hat{T}_\s^{(\pm)} (\xi,t) = \hat{\Theta}_0^{(\pm)} (\xi,t) \label{Theta} \\
[\hat{\Theta}_u^{(+)} (\xi,t) ,\hat{\Theta}_v^{(+)} (\eta,t)] =i\hat{\Theta}_{u+v}^{(+)} (\eta,t) \pl_\xi \de_{\frac{u}{2}} (\xi - \eta) 
   - i\hat{\Theta}_{u+v}^{(+)} (\xi,t) \pl_\eta \de_{\frac{v}{2}} (\eta -\xi) \nn \\
\bigspc -i \frac{2c_g}{24\pi} \de_{u+v ,0\,\text{mod }2} (\pl_\xi^3 +\pl_\xi ) \de_{\frac{u}{2}} (\xi-\eta) \label{OrbVAlg2} 
\end{gather}
\end{subequations}
in terms of the twisted stress tensors $\hat{\Theta}_u$.

The orientation-orbifold current algebra $\gfrakh_O (\s)$ in Eq.~\eqref{Eq 4.1a} is a particular type of ``doubly-twisted" current algebra \cite{TVME,Coset}. 
These current algebras are in fact isomorphic to the left-mover twisted current algebras of the {\it generalized $\Zint_2$ permutation orbifold} on 
$g\oplus g$, whose automorphism group $H'$ includes a general Lie automorphism on $g$:
\begin{gather}
\frac{A_{g\oplus g} (H')}{H'} ,\quad H' \subseteq \Zint_2 \times H ,\quad H \subset Aut(g) \,. \label{Z2}
\end{gather}
(In the WZW orientation orbifolds, the two copies of $g$ are the left- and right-mover sectors of the closed string theory). The doubling $\hat{c}=2c_g$ 
of the central charge in the Virasoro algebra \eqref{Eq 4.1e} and the orbifold Virasoro algebra \eqref{OrbVirAlg}, \eqref{OrbVAlg2} are easily understood in 
this picture. 

The orientation-orbifold operator algebra \eqref{43} also holds in the (left-mover sector of the) generalized $\Zint_2$ permutation orbifold, but the 
orientation orbifolds have a 
{\it distinct} representation theory in terms of the twisted affine primary fields $\tilde{\hg}$. For the commutators with the twisted current modes,
we read again from Ref.~\cite{Orient1}
\begin{subequations}
\label{45}
\begin{gather}
[\hj_\nrmu (m\!+\!\nrrs \!+\!\srac{u}{2}) ,\tilde{\hg}(\st,\xi,t)] =\tilde{\hg}(\st,\xi,t) \Big{(} \st_\nrmu(T,\s) e^{i(m+\nrrsf +\frac{u}{2})(t+\xi)} + \bigspc \quad \quad \nn \\
\bigspc \bigspc \bigspc \bigspc \quad \quad + \bar{\st}_\nrmu (\bT,\s) e^{i(m+\nrrsf +\frac{u}{2})(t-\xi)} \Big{)} \label{Eq 4.1b} 
\end{gather}
\begin{gather}
\st_\nrmu (T,\s) \equiv \st_\nrm (T,\s) \tau_u ,\quad \bar{\st}_\nrmu (\bT,\s) \equiv \st_\nrm (\bT,\s) (-1)^u \tau_u ,\quad u \in \{0,1\} \label{bst-nrmu} \\
\tau_0 =\one_2 ,\quad \vec{\tau} = \,\text{Pauli matrices}
\end{gather}
\end{subequations}
where these results are given in the one-sided notation. The matrices $\st_\nrmu (T,\s)$, \linebreak \noindent $\bar{\st}_\nrmu (\bT,\s)$ are the total 
twisted representation matrices, while $\st_\nrm (T,\s), \st_\nrm (\bT,\s)$ are ordinary twisted representation matrices \cite{Big} constructed from the Lie 
part $h_\s$ of the orientation-reversing automorphism:
\begin{subequations}
\label{OR-Auto}
\begin{gather}
\hat{\omega} (\hat{h}_\s) =\tau_1 \otimes \ws ,\quad \hat{W} (\hat{h}_\s ;T) =\tau_1 \otimes W(h_\s ;T) ,\quad h_\s \in Aut (g) \label{45a} \\
\ws U\hc (\s) =U\hc (\s) E(\s) ,\quad W(h_\s ;T) U\hc (T,\s) =U\hc (T,\s) E(T,\s) \label{Heigen} \\
\st_\nrm (T,\s) =\schisig_\nrm U(\s)_\nrm{}^a U(T,\s) T_a U\hc (T,\s) \,.
\end{gather}
\end{subequations}
Here $\tau_1$ is world-sheet orientation reversal and $\hat{\omega}$ and $\hat{W}$ are the actions of the total orientation-reversing automorphism 
$\hat{h}_\s$, as discussed in Refs.~\cite{Orient1,Orient2}. The relations in Eq.~\eqref{Heigen} are known in closed-string orbifold theory as the 
$H$-eigenvalue problem \cite{Dual,More} and its extended form \cite{Big}, which encode the action of $h_\s$ in various representations of $g$.

We will also need the commutators of the twisted affine primary fields with the Virasoro generators
\begin{subequations}
\begin{gather}
[\hat{L}_u (m\!+\!\srac{u}{2}) ,\tilde{\hg}(\st,\xi,t)] =\tilde{\hg} (\st,\xi,t) \tau_u \Big{(} \big{(} -\!\srac{i}{2} \overleftarrow{\pl_+} 
    +(m\!+\!\srac{u}{2}) \Dg(\st) \big{)} e^{i(m+\frac{u}{2})(t+\xi)} \bigspc \bigspc \nn \\
  \bigspc \bigspc \bigspc + \big{(} -\!\srac{i}{2} \overleftarrow{\pl_-} +(m\!+\!\srac{u}{2}) \Dg (\bar{\st}) \big{)} e^{i(m+\frac{u}{2})(t-\xi)} \Big{)}
\end{gather}
\begin{gather}
[L_\s (m) ,\tilde{\hg}(\st,\xi,t)] =\tilde{\hg} (\st,\xi,t) \Big{(} \big{(} -\!\srac{i}{2} \overleftarrow{\pl_+} +m\Dg(\st) \big{)} e^{im(t+\xi)} \bigspc \bigspc \nn \\
 \bigspc \bigspc \bigspc \quad \quad  + \big{(} -\!\srac{i}{2} \overleftarrow{\pl_-} +m\Dg (\bar{\st}) \big{)} e^{im(t-\xi)} \Big{)} \\
 \quad \quad = e^{imt} \left( -i\cos (m\xi) \pl_t +\sin(m\xi) \pl_\xi \right) \tilde{\hg}(\st,\xi,t) \quad \nn \\
 \bigspc \bigspc \quad \quad +me^{imt} \tilde{\hg}(\st,\xi,t) \left( e^{im\xi} \Dg (\st) +e^{-im\xi} \Dg (\bar{\st}) \right) \label{Eq 4.1g} \\
\Dg (\st) = \sum_{u=0}^1 {\cL}_{\sgb_O (\s)}^{\nrmu;\mnrn ,-u}(\s) \st_\nrmu (T,\s) \st_{\mnrn ,-u}(T,\s) \quad \quad \nn \\
  = {\cL}_{\sgb(\s)}^{\nrm;\mnrn}(\s) \st_\nrm (T,\s) \st_\mnrn (T,\s) \\
\Dg (\bar{\st}) = \sum_{u=0}^1 {\cL}_{\sgb_O (\s)}^{\nrmu;\mnrn ,-u}(\s) \bar{\st}_\nrmu (\bT,\s) \bar{\st}_{\mnrn ,-u}(\bT,\s) = \Dg (\st) \quad \quad \\
{\cL}_{\sgb_O (\s)}^{\nrmu;\nsnv}(\s) = \srac{1}{2} \de_{u+v,0\,\text{mod }2} {\cL}_{\sgb(\s)}^{\nrm;\nsn}(\s)  \label{total-cL}
\end{gather}
\end{subequations}
which are similarly read off from Ref.~\cite{Orient1}. The six-index tensor ${\cL}^{\bullet}$ in Eq.~\eqref{total-cL} is the total twisted inverse inertia 
tensor of the open-string orientation-orbifold sector, while the four-index tensor ${\cL}^{\bullet}$ is the ordinary twisted inverse inertia tensor above --
which is associated to the Lie part $h_\s$ of the orientation-reversing automorphism.

Fractional-moded Virasoro generators are not generic in twisted open strings -- and hence cannot be included explicitly in our general construction.
For open strings, so far as we know, fractional-moded Virasoro generators are specific to the open-string sectors of the orientation orbifolds and the 
permutation-twisted open strings (see Subsec.~$3.4$) -- where they can also be added consistently by hand.

Except for the relations involving the half-integer moded Virasoro generators $\hat{L}_1 (m\!+\!\srac{1}{2})$, all the orientation-orbifold 
results above can in fact be put into the form of our general open-string construction. To do this, one needs the {\it super-index notation} 
$\hat{n}(r)\hat{\m}$ introduced in App.~B of Ref.~\cite{Orient1}:
\begin{subequations}
\label{Eq4.2}
\begin{gather}
\nrm \rightarrow \hat{n}(r) \hat{\m} =\nrmu ,\quad \nrrs \rightarrow \srac{\hat{n}(r)}{\hat{\r}(\s)} =\nrrs \!+\!\srac{u}{2} \label{Eq 4.2a} \\
\sbnrrs \rightarrow \srac{\bar{\hat{n}}(r)}{\hat{\r}(\s)} = \nrrs \!+\!\srac{u}{2} -\lfloor \nrrs \!+\!\srac{u}{2} \rfloor \\
\left( \frac{w}{z} \right)^{\bnrrs} \rightarrow \left( \frac{w}{z} \right)^{\frac{\bar{\hat{n}}(r)}{\hat{\r}(\s)}} = \left( \frac{w}{z} \right)^{\bar{y}(r,u)}
   f(z,w;\bar{y}(r,u)) 
\end{gather}
\begin{gather}
\bar{y}(r,u) \equiv \sbnrrs +\srac{\bar{u}}{2} ,\quad f (z,w;\bar{y}) \equiv 1 +\frac{z-w}{w} \theta (\bar{y} \geq 1) \label{48d} \\
\sG_{\nrm;\nsn}(\s) \rightarrow \sG_{\hat{n}(r)\hat{\m} ;\hat{n}(s)\hat{\n}} (\s) =\sG_{\nrmu;\nsnv}(\s) =2\de_{u+v,0\,\text{mod }2} \sG_{\nrm;\nsn}(\s) \label{Eq 4.3c} \\
{\cL}_{\sgb(\s)}^{\nrm;\nsn}(\s) \rightarrow {\cL}_{\sgb_O (\s)}^{\hat{n}(r)\hat{\m} ;\hat{n}(s)\hat{\n}} (\s) ={\cL}_{\sgb_O (\s)}^{\nrmu;\nsnv}(\s) =
  \srac{1}{2} \de_{u+v,0\,\text{mod }2} {\cL}_{\sgb(\s)}^{\nrm;\nsn}(\s) \,.
\end{gather}
\end{subequations}
According to the super-index map $\rightarrow$ above, one obtains the orientation-orbifold results from our general development as follows:
Take any $\nrm$ result of this paper, rewrite it as an $\hat{n}(r)\hat{\m}$ result, and then expand into the $\nrmu$ notation of the orientation orbifolds.
For instance, Eqs.~($4.10$e,f) illustrate this procedure in the case of the twisted tangent-space metric and inverse inertia tensor of this paper. The final
(four-index) quantities $\sG_{\bullet}(\s)$ and ${\cL}^{\bullet}(\s)$ in Eq.~($4.10$e,f) are the ordinary twisted tensors associated to the Lie part $h_\s$ 
of the orientation-reversing automorphism. We will also need the super-index form of the extended $H$-eigenvalue problem in the orientation orbifolds 
\cite{Orient1}
\begin{subequations}
\begin{gather}
\hat{W} (\hat{h}_\s ;T) \hat{U}^{\hc} (T,\s) = \hat{U}^{\hc} (T,\s) \hat{E} (T,\s) \\
E(T,\s) \rightarrow \hat{E}(T,\s) = \tau_3 E(T,\s)  \label{OO-ExtHEig} 
\end{gather}
\end{subequations}
where $\hat{W}$ is the action in rep $T$ of the total orientation-reversing automorphism $\hat{h}_\s$ (see Eq.~\eqref{45a}).

As another example, we give the super-index form of the (ordinary) orientation-orbifold stress tensors in Eq.~\eqref{Theta}
\begin{gather}
\hat{\Theta}_0 (\pm \xi,t) =\hat{T}_\s^{(\pm)} (\xi,t) =\srac{1}{2\pi} {\cL}_{\sgb(\s)}^{\hat{n}(r)\hat{\m} ;\hat{n}(s)\hat{\n}} (\s)
   :\!\hj_{\hat{n}(r)\hat{\m}}^{(\pm)} (\xi,t) \hj_{\hat{n}(s)\hat{\n}}^{(\pm)} (\xi,t) \!: \nn \\
\bigspc =\srac{1}{4\pi} {\cL}_{\sgb(\s)}^{\nrm;\mnrn}(\s) \sum_{u=0}^1 :\! \hj_{\nrmu}^{(+)} (\xi,t) \hj_{\mnrn ,-u}^{(+)} (\xi,t) \!: 
\end{gather}
which is now understood as nothing but the super-index map from the general open-string stress tensors in Eq.~\eqref{Eq3.1}. Similarly, the 
orientation-orbifold operator algebra in Eqs.~\eqref{Eq 4.1a}, \eqref{43} and \eqref{Eq 4.1g} follows by the super-index map as a special case of the 
general twisted operator algebra in Eqs.~\eqref{Eq3.1} and \eqref{Eq3.2}. 

To complete the identification of the open-string orientation-orbifold sectors in our general construction, we focus on the $\hj ,\hg$ commutator in 
Eq.~\eqref{Eq 4.1b}, which may be compared to the image of Eq.~\eqref{311} under the super-index map \eqref{Eq4.2}:
\begin{gather}
[\hj_{\hat{n}(r)\hat{\m}} (m\!+\!\srac{\hat{n}(r)}{\hat{\r}(\s)}) ,\tilde{\hg}(\st,\xi,t)] = \tilde{\hg}(\st,\xi,t) \Big{(} \st_{\hat{n}(r)\hat{\m}} (T,\s)
   e^{i(m+\frac{\hat{n}(r)}{\hat{\r}(\s)})(t+\xi)} + \bigspc \nn \\
\bigspc \bigspc \bigspc \bigspc + \bar{\st}^\pom_{\hat{n}(r)\hat{\m}} (T,\s) e^{i(m+\frac{\hat{n}(r)}{\hat{\r}(\s)})(t-\xi)} \Big{)} \,.
\end{gather}
In this comparison, one sees first that
\begin{gather}
\st_{\hat{n}(r)\hat{\m}}(T,\s) = \st_\nrmu (T,\s) = \st_\nrm (T,\s) \tau_u 
\end{gather}
and more importantly, with Eq.~\eqref{Eq3.37}, we may now identify the T-duality automorphism $\pom$
\begin{subequations}
\label{Eq4.3}
\begin{gather}
\bar{\st}^\pom_{\hat{n}(r)\hat{\m}} (T,\s) = \bar{\st}^\pom_\nrmu (T,\s) = -\pom(n(r),u,\s)_\m{}^\n \st_{\nrn u}(T,\s)^t  \label{4.3a} \\
\bar{\st}_\nrmu (\bT,\s) = \st_\nrm (\bT,\s) (-1)^u \tau_u = -\st_\nrmu(T,\s)^t (-1)^u \tau_u \label{4.3b}  \\
\bar{\st}^\pom_{\hat{n}(r)\hat{\m}} (T,\s) = \bar{\st}_\nrmu (\bT,\s)  \\
\Rightarrow \pom (n(r),u,\s)_\m{}^\n = (-1)^u \de_\m{}^\n  ,\quad \bar{u}\in \{ 0,1\}  \label{48b}
\end{gather}
\end{subequations}
in the open-string orientation-orbifold sectors. Here Eq.~\eqref{4.3a} repeats the super-index map, while Eq.~\eqref{4.3b} appeared earlier in 
Eq.~\eqref{bst-nrmu}. Finally, it is easily checked that this $\pom$ is indeed a mode-number-independent automorphism because the transformed current
\begin{gather}
\bkspc \hj_\nrmu (m\!+\!\nrrs \!+\!\srac{u}{2})^\prime \!\equiv \pom(n(r),\!u,\s)_\m{}^\n \hj_{\nrn u} (m\!+\!\nrrs \!+\!\srac{u}{2}) \!=\!(-1)^u \hj_\nrmu 
   (m\!+\!\nrrs \!+\!\srac{u}{2}) \label{Eq4.4}
\end{gather}
satisfies the doubly-twisted current algebra $\gfrakh_O (\s)$ in Eq.~\eqref{Eq 4.1a}.

In summary, the results above tell us that the open-string sectors of the orientation orbifolds can be viewed as twisted open strings built according to 
the construction of this paper, and in this case:

\noindent $\bullet$ the initial left-mover data is appropriated from the generalized $\Zint_2$ permutation orbifold $A_{g\oplus g}(H')/H'$ in Eq.~\eqref{Z2},

\noindent $\bullet$ the T-duality automorphism $\pom$ is that given in Eq.~\eqref{48b}.

\noindent Using this characterization, the following subsection discusses the classical description and T-dual partners of the open-string orientation-orbifold
sectors, and another application of this characterization is found in Sec.~5.

\subsection{Classical Description of the WZW Orientation Orbifolds}

Refs.~\cite{Orient1,Orient2} described certain aspects of the classical theory of open-string WZW orientation-orbifold sectors, including in particular the 
classical equations of motion of the group orbifold elements and the action formulation of each sector on the solid half cylinder. Using Sec.~2, the 
super-index map of the previous subsection and $\pom =(-1)^u$, we may now read off many new results in this classical theory.

We begin with the {\it phase-space} description of the open-string orientation-orbifold sectors
\begin{subequations}
\label{417}
\begin{gather}
\hj_\nrmu^{(+)}(\xi) = 2\pi \hei_\nrmu{}^{\ntd w} \hp_{\ntd w} (\hB)+ \pl_\xi \hx_\s^{\ntd w} \he_{\ntd w}{}^{\nsn u} \sG_{\nsn;\nrm}(\s) \\
\hj_\nrmu^{(-)}(\xi) =(-1)^u\Big{(} 2\pi \hebi_\nrmu{}^{\ntd w} \hp_{\ntd w} (\hB)+ \pl_\xi \hx_\s^{\ntd w} \heb_{\ntd w}{}^{\nsn u} \sG_{\nsn;\nrm}(\s) \Big{)}\end{gather}
\begin{align}
\hat{H}_\s &=\int_0^\pi \!\!d\xi \, \Big{(} \hat{T}_\s^{(+)} (\xi,t) +\hat{T}_\s^{(-)} (\xi,t) \Big{)} \bigspc \bigspc \bigspc \bigspc \bigspc \quad \nn \\
&=\srac{1}{4\pi} \!\int_0^\pi \!\!d\xi \, \sG^{\nrmu;\nsnv}(\s) \Big{(} \hj^{(+)}_\nrmu(\xi,t) \hj^{(+)}_\nsnv (\xi,t)+ \hj^{(-)}_\nrmu(\xi,t) 
   \hj^{(-)}_\nsnv (\xi,t) \Big{)} \nn \\
&=\srac{1}{8\pi} \!\int_0^\pi \!\!\!d\xi \, \sG^{\nrm;\nsn}(\s) \sum_{u=0}^1 \Big{(} \hj^{(+)}_\nrmu(\xi,t) \hj^{(+)}_{\nsn u} (\xi,t) \!+ \!
  \hj^{(-)}_\nrmu(\xi,t) \hj^{(-)}_{\nsn u} (\xi,t) \Big{)} 
\end{align}
\begin{align}
&= \!\int_0^\pi \!\!d\xi \Big{(} 2\pi \hG^{\nrmu;\nsnv}(\hx(\xi)) \hp_\nrmu (\hB,\xi) \hp_\nsnv (\hB,\xi) +\bigspc \bigspc \nn \\
&\bigspc \bigspc \bigspc  \quad +\srac{1}{8\pi} \pl_\xi \hx_\s^\nrmu (\xi) \hx_\s^\nsnv (\xi) \hG_{\nrmu;\nsnv}(\hx(\xi)) \Big{)} \\
& \bigspc \hG_{\nrmu;\nsnv} (\hx) = \he(\hx)_\nrmu{}^{\ntd w} \he(\hx)_\nsnv{}^{\nue y} \sG_{\ntd w;\nue y}(\s) 
\end{align}
\begin{gather}
\{ \hj_\nrmu^{(+)}(\xi,t) ,\hx_\s^\nsnv (\eta,t) \} =-\tp \Big{(} \hei(\eta)_\nrmu{}^{\!\!\nsnv} \de_{\bar{y}(r,u)}(\xi -\eta) \bigspc \bigspc \bigspc \nn \\
  \bigspc \bigspc \bigspc +(-1)^u \hebi (\eta)_\nrmu{}^{\!\!\nsnv} \de_{\bar{y}(r,u)} (\xi+\eta) \Big{)} \\
\{ \hj_\nrmu^{(-)}(\xi,t) ,\hx_\s^\nsnv (\eta,t) \} =-\tp \Big{(} (-1)^u \hebi(\eta)_\nrmu{}^{\!\!\nsnv} \de_{-\bar{y}(r,u)}(\xi -\eta) \bigspc \bigspc  \nn \\
  \bigspc \bigspc \bigspc +\hei (\eta)_\nrmu{}^{\!\!\nsnv} \de_{-\bar{y}(r,u)} (\xi+\eta) \Big{)} \,.
\end{gather}
\end{subequations}
which are special cases of the general results in Subsecs.~$2.2 ,\, 2.3$ and $2.4$. The phase-modified delta functions $\de_{\bar{y}} (\xi \pm \eta)$ 
are defined as in Eq.~\eqref{Eq2.2} with $\nrrs \rightarrow \bar{y}$.

The following {\it coordinate-space} results can similarly be read from the general results of Subsec.~$2.6$:
\begin{subequations}
\label{418}
\begin{gather}
\hp_\nrmu (\hB,\xi,t) =\srac{1}{4\pi} \hG_{\nrmu;\nsnv} (\hx(\xi,t)) \pl_t \hx_\s^\nsnv (\xi,t) \\
\hj_\nrmu^{(+)} (\xi,t,\s) = \pl_+ \hx_\s^\nsnv (\xi,t) \he(\xi,t)_\nsnv{}^{\ntd u} \sG_{\ntd ;\nrm}(\s) \\
\hj_\nrmu^{(-)} (\xi,t,\s) = (-1)^u \pl_- \hx_\s^\nsnv (\xi,t) \heb(\xi,t)_\nsnv{}^{\ntd u} \sG_{\ntd ;\nrm}(\s) \\
\hat{S}_\s =\int \!\!dt \int_0^\pi \!\!d\xi \hat{\cL}_\s ,\quad \hat{\cL}_\s =\srac{1}{8\pi} ( \hG_{\nrmu ;\nsnv} +\hB_{\nrmu ;\nsnv} ) \pl_+ \hx_\s^\nrmu \pl_- \hx_\s^\nsnv \label{OO-L-Dens} \\
\{ \hx_\s^\nrmu (\xi,t) ,\hx_\s^\nsnv (\eta,t) \} =\left\{ \begin{array}{ll} -i\pi\hPs_\s^{\nrmu;\nsnv} (0,0) &\text{if } \xi=\eta=0, \\
   i\pi \hPs_\s^{\nrmu;\nsnv} (\pi,\pi) & \text{if } \xi=\eta=\pi, \\ 0 & \text{otherwise} \end{array} \right. \label{OO-NC-Geom}
\end{gather}
\begin{align}
\!\!&\hPs_\s^{\nrmu;\nsnv}(\xi,\xi) = -\hPs_\s^{\nsnv;\nrmu} (\xi,\xi) \bigspc \bigspc \bigspc \\
&\quad =\sG^{\ntd w;\nue y}(\s) e^{-2i(\frac{n(t)}{\r(\s)} +\frac{w}{2})\xi} \Big{(} (-1)^w \hebi (\xi)_{\ntd w}{}^{\!\!\!\nrmu} \hei(\xi)_{\nue y}
  {}^{\!\!\!\nsnv} + \nn \\
&\bigspc \bigspc \bigspc + (-1)^{w+1} \hei(\xi)_{\nue y}{}^{\!\!\!\nrmu} \hebi (\xi)_{\ntd w}{}^{\!\!\!\nsnv} \Big{)} 
\end{align}
\begin{align}
&\quad =\srac{1}{2} \sG^{\ntd;\nue}(\s) \sum_{w=0}^1 e^{-2i(\frac{n(t)}{\r(\s)} +\frac{w}{2})\xi} \Big{(} (-1)^w \hebi (\xi)_{\ntd w}{}^{\!\!\!\nrmu} 
  \hei(\xi)_{\nue ,-w}{}^{\!\!\!\nsnv} + \quad \nn \\
&\bigspc \bigspc \bigspc + (-1)^{w+1} \hei(\xi)_{\nue ,-w}{}^{\!\!\!\nrmu} \hebi (\xi)_{\ntd w}{}^{\!\!\!\nsnv} \Big{)} \,.
\end{align}
\end{subequations} 
In Eq.~\eqref{OO-L-Dens}, we have given the twisted sigma-model form of the {\it complete} action \cite{Orient2} for this case, in agreement with the 
general bulk Lagrange density in Eq.~\eqref{LDens}. We remind the reader that the complete open-string orientation-orbifold action in terms of group 
orbifold elements on the solid half cylinder is also given in Ref.~\cite{Orient2}.

Moreover, the following equivalent descriptions of the {\it WZW orientation-orbifold branes}
\begin{subequations}
\begin{gather}
\hj^{(+)}_\nrmu (0,t,\s) =\hj^{(-)}_\nrmu (0,t,\s) ,\quad \hj^{(+)}_\nrmu (\pi,t,\s) =e^{-\tp (\nrrsf +\frac{u}{2})} \hj_\nrmu^{(-)} (\pi,t,\s) \\
\pl_t \hx_\s^\nrmu (\xi,t) \Big{(} (-1)^v \heb (\xi,t)_\nrmu{}^\nsnv -\he (\xi,t)_\nrmu{}^\nsnv \Big{)} \bigspc \bigspc \quad \quad \nn \\
  \pl_\xi \hx_\s^\nrmu (\xi,t) \Big{(} (-1)^v \heb (\xi,t)_\nrmu{}^\nsnv +\he (\xi,t)_\nrmu{}^\nsnv \Big{)} \,\,\text{at } \xi =0 \\
\pl_t \hx_\s^\nrmu (\xi,t) \Big{(} e^{\tp \nsrsf} \heb (\xi,t)_\nrmu{}^\nsnv -\he (\xi,t)_\nrmu{}^\nsnv \Big{)} \bigspc \bigspc \quad \quad \nn \\
  \pl_\xi \hx_\s^\nrmu (\xi,t) \Big{(} e^{\tp \nsrsf} \heb (\xi,t)_\nrmu{}^\nsnv +\he (\xi,t)_\nrmu{}^\nsnv \Big{)} \,\,\text{at } \xi =\pi 
\end{gather}
\begin{gather}
\tilde{W}(\st,\s) =\tau_3 : \quad \st_\nrmu^\pom = \st_\nrm (-1)^u \tau_u =\tau_3 \st_\nrm \tau_u \tau_3 =\tau_3 \st_\nrmu \tau_3 \label{OO-Tw-Link} \\
\hj^{(+)} (\st,\xi,t,\s) =\hj_\nrmu^{(+)} (\xi,t,\s) \sG^{\nrmu ;\nsnv} (\s) \st_\nsnv \\
\hj^{(-)} (\st^\pom ,\xi,t,\s) =\hj_\nrmu^{(-)} (\xi,t,\s) \sG^{\nrmu ;\nsnv} (\s) (-1)^v \st_\nsnv \\
\sG^{\nrmu;\nsnv}(\s) = \srac{1}{2} \de_{u+v ,0\,\text{mod 2}} \sG^{\nrm;\nsn}(\s) 
\end{gather}
\begin{gather}
\hj^{(+)} (\st,0,t,\s) =\tau_3 \hj^{(-)} (\st^\pom,0,t,\s) \tau_3 \\
\hj^{(+)} (\st,\pi,t,\s) =E(T,\s) \hj^{(-)} (\st^\pom,\pi,t,\s) E(T,\s)^\ast \\
\hg^{-1}(\st,\xi,t,\s) \pl_+ \hg(\st,\xi,t,\s) =\bigspc \bigspc \bigspc \bigspc \nn \\
   \bigspc \quad \quad =\left\{ \begin{array}{cc} \tau_3 \hg(\st,\xi,t,\s) \pl_- \hg^{-1} (\st,\xi,t,\s) \tau_3 & \text{at }\xi=0 , \\
   E(T,\s) \hg(\st,\xi,t,\s) \pl_- \hg^{-1} (\st,\xi,t,\s) E(T,\s)^\ast & \text{at }\xi= \pi \end{array} \right. \label{OO-Branes}
\end{gather}
\end{subequations}
follow straightforwardly from the general results of Subsec.~$2.7$. We note in particular the simple solution $\tilde{W}$ in Eq.~\eqref{OO-Tw-Link} obtained 
in this case from the twisted linkage relation in Eq.~\eqref{Tw-Link}. Using this $\tilde{W}$ and $\hat{E}(T,\s) =\tau_3 E(T,\s)$ from Eq.~\eqref{OO-ExtHEig}, 
the $\hg$ description of the orientation-orbifold branes in Eq.~\eqref{OO-Branes} follows as a special case of the general WZW brane description in 
Eq.~\eqref{GOE-BCs}. The quantity $E(T,\s)$ is the eigenvalue matrix of the extended $H$-eigenvalue problem \cite{Big,Orient1} associated to the Lie part (see 
Eq.~\eqref{45a}) of the automorphism in rep $T$.

The description of the WZW orientation-orbifold branes in Eqs.~($4.19$h,i,j) is exactly that given in Ref.~\cite{Orient2} under the identifications:
\begin{gather}
\hj^{(+)} (\st) \rightarrow \hj(\st) ,\quad \hj^{(-)} (\st^\pom) \rightarrow \hjb (\st) \,.
\end{gather}
Aside from Eqs.~($4.19$h,i,j), the classical description of the open-string orientation-orbifold sectors in this subsection is new.

We finally consider {\it T-duals} to the open-string sectors of the general WZW orientation orbifolds: 
\begin{gather}
\frac{A_g(H_-)}{H_-} ,\quad H_- \subset \Zint_2 (\text{world-sheet}) \times H ,\quad H \subset Aut(g) \,.
\end{gather}
According to the characterization at the end of Subsec.~$4.1$, the open-string orientation-orbifold sectors are constructed using the left-mover data of the 
generalized (closed-string) $\Zint_2$ permutation orbifold $A_{g\oplus g}(H')/H'$ in Eq.~\eqref{Z2}. It follows that the open-string sectors of the 
WZW orientation orbifolds are T-dual (with $\pom =(-1)^u$) to the corresponding sectors of the {\it generalized open-string permutation orbifold}
\begin{gather}
\frac{A_{g\oplus g}^{open} (H')}{H'} ,\quad H' \subset \Zint_2 \times H ,\quad H \subset Aut(g)
\end{gather}
which are constructed using the {\it same} left-mover data, but with T-duality automorphism $\pom =1$. In particular, the twisted current algebra of 
$A_{g\oplus g}^{open}(H')/H'$ is the same $\gfrakh_O (\s)$ in Eq.~\eqref{Eq 4.1a}, while the $\hj \hg$ commutators in these sectors differ from those of 
Eq.~\eqref{45} only by the omission of the $(-1)^u$ factor.

We will illustrate this T-duality with a simple example, by comparing the branes in the single twisted sector of the simplest open-string permutation orbifold 
(see Ref.~\cite{TwGiu} and Subsec.~$3.4$)
\begin{gather}
\frac{A_{g\oplus g}^{open}(\Zint_2)}{\Zint_2} \label{Z2Perm-Orb}
\end{gather}
with those of the single twisted sector of the simplest orientation orbifold
\begin{gather}
\frac{A_{g}(H_-)}{H_-} ,\quad H_- = \,\Zint_2 (\text{world-sheet}) \,. \label{Basic-OO}
\end{gather}
The untwisted sectors of these two orbifolds are respectively an open and a closed WZW string. Their twisted sectors are T-dual with the same twisted current
algebra
\begin{subequations}
\begin{gather}
[\hj_{au} (m\!+\!\srac{u}{2}), \hj_{bv} (n\!+\!\srac{v}{2})] =if_{ab}{}^c \hj_{c,u+v} (m\!+\!n\!+\srac{u+v}{2}) +2(m\!+\!\srac{u}{2}) \de_{m+n+\frac{u+v}{2},0} G_{ab} \\
a,b,c =1,\ldots ,\text{dim }g ,\quad \bar{u},\bar{v} \in \{0,1 \}
\end{gather}
\end{subequations}
and the same Hamiltonian, but representation theories which differ by the T-duality automorphism:
\begin{subequations}
\begin{gather}
[\hj_{au} (m\!+\!\srac{u}{2}) ,\tilde{\hg}(\st,\xi,t)] = \tilde{\hg}(\st,\xi,t) \Big{(} \st_{au} e^{i(m+\frac{u}{2})(t+\xi)} + (\pom \bar{\st})_{au}
   e^{i(m+\frac{u}{2}) (t-\xi)} \Big{)} \\
\st_{au} = T_a \tau_u ,\quad \bar{\st}_{au} = \bT_a \tau_u ,\quad [T_a,T_b ]=if_{ab}{}^c T_c ,\quad [\bT_a ,\bT_b] =if_{ab}{}^c \bT_c \\
\frac{A_{g\oplus g}^{open}(\Zint_2)}{\Zint_2} \,: \quad \pom =\tilde{W}(\st) =\one ,\quad E(T,\s) =E(\s) =\tau_3 \bigspc \bigspc  \\
\frac{A_{g}(\Zint_2 (\text{w.~s.}))}{\Zint_2 (\text{w.~s.})} \,: \quad \hat{\omega} (\hat{h}_\s) =\tau_1 ,\quad \pom =(-1)^u ,\quad \tilde{W}(\st) =\tau_3
  ,\quad E(T,\s) =\one \,.
\end{gather}
\end{subequations}
Then using Eqs.~\eqref{Perm-Branes} and \eqref{OO-Branes} for the branes, we find the boundary conditions
\begin{subequations}
\label{Tdual-Branes}
\begin{gather}
\frac{A_{g\oplus g}^{open}(\Zint_2)}{\Zint_2} \,: \quad \hg^{-1} (\st) \pl_+ \hg(\st) = \left\{ \begin{array}{cc} \hg(\st) \pl_- \hg^{-1}(\st) & \text{at }
   \xi=0 \\  \tau_3 \hg(\st) \pl_- \hg^{-1}(\st) \tau_3 & \text{at } \xi=\pi \end{array} \right. \\
\frac{A_{g}(\Zint_2 (\text{w.~s.}))}{\Zint_2 (\text{w.~s.})} \,: \quad \hg^{-1} (\st) \pl_+ \hg(\st) = \left\{ \begin{array}{cc} \tau_3 \hg(\st) \pl_- 
   \hg^{-1}(\st) \tau_3 & \text{at } \xi=0 \\ \hg(\st) \pl_- \hg^{-1}(\st) & \text{at } \xi=\pi \end{array} \right. \
\end{gather}
\end{subequations}
which tell us that T-duality reverses the branes in this case. Interestingly, the action of T-duality is not so simple for the corresponding 
non-commutative geometry in Eqs.~\eqref{Tw-NC-Geom} and \eqref{418}, which we leave as an exercise for the reader.

\subsection{Subexample: Free-Bosonic Orientation Orbifolds}

Certain aspects of the open-string sectors of the free-bosonic orientation orbifolds 
\begin{gather}
\frac{A_g (H_-)}{H_-} ,\quad H_- \subseteq \Zint_2 (\text{w.~s.}) \times H ,\quad H \subset Aut(g) ,\quad g \,\text{ abelian}
\end{gather}
were presented in Refs.~\cite{Orient1,Orient2}, including their quasi-canonical algebra and action formulation:
\begin{gather}
\hat{S}_\s = \srac{1}{4\pi} \sG_{\nrm;\mnrn}(\s) \int \!\!dt \int_0^\pi \!\!d\xi \sum_{u=0}^1 \pl_+ \hx^{\nrmu}_\s \pl_- \hx^{\mnrn, -u}_\s \,.
\end{gather}
Drawing on the development above, we turn now to a more complete discussion of these open-string sectors.

In the abelian limit \eqref{Ab-Data} of the results \eqref{417}, \eqref{418}, we find the phase-space description of the open-string sectors of the 
free-bosonic orientation orbifolds \cite{Orient1,Orient2}, for example:
\begin{subequations}
\begin{gather}
\hj_\nrmu^{(+)}(\xi) =2\pi \hp_\nrmu^\s (\xi) +\pl_\xi \hx_\s^{\nsn u} \sG_{\nsn;\nrm}(\s) \\
\hj_\nrmu^{(-)}(\xi) =(-1)^u \Big{(} -2\pi \hp_\nrmu^\s (\xi) +\pl_\xi \hx_\s^{\nsn u} \sG_{\nsn;\nrm}(\s) \Big{)} \\
\!\!\![ \hj_\nrmu^{(+)}(\xi,t) ,\hx_\s^\nsnv (\eta,t) ] =-\tp \de_\nrmu{}^{\!\!\!\nsnv} \Big{(} \de_{\bar{y}(r,u)}(\xi-\eta) \!+(-1)^{u+1} 
   \de_{\bar{y}(r,u)}(\xi+\eta) \Big{)}
\end{gather}
\begin{gather}
[\hj_\nrmu^{(-)}(\xi,t) ,\hx_\s^\nsnv (\eta,t)] =\bigspc \bigspc \bigspc \bigspc \quad \quad \nn \\
  \quad \quad =-\tp \de_\nrmu{}^{\!\!\!\nsnv} \Big{(} (-1)^{u+1} \de_{-\bar{y}(r,u)}(\xi-\eta) +\de_{-\bar{y}(r,u)}(\xi+\eta) \Big{)} \\
\pl_t \hx_\s^{\nrm 0}(\xi,t) = \pl_\xi \hx_\s^{\nrm 1} (\xi,t) =0 \,\,\,\text{at } \xi=0 \\
\cos (\nrrs \pi) \pl_t \hx_\s^\nrmu (\xi,t) -i\sin (\nrrs \pi) \pl_\xi \hx_\s^\nrmu (\xi,t) =0 \,\,\,\text{at } \xi=\pi \label{Eq 4.8f} 
\end{gather}
\begin{gather}
\hp_\nrmu^\s (\xi,t) =\srac{1}{2\pi} \sG_{\nrm;\nsn}(\s) \pl_t \hx_\s^{\nsn ,-u} (\xi,t) \\
\hPs_\s^{\nrmu;\nsnv}(\xi,\xi) = i\sG^{\nrm;\nsn}(\s) (-1)^u \sin \!\Big{(} 2(\nrrs \!+\!\srac{u}{2}) \xi \Big{)} \,\de_{u+v,0\,\text{mod }2}
\,.
\end{gather}
\end{subequations}
Here we have used the standard quantization $\{ \,,\,\} \rightarrow [\,,\,]$ in Eq.~\eqref{319} to map the classical results into their operator analogues. 
We emphasize that each of these twisted open-string sectors collects a set $\{ \nrmu \}$ of boundary conditions ($4.30$e,f) whose corresponding set of twisted 
free bosons are sufficient to realize the orbifold Virasoro algebra in Eq.~\eqref{OrbVirAlg}. 

We turn next to consider the open-string free-boson mode expansions obtained from the general mode expansions \eqref{Ab-hx-Mode} via the super-index
map \eqref{Eq4.2}. For this, it is helpful to recall the dictionary \cite{Orient1}
\begin{subequations}
\begin{gather}
\{ \hx^{0\m} \} \rightarrow \{ \hx^{\hat{n}(r)\hat{\m}} |\bar{\hat{n}}(r)=0 \} = \{ \hx^{0\m0} ,\hx^{\frac{\r(\s)}{2},\m 1} \} \label{Eq 4.5a} \\
\{ \hx^\nrm |\bar{n}(r) \!\neq \!0 \} \rightarrow \{ \hx^{\hat{n}(r)\hat{\m}} |\bar{\hat{n}}(r) \!\neq \!0 \} = \{ \hx^{\nrm 0} |\bar{n}(r) \!\neq \!0 \} \cup 
   \{ \hx^{\nrm 1} |\bar{n}(r) \!\neq \!\srac{\r(\s)}{2} \} \label{425b}
\end{gather}
\end{subequations}
which follows from the super-index map. As seen in Eq.~\eqref{Eq 4.5a}, the bosons with super-index $\bar{\hat{n}}(r) =0$ split into two cases when the 
order $\r(\s)$ of $\ws$ is even (because there are two solutions of this condition in terms of $\bar{n}(r)$ and $u$). Similarly, the bosons with $\bar{\hat{n}}
(r) \neq 0$ split into the two cases shown in Eq.~\eqref{425b}. Then the general mode expansions above take the specific form:
\begin{subequations}
\label{Ori-Orb-hx}
\begin{align}
&\quad \quad \hx^{0\m0}(\xi,t) =\hat{q}^{0\m 0} +\sG^{0\m 0;0\n 0}(\s) \Big{(} 2\hj_{0\n 0}(0)\xi +2 \sum_{m\neq 0} e^{-imt} \sin (m\xi) 
   \frac{\hj_{0\n 0}(m)}{m} \Big{)} \\
&\hx^{\frac{\r(\s)}{2},\m 1}(\xi,t) =\hat{q}^{\frac{\r(\s)}{2},\m 1} +\sG^{\frac{\r(\s)}{2},\m 1 ;\frac{\r(\s)}{2},\n 1} \Big{(} 2\hj_{\frac{\r(\s)}{2},\n 1} 
  (0)t + \nn \\
& \bigspc \bigspc \bigspc +2i \sum_{m\neq -1} e^{-i(m+1)t} \cos ((m+1)\xi) \frac{\hj_{\frac{\r(\s)}{2},\m 1}(m+1)}{m+1} \Big{)} \\ \vspace{-0.2in}
&\foot{\bar{n}(r)\neq 0\!:} \nn \\
&\quad \hx^{\nrm 0}(\xi,t) =\hat{q}^{\nrm 0} +\nn \\
& \bigspc +2\sG^{\nrm 0;\nsn 0}(\s) \sum_{m\in \Zint} e^{-i(m+\nsrsf)t} \sin ((m\!+\!\nsrs) \xi) \frac{\hj_{\nsn 0}(m \!+\!\nsrs)}{m\!+\!\nsrs} 
\end{align}\vspace{-0.2in}
\begin{align}
&\foot{\bar{n}(r)\neq \srac{\r(\s)}{2} \!:} \nn \\
& \,\, \hx^{\nrm 1}(\xi,t) =\hat{q}^{\nrm 1} +\nn \\
& \quad  +2i\sG^{\nrm 1;\nsn 1}(\s) \sum_{m\in \Zint} e^{-i (m+\nsrsf +\srac{1}{2})t} \cos ((m\!+\!\nsrs \!+\!\srac{1}{2})\xi ) \frac{\hj_{\nsn 1}
  (m\!+\!\nsrs \!+\!\srac{1}{2})}{m\!+\!\nsrs \!+\!\srac{1}{2}} \\
&\bigspc \bigspc \sG^{\nrmu ;\nsnv}(\s) =\srac{1}{2} \de_{u+v,0\,\text{mod }2} \sG^{\nrm;\nsn}(\s) \,. \label{Eq 4.6f}
\end{align}
\end{subequations}
Here $\{ \hat{q}^\nrmu \}$ are the bosonic zero modes and Eq.~\eqref{Eq 4.6f} expresses the total inverse twisted metric (see Eq.~\eqref{Eq 4.3c}) in 
terms of the ordinary inverse twisted metric of space-time orbifold theory. 

Similarly, the general quasi-canonical algebra \eqref{Ab-Alg} reduces in this case to the following:
\begin{subequations}
\label{OOAbAlg}
\begin{align}
& [\hat{p}^\s_\nrmu (\xi,t) ,\hat{p}^\s_{\nsn v} (\eta,t)] = \nn \\
&\bigspc \quad \quad \quad =\frac{(-1)^{u+1}}{4\pi} \sG_{\nrmu ;\nsn v}(\s) \pl_\xi \Big{(} \sin ((\xi +\eta) \bar{y}(r,u))\, 
   \de(\xi+\eta) \Big{)} \label{pp-Comm} \\
&[\hx_\s^\nrmu (\xi,t) ,\hat{p}^\s_{\nsn v}(\eta,t)] = \bigspc \bigspc \bigspc \bigspc \nn \\
& \bigspc \quad = i\de_{\nsn v}{}^{\!\!\nrmu} \Big{(} \de(\xi -\eta) +(-1)^{u+1} \cos ((\xi+\eta) \bar{y}(r,u)) \,\de(\xi+\eta) \Big{)} \label{xp-Comm} 
\end{align}
\begin{align}
& [\hx_\s^\nrmu (\xi,t) ,\hx_\s^{\nsn v}(\eta,t)] = -2\pi \sG^{\nrmu ;\nsn v}(\s) \left\{ \begin{array}{cc} 0 & \text{if } \xi=\eta=0, \\ \sin (2\pi \nrrs) &
   \text{if } \xi=\eta =\pi , \\ 0 & \text{otherwise,} \end{array} \right. \label{xx-Comm} \\
&\sG^{\nrmu ;\nsn v}(\s) =\srac{1}{2} \de_{u+v,0\, \text{mod }2} \sG^{\nrm;\nsn}(\s) ,\quad \bar{y}(r,u) =\srac{\bar{n}(r)}{\r(\s)} +\srac{\bar{u}}{2} \,.
\end{align}
\end{subequations}
This is precisely the algebra given for the free-boson orientation orbifolds in Ref.~\cite{Orient1}. 

This concurrence is somewhat surprising however, because the mode expansions in Eq.~\eqref{Ori-Orb-hx} are in fact slightly more general than those of 
the open-string orientation-orbifold sectors \cite{Orient1}. While the $\hj$ terms of both expansions are identical, the expressions here contain the 
{\it extra} zero-mode terms $\hat{q}^{\nrm 0}$, which are constrained to vanish in the orientation orbifolds by the {\it world-sheet parity}\footnote{Like the 
twisted Virasoro operators, the world-sheet parity \eqref{hx-WSP} is specific to the open-string sectors of the orientation orbifolds \cite{Orient1,Orient2}. 
Extra relations of this type are discussed more generally in the following subsection.}
\begin{gather}
\hx^\nrmu (-\xi,t) =(-1)^{u+1} \hx^\nrmu (\xi,t) \,\,\Rightarrow \,\, \hat{q}^{\nrm 0}=0 \label{hx-WSP}
\end{gather}
of the twisted free bosons. The resolution of this puzzle is that the additional constraint $\hat{q}^{\nrm 0}=0$ does not in fact change the quasi-canonical 
algebra \eqref{Ab-Alg} because the quantities $\hat{q}^{\nrm 0}$ commute with all operators:
\begin{subequations}
\begin{gather}
[\hj_\nrmu (m\!+\!\nrrs \!+\!\srac{u}{2}) ,\hat{q}^{\nsn 0}] =-i(\one -(-1)^0 \one )_\m{}^\n \de_{m+\nrrsf+\frac{u}{2},0} \de_{n(s),0\,\text{mod }\r(\s)} =0 \\
[\hat{q}^\nrmu ,\hat{q}^{\nsn 0} ] =0 \,.
\end{gather}
\end{subequations}
These vanishing commutators are easily seen by using the super-index map and $\pom =(-1)^u$ in the general algebra \eqref{Jq-Comm}, \eqref{qq-Comm}.

In a broader context, we note that the extra $\hat{q}^{\nrm 0}$ terms in the expansions \eqref{Ori-Orb-hx} show that the open-string orientation-orbifold 
sectors are embedded in a larger class of ``nearby" twisted open strings.

Finally, we compare these free-bosonic open-string orientation-orbifold sectors at $\pom =(-1)^u$ to the T-dual sectors of the corresponding generalized
open-string permutation orbifolds
\begin{gather}
\frac{A_{g\oplus g}^{open} (H')}{H'} ,\quad H' \subseteq \Zint_2 \times H ,\quad H\subset Aut(g) ,\quad g \, \text{ abelian} \label{OpenZ2}
\end{gather}
which live in our construction at $\pom =\thickone$. (The corresponding T-duality for non-abelian $g$ was discussed in the previous subsection.) For ease of 
comparison we present the branes and the non-commutative geometry of the free-bosonic open-string orbifolds in the same $\{ \nrmu \}$ notation, where 
$\bar{u} =0,1$ is the spectral index associated to the $\Zint_2$ in Eq.~\eqref{OpenZ2}. This notation allows us to use the super-index map \eqref{Eq4.2} with 
Eqs.~\eqref{Ab-hx-BCs} and \eqref{Ab-xx-Comm} to obtain the following results at $\pom =1$:
\begin{subequations}
\begin{gather}
\pl_t \hx^\nrmu (\xi,t) =0 \,\,\,\text{at } \xi =0 \\
\left. \begin{array}{c} \cos (\pi \nrrs) \pl_t \hx^{\nrm 0}(\xi,t) -i\sin (\pi \nrrs) \pl_\xi \hx^{\nrm 0}(\xi ,t)=0 \\
  i\sin (\pi \nrrs) \pl_t \hx^{\nrm 1} (\xi,t) -\cos (\pi \nrrs) \pl_\xi \hx^{\nrm 1}(\xi,t) =0 \end{array} \right\} \quad \text{at } \xi =\pi \\
[ \hx^\nrmu (\xi,t) ,\hx^\nsnv (\eta,t) ] = (-1)^{u+1} 2\pi \sG^{\nrmu ;\nsnv} (\s) \left\{ \begin{array}{cc} 0 & \text{at } \xi =\eta =0 , \\
\sin (\srac{2\pi n(r)}{\r(\s)}) & \text{at } \xi =\eta =\pi , \\ 0 & \text{otherwise.} \end{array} \right.
\end{gather}
\end{subequations}
These results are very different from the T-dual orientation-orbifold results at $\pom =(-1)^u$ in Eqs.~($4.30$e,f) and \eqref{xx-Comm}. We note however that,
in the case of trivial $H$ (i.~e.~$\bar{n}(r) =0$), T-duality simply reverses the branes
\begin{subequations}
\begin{gather}
\pom =\one : \,\, u=0 \,\text{ is } \, DD \quad ; \quad u=1 \,\text{ is }\, DN \quad \\
\pom =(-1)^u : \,\, u=0 \,\text{ is }\, DD \quad ; \quad  u=1 \,\text{ is }\, ND \quad \quad
\quad 
\end{gather}
\end{subequations}
as noted above for the more general non-abelian case in Eq.~\eqref{Tdual-Branes}.

\subsection{The Twisted KZ Systems of the WZW Orientation Orbifolds}

We return now to the open-string sectors of the general WZW orientation orbifold, as special cases of our operator results in Subsecs.~$3.2$ and $3.3$.

We begin with the one-sided form \eqref{One-S-TVOE} of the twisted vertex operator equations, which, under the super-index map, takes the following form:
\begin{subequations}
\label{OO-TVOE}
\begin{gather}
\srac{1}{2} \pl_+ \tilde{\hg}_R(\st,\bz,z) =\sum_{u=0}^1 i {\cL}_{\sgb(\s)}^{\nrm;\mnrn}(\s) \Big{(} :\!\hj_\nrmu^{(+)} (\xi,t) \tilde{\hg}_R(\st,\bz,z) \!:_M \bigspc \nn\\
  \quad \quad -(\bar{y}(r,u) -\theta (\bar{y}(r,u) \geq 1)) \tilde{\hg}_R(\st,\bz,z) \st_\nrmu \nn \\
 \bigspc +\left( \frac{\bz}{z} \right)^{\bar{y}(r,u)} \frac{f(z,\bz;\bar{y}(r,u))}{1-\bz/z} \tilde{\hg}_R(\st,\bz,z) \bar{\st}_\nrmu^\pom \Big{)} \st_{\mnrn u}
\end{gather}
\begin{gather}
\srac{1}{2} \pl_- \tilde{\hg}_R(\st,\bz,z) =\sum_{u=0}^1 i {\cL}_{\sgb(\s)}^{\nrm;\mnrn}(\s) \Big{(} :\!\hj_\nrmu^{(-)} (\xi,t) \tilde{\hg}_R(\st,\bz,z) \!:_M \bigspc \nn\\
  \quad \quad -(\bar{y}(r,u) -\theta (\bar{y}(r,u) \geq 1)) \tilde{\hg}_R(\st,\bz,z) \bar{\st}^\pom_\nrmu \nn \\
 \bigspc +\left( \frac{z}{\bz} \right)^{\bar{y}(r,u)} \frac{f(\bz,z;\bar{y}(r,u))}{1-z/\bz} \tilde{\hg}_R(\st,\bz,z) \st_\nrmu \Big{)} \bar{\st}_{\mnrn u}^\pom  \\
z= e^{i(t+\xi)} ,\quad \bz =e^{i(t-\xi)} \\
\left( \frac{z}{\bz} \right)^{\bar{y}(r,u)} \equiv e^{2i \bar{y}(r,u) \xi} ,\quad \left( \frac{\bz}{z} \right)^{\bar{y}(r,u)} \equiv e^{-2i \bar{y}(r,u) \xi} \,.
\end{gather}
\end{subequations}
Here $\tilde{\hg}_R$ is the reduced twisted affine primary field defined in Eq.~\eqref{hg-Red}, and the function $f(\scrs{\bullet} ;\bar{y})$ is defined in 
Eq.~\eqref{48d}.

We may similarly obtain the twisted open-string KZ equations for the WZW orientation orbifolds: 
\begin{subequations}
\label{Eq4.24}
\begin{align}
&\!\!\pl_i \tilde{\hat{F}}_\s (\st,\bz,z) \!=\! \tilde{\hat{F}}_\s (\st,\bz,z) \hat{W}_i (\st,\bz,z,\s) ,\,\,\,
   \bpl_i \tilde{\hat{F}}_\s (\st,\bz,z) \!=\! \tilde{\hat{F}}_\s (\st,\bz,z) \hat{\bar{W}}_{\!i} (\st,\bz,z,\s) 
\end{align}
\begin{align}
\hat{W}_i (\st,\bz,z,\s) \equiv & {\cL}_{\sgb(\s)}^{\nrm;\mnrn}(\s) \sum_{u=0}^1 \BIG{[} \BIG{(} \sum_{j\neq i} \left( \frac{z_j}{z_i} 
  \right)^{\!\!\bar{y}(r,u)} \!\frac{f (z_i,z_j ;\bar{y}(r,u))}{z_i -z_j} \st_\nrmu^{(j)} \bigspc \nn \\
&\quad \quad \quad +\sum_j \left( \frac{\bz_j}{z_i} \right)^{\!\!\bar{y}(r,u)} \!\frac{f(z_i,\bz_j ;\bar{y}(r,u))}{z_i -\bz_j} \bar{\st}_\nrmu^{(j),\pom} \BIG{)}
\otimes \st_{\mnrn ,-u}^{(i)} \nn \\
&\quad \quad \quad -\frac{1}{z_i} \left( \bar{y}(r,u) -\theta(\bar{y}(r,u) \geq 1) \right) \st_\nrmu^{(i)} \st_{\mnrn ,-u}^{(i)} \BIG{]} \label{Eq 4.24b} 
\end{align}
\begin{align}
\hat{\bar{W}}_{\!i} (\st,\bz,z,\s) \equiv & {\cL}_{\sgb(\s)}^{\nrm;\mnrn}(\s) \sum_{u=0}^1 \BIG{[} \BIG{(} \sum_{j\neq i} \left( \frac{\bz_j}{\bz_i} 
  \right)^{\!\!\bar{y}(r,u)} \!\frac{f (\bz_i,\bz_j ;\bar{y}(r,u))}{\bz_i -\bz_j} \bar{\st}_\nrmu^{(j),\pom} \nn \\
&\quad \quad \quad +\sum_{j} \left( \frac{z_j}{\bz_i} \right)^{\!\!\bar{y}(r,u)} \!\frac{f (\bz_i,z_j ;\bar{y}(r,u))}{\bz_i -z_j} \st_\nrmu^{(j)} \BIG{)} 
   \otimes \bar{\st}_{\mnrn ,-u}^{(i),\pom} \nn \\
&\quad \quad \quad -\frac{1}{\bz_i} \left( \bar{y}(r,u) -\theta(\bar{y}(r,u) \geq 1) \right) \bar{\st}_\nrmu^{(i),\pom} \bar{\st}_{\mnrn ,-u}^{(i),\pom} \BIG{]} \label{Eq 4.24c}
\end{align}
\begin{gather}
\tilde{\hat{F}}_\s (\st,\bz,z) \left( \sum_{i=1}^n (\st_{0,\m,0}^{(i)} \!\otimes \!\one +\one \!\otimes \!\bar{\st}_{0,\m,0}^{(i),\pom}) \right) =0 ,\quad \forall \m   \label{Eq 4.24d} \\
\tilde{\hat{F}}_\s (\st,\bz,z) \left( \sum_{i=1}^n (\st_{\r(\s)/2,\m,1}^{(i)} \!\otimes \!\one +\one \!\otimes \!\bar{\st}_{\r(\s)/2,\m,1}^{(i),\pom}) \right) =0 ,\quad \forall \m ,\quad 
   \text{for } \r(\s) \text{ even} \label{Eq 4.24e} \\
\st^{(i)}_\nrmu \!\equiv \st_\nrm (T^{(i)}\!,\s) \tau_u ,\quad \bar{\st}^{(i),\pom}_\nrmu \!\equiv \st_\nrm(\bT^{(i)}\!,\s) (-1)^u \tau_u ,\quad \pl_i \equiv \frac{\pl}{\pl z_i} ,\quad \bpl_i
  \equiv \frac{\pl}{\pl \bz_i} \\
\left( \frac{z_j}{z_i} \right)^{\bar{y}(r,u)} =e^{i{\bar{y}(r,u)} (t_j+\xi_j -t_i -\xi_i)} 
  ,\quad \left( \frac{\bz_j}{z_i} \right)^{\bar{y}(r,u)} =e^{i{\bar{y}(r,u)} (t_j -\xi_j -t_i -\xi_i)} \\
\left( \frac{z_j}{\bz_i} \right)^{\bar{y}(r,u)} =e^{i{\bar{y}(r,u)} (t_j+\xi_j -t_i +\xi_i)} 
  ,\quad \left( \frac{\bz_j}{\bz_i} \right)^{\bar{y}(r,u)} =e^{i{\bar{y}(r,u)} (t_j -\xi_j -t_i +\xi_i)} \\
\bar{y}(r,u) =\sbnrrs +\srac{\bar{u}}{2} ,\quad f (z,w;\bar{y}) = 1 +\frac{z-w}{w} \theta (\bar{y} \geq 1) \,.
\end{gather}
\end{subequations}
Here the correlators $\tilde{\hat{F}}_\s$ of the reduced fields $\tilde{\hg}_R$ are defined in Eq.~\eqref{Fred}. Except for the phases in Eqs.~($4.40$g,h), 
this system matches the one-sided form of the twisted open-string KZ system given in Eq.~$(4.24)$ of Ref.~\cite{Orient1}. The issue of phase conventions will be
further discussed in the following subsection.

Aside from the half-integer moded Virasoro generators in Subsec.~$4.1$, there are other relations which are specific to the open-string WZW 
orientation-orbifold sectors (and hence are not included in the general development of this paper). In particular, it is known from 
Refs.~\cite{Orient1,Orient2} that the orientation-orbifold vertex operator equations \eqref{OO-TVOE} and twisted KZ equations \eqref{Eq4.24} are consistent 
with the following special relations on the twisted affine primary fields $\tilde{\hg}$
\begin{subequations}
\label{Constraints}
\begin{gather}
\tilde{\hg} (\st(\bT),\xi,t)^t =\tilde{\hg} (\st(T),-\xi,t) \tau_3 \otimes \tau_3 \label{hg-WSP} \\
\tilde{\hg} (\st(T),\xi,t) \tau_1 \otimes \one = \tilde{\hg}(\st(T),\xi,t) \one \otimes \tau_1 \label{hg-2Comp} 
\end{gather}
\end{subequations}
where $\vec{\tau}$ are the Pauli matrices. In these results, Eq.~\eqref{hg-WSP} describes the behavior of the twisted affine primary fields under world-sheet 
parity and Eq.~\eqref{hg-2Comp} specifies their two-component structure. The free-boson world-sheet parity \eqref{hx-WSP} is equivalent to \eqref{hg-WSP} 
in the limit of abelian $g$.

\subsection{The Singularity Prescription Revisited: Euclidean vs. Minkowski}

In this subsection, we will contrast the Euclidean-space phase and singularity prescription of Ref.~\cite{Orient1} with the Minkowski-space phase and 
singularity prescription of this paper. Of course either prescription can be rotated to the other space, but the two prescriptions are in fact different. As 
we shall see in an instructive consistency check below, the Minkowski-space prescription obtained in this paper for all twisted open WZW strings is the 
correct one for the open-string sectors of orientation orbifolds. 

In Ref.~\cite{Orient1}, when we performed the sums on $u$ in the twisted Euclidean vertex operator equations and/or the twisted Euclidean KZ equations, we 
found structures of the form
\begin{subequations}
\label{Eq4.11}
\begin{gather}
\sum_{u=0}^1 (-1)^u \left( \frac{\bz}{z} \right)^{\!\!\bar{y}(r,u)} \!\frac{f (z,\bz ; \bar{y}(r,u))}{z -\bz} =\left( \frac{\bz}{z} 
   \right)^{\frac{\bar{n}(r)}{\r(\s)}} 
\left\{ \begin{array}{ll} \frac{1 -(\bz/z )^{\frac{1}{2}}}{z-\bz} & \text{when } \srac{\bar{n}(r)}{\r(\s)} < \srac{1}{2} \\ 
    \frac{1 -(z/\bz )^{\frac{1}{2}}}{z-\bz} & \text{when } \srac{\bar{n}(r)}{\r(\s)} \geq \srac{1}{2} \end{array} \right. \label{Eq 4.11a} \\
\bigspc \bigspc =\left\{ \begin{array}{ll} \frac{1}{2z} +\Ord (\bz -z) &\text{when } \srac{\bar{n}(r)}{\r(\s)} < \srac{1}{2} \\
   -\frac{1}{2z} +\Ord (\bz -z) & \text{when } \srac{\bar{n}(r)}{\r(\s)} \geq \srac{1}{2} \end{array} \right.  \label{Eq 4.11b}
\end{gather}
\end{subequations}
which describe the interaction of a charge at $z$ with its image charge at $\bz$. As shown in Eq.~\eqref{Eq 4.11b}, we posited there (in parallel with the
standard Euclidean treatment of $(z/w)^{1/2}$) that such terms were non-singular as $\bz \rightarrow z$ because $(\bz /z)^{1/2} \rightarrow 1$. In proposing 
this Euclidean-space prescription, we were motivated by common practice in untwisted open-string theory \cite{Cardy}, but in that case fractional powers are 
not encountered. Similarly, the Minkowski-space prescription of this paper is based on the evaluation of a conditionally convergent sum (see Eq.~$(4.13)$ 
of Ref.~\cite{TwGiu}). Both prescriptions therefore merit consistency checks.

The Minkowski-space prescription in Eqs.~($4.40$g,h) defines the factors in Eq.~\eqref{Eq 4.11a} as follows:
\vspace{-0.1in}
\begin{subequations}
\label{Mink-Rx}
\begin{gather}
z \rightarrow e^{i(t+\xi)} ,\quad \bz \rightarrow e^{i(t-\xi)} \\
\frac{1 -(\bz/z)^{\frac{1}{2}}}{z-\bz}= \frac{1- e^{-i\xi}}{2ie^{it}\sin \xi} ,\quad \frac{1 -(z/\bz)^{\frac{1}{2}}}{z-\bz} 
  =\frac{1- e^{i\xi}}{2ie^{it}\sin \xi} \,.  \label{Mink-Rx2}
\end{gather} 
\end{subequations}
This evaluation, although non-singular at $\xi =0$, is {\it singular} as $\xi \rightarrow \pi$ (that is, as $\bz \rightarrow z \!\in \!\Real^-$). Both 
prescriptions are in agreement that there is no singularity at $\xi =0$, however owing to fractional powers, the two prescriptions differ near $\xi =\pi$. 
Which prescription is correct?

To resolve this issue, we will consider an example where we can be completely explicit. As background for this example (see Refs.~\cite{Orient1,Orient2}), we 
begin with the untwisted free-bosonic closed-string conformal field theory
\begin{subequations}
\begin{gather}
[J_\m (m) ,J_\n(n)] =[\bJ_\m (m),\bJ_\n (n)] =mG_{\m \n} \de_{m+n,0} \\
[J_\m (m), g(T,\xi,t)] \!=\!g(T,\xi,t) T_\m e^{im(t+\xi)} ,\,\,\,\, [\bJ_\m (m) ,g(T,\xi,t)] \!=\!-T_\m g(T,\xi,t) e^{im(t-\xi)} \\
[T_\m ,T_\n ]=0 ,\quad c =\bar{c} =\text{dim }[\m ]
\end{gather}
\end{subequations}
where $T_\m$ are the abelian ``momenta", and we choose the following simple orientation-reversing automorphism:
\begin{subequations}
\begin{gather}
\hat{w}(\hat{h}_\s) = \tau_1 \otimes (-\one ) ,\quad \ws = -\one \\
J_\m (m)' =-\bJ_\m (m) ,\quad \bJ_\m (m)' =-J_\m (m) ,\quad g(T,\xi,t)' =g(\bT ,-\xi,t)^t \,.
\end{gather}
\end{subequations}
Twisting this orientation-reversing automorphism gives the following data for the corresponding open-string orientation-orbifold sector \cite{Orient1}
\begin{subequations}
\label{Ab-OO-Data}
\begin{gather}
\bar{n}(r) \!=\!1 ,\,\,\, \r(\s) \!=\!2 ,\,\,\,\, \sG_{1\m u;1\n v}(\s) \!=2\de_{u+v,0\,\text{mod }2} G_{\m \n} ,\,\,\,\, 
    {\cL}_{\sgb_O (\s)}^{1\m u;1\n v}(\s) \!=\!\srac{1}{4} \de_{u+v,0\,\text{mod }2} G^{\m \n}  \\ 
\st_{1\m u} =T_\m \tau_u ,\quad \st^\pom_{1\m u} =(-1)^u T_\m \tau_u ,\quad [\st_{1\m u} ,\st_{1\n v}] =[\st^\pom_{1\m u} ,\st^\pom_{1\n v}] =0 \\
 \Dg(\st(T)) \rightarrow \Delta (T)  =\srac{1}{2} G^{\m \n} T_\m T_\n  \label{TwFrB} 
\end{gather}
\end{subequations}
where $\st$ and $\st^\pom$ are the twisted abelian ``momenta". For this sector, the action and orbifold Virasoro generators are given by
\begin{subequations}
\begin{gather}
\hat{S}_\s = \srac{1}{4\pi} G_{\m \n} \int \!\!dt \int_0^\pi \!\!d\xi \sum_{u=0}^1 \pl_+ \hx^{1\m u} \pl_- \hx^{1\n ,-u} \\
\bkspc \hat{L}_u (m \!+\!\srac{u}{2}) = \srac{1}{4} G^{\m \n} \sum_{v=0}^1 \sum_{p\in \Zint} :\! \hj_{1\m v} (p\!+\!\srac{1+v}{2}) \hj_{1\n ,u-v} (m\!-\!p +\!
  \srac{u-v-1}{2}) \!:_M + \de_{m+\frac{u}{2},0} \frac{\text{dim }[\m ]}{16} \\
\hat{c}  = 2\text{dim } [\m ] 
\end{gather}
\end{subequations}
where the $u=0$ and $1$ coordinates are respectively DN and NN.

Substitution of the data \eqref{Ab-OO-Data} into the general results above gives the following quasi-canonical algebra and mode expansions
\begin{subequations}
\begin{gather}
[\hp_{1\m u} (\xi,t) ,\hp_{1\n v}(\eta,t)] = \frac{(-1)^{u+1}}{2\pi} G_{\m \n} \de_{u+v,0\,\text{mod }2} \pl_\xi \Big{(} \sin (\srac{1+\bar{u}}{2} (\xi +\eta))
   \de (\xi +\eta) \Big{)} \\
\!\![\hx^{1\m u}(\xi,t) ,\hp_{1\n v} (\eta,t)] \!= i \de_\n{}^\m \de_{u+v,0\,\text{mod }2} \Big{(} \de(\xi-\eta) \!+\!(-1)^{u+1} \cos (\srac{1+\bar{u}}{2} (\xi +\eta))
   \de (\xi +\eta) \Big{)} \\
[\hx^{1\m u}(\xi,t) ,\hx^{1\n v} (\eta,t)] =0
\end{gather}
\begin{gather}
\hx_{1\m u} (\xi,t) \equiv \sG_{1\m u;1\n v} (\s) \hx^{1\n v}(\xi,t) = 2G_{\m \n} \hx^{1\n ,-u} (\xi,t) \\
\text{DN:}\,\,\,\, \hx_{1\m 0}(\xi,t) =2\sum_{m\in \Zint} \frac{\hj_{1\m 0}(m\!+\!\srac{1}{2})}{m\!+\!\srac{1}{2}} e^{-i(m+\frac{1}{2})t} 
   \sin ((m\!+\!\srac{1}{2})\xi) \\
\quad \text{NN:}\,\,\,\, \hx_{1\m 1}(\xi,t) =\hat{q}_{1\m 1} +2\hj_{1\m 1}(0)t +2i \sum_{m\neq 0} \frac{\hj_{1\m 1}(m)}{m} e^{-imt} \cos (m\xi) \label{NN}
\end{gather}
\begin{gather}
[\hj_{1\m u}( m\!+\!\srac{u+1}{2}) ,\hj_{1\n v}( n\!+\!\srac{v+1}{2})] =2(m\!+\!\srac{u+1}{2}) G_{\m \n} \de_{m+n+1+\frac{u+v}{2},0} \\
[\hj_{1\m u}(m\!+\!\srac{u+1}{2}) ,\hat{q}_{1\n 1}] =-4i \de_{m+\frac{u+1}{2},0} G_{\m \n} ,\quad [\hat{q}_{1\m 1},\hat{q}_{1\n 1}]=0 \\
[\hj_{1\m u}(m\!+\!\srac{u+1}{2}) ,\hg(\st,\xi,t)] = \hg(\st,\xi,t) T_\m \tau_u e^{i(m+\frac{u+1}{2})(t+\xi)} \bigspc \bigspc \bigspc  \nn \\ 
  \bigspc \bigspc \bigspc \bigspc \quad \quad +(-1)^{u+1} T_\m \tau_u \hg(\st,\xi,t) e^{i(m+\frac{u+1}{2})(t-\xi)}  
\end{gather}
\end{subequations}
in agreement with the results of Ref.~\cite{Orient1}.

For our consistency checks, we will need two further results which are similarly read from the data in Eq.~\eqref{Ab-OO-Data} and the general 
results above. First we give the twisted vertex operator equations for this example
\begin{subequations}
\label{Ab-TVOE}
\begin{gather}
\pl_t \hg_R(\st(T),\xi,t) = \srac{i}{2} G^{\m\n} \sum_{u=0}^1 :\left( \hj^{(+)}_{1\m u}(\xi,t) -(-1)^u \hj_{1\m u}^{(-)} (\xi,t) \right) \hg_R(\st(T),\xi,t)
  T_\n \tau_u \!:_M \\
\pl_\xi \hg_R(\st(T),\xi,t) = \srac{i}{2} G^{\m\n} \sum_{u=0}^1 :\!\left( \hj^{(+)}_{1\m u}(\xi,t) +(-1)^u \hj_{1\m u}^{(-)} (\xi,t) \right) \hg_R(\st(T),\xi,t)
  T_\n \tau_u \!:_M \quad \quad \quad \nn \\
-\hg_R(\st(T),\xi,t) \Delta (T) \frac{(1-\cos \xi)}{\sin \xi} \label{4.20b} \\
\hj^{(\pm)}_{1\m u} (\xi,t) = \sum_{m\in \Zint} \hj_{1\m u} (m\!+\!\srac{u+1}{2} ) e^{-i(m+\frac{u+1}{2})(t\pm \xi)} \\
\hg_R (\st(T),\xi,t) \equiv e^{-2it\Delta(T)} \hg(\st(T),\xi,t) \label{hg-Resc} \\
 [T_\m ,\hg] =[\tau_u ,\hg] =0
\end{gather}
\end{subequations}
where $\hg_R$ is the two-sided form of the reduced primary field \eqref{hg-Red}. The one-sided form of this system can be immediately read from 
Eqs.~\eqref{OO-TVOE} and \eqref{Ab-OO-Data}, but the two-sided form in Eq.~\eqref{Ab-TVOE} is more convenient for the purposes of this discussion. We note
in particular that the singularity at $\xi =\pi$ of the last term of \eqref{4.20b} is a consequence of the Minkowski-space prescription \eqref{Mink-Rx},
whereas in the Euclidean-space prescription (rotated to Minkowski space) this term would be non-singular. Except for this issue, Eq.~\eqref{Ab-TVOE} is 
exactly the Minkowski-space version of the Euclidean twisted vertex operator equations in Ref.~\cite{Orient1}.

The second result needed is the $L_\s ,\hg$ commutator in Eq.~\eqref{Eq 3.3b}, which reduces in this case to the following:
\begin{gather}
[L_\s (m) ,\hg_R (\st(T),\xi,t)] = e^{imt} \Big{(} \cos (m\xi) ( -i\pl_t +2(m\!+\!1) \Delta(T) ) \bigspc \quad \quad \nn \\
  \bigspc \bigspc \bigspc \bigspc +\sin (m\xi) \pl_\xi \Big{)} \hg_R(\st(T),\xi,t) \,. \label{Eq4.18}
\end{gather}
It is easily checked from the corresponding result in Ref.~\cite{Orient1} that this equation holds in Minkowski space independent of which singularity 
prescription is used.

For the computations below, we reexpress the integer-moded coordinates in Eq.~\eqref{NN} in terms of the conventional NN coordinates:
\begin{subequations}
\begin{gather}
J_\m (m) \equiv \srac{1}{\sqrt{2}} \hj_{1\m 1}(m) ,\quad q_\m \equiv \srac{1}{2\sqrt{2}} \hat{q}_{1\m 1} \\
\srac{1}{2} \hx_{1\m 1}(\xi,t) =\sqrt{2} \Big{(} q_\m +tJ_\m (0) +i\sum_{m\neq 0} \frac{J_\m (m)}{m} e^{-imt}\cos (m\xi) \Big{)} \label{FV-hx} \\
[q_\m ,J_\n (0)] =iG_{\m \n} ,\quad [J_\m (m), J_\n (n)] =mG_{\m \n} \de_{m+n,0} \,.
\end{gather}
\end{subequations}
After some algebra, we then verify that the following normal-ordered exponential 
\begin{subequations}
\label{AbOO-TVO}
\begin{gather}
\hg_R(\st(T),\xi,t) =\hat{\hg}(\st(T),\xi,t) (1+\cos \xi)^{\Delta (T)} \label{TVO-1} \\
\hat{\hg} (\st(T),\xi,t) \equiv \,: \!e^{i \hx_{1\m u} (\xi,t) \sG^{1\m u;1\n v}(\s) \st_{1\n v}} \!: \,=\, :\!e^{\frac{i}{2} \hx_{1\m u}(\xi,t) G^{\m \n} T_\n 
  \tau_u} \!: \,\equiv \bigspc \quad \quad \quad \\
\equiv \exp [i\sqrt{2} T \cdot q \tau_1] \exp [i \sqrt{2}T \cdot J(0) t\tau_1] \times \bigspc \bigspc \bigspc \nn \\
\times \exp [\sqrt{2} T \!\cdot \!\sum_{m\geq 1} \frac{J(-m)}{m} e^{imt} \cos (m\xi) \tau_1 ] \exp [-\sqrt{2} T \!\cdot \!\sum_{m\geq 1} \frac{J(m)}{m} 
   e^{-imt} \cos (m\xi) \tau_1 ] \times \bigspc \nn \\
\times \exp [iT \!\cdot \!\!\sum_{m \leq -1} \!\frac{\hj_{10} (m+\frac{1}{2})}{m+\frac{1}{2}} e^{-i(m+\frac{1}{2})t} \sin ((m\!+\!\srac{1}{2})\xi )] \times \nn \\
\quad \quad \quad \times \exp [iT\!\cdot \!\sum_{m \geq 0} \frac{\hj_{10} (m+\frac{1}{2})}{m+\frac{1}{2}} e^{-i(m+\frac{1}{2})t} \sin ((m\!+\!\srac{1}{2})\xi ) ] \label{TVO-2} \\
T\cdot q \equiv G^{\m \n} T_\m q_\n ,\quad T\cdot J \equiv G^{\m \n} T_\m J_\n ,\quad T \cdot \hj_{10} \equiv G^{\m \n} T_\m \hj_{1\n 0} 
\end{gather}
\end{subequations}
is the solution of the twisted vertex operator equations \eqref{Ab-TVOE}. The conformal weight $\Delta (T)$ which appears here is defined in Eq.~\eqref{TwFrB}.
The $(1+\cos \xi )^{\Delta (T)}$ factor in \eqref{TVO-1} arises directly by integration of the last term in the vertex operator equation \eqref{4.20b}. Such 
$\xi$-dependent factors are generally present (see Eq.~$(7.39)$ of Ref.~\cite{Giusto}) in the solution of open-string vertex operator equations, while the 
remaining $\hat{\hg}$ factor is closely related to ordinary vertex operators: At $\xi=0$ (and $z \equiv e^{it}$), the half-integral DN modes in \eqref{TVO-2} 
decouple and the vertex operator $\hat{\hg}$ becomes an ordinary untwisted vertex operator with momentum $\sqrt{2} T\tau_1$. 

We are interested however in the problematic limit $\xi \rightarrow \pi$, where the vertex operator $\hat{\hg}$ becomes:
\begin{subequations}
\begin{gather}
\xi \rightarrow \pi ,\quad z \equiv e^{it} \\
\hat{\hg} (\st(T),\xi,t) \rightarrow \hat{\hg} (\st(T),\pi,z) = \exp [i\sqrt{2} T \!\cdot \!q \tau_1 ] \exp [\sqrt{2} \ln{z}\, T \!\cdot \!J(0) \tau_1] \times \bigspc \quad \quad \nn \\
 \times \exp [\sqrt{2} T \!\cdot \!\sum_{m\geq 1} \frac{J(-m)}{m} (-1)^m z^m \tau_1 ] \exp [-\sqrt{2} T \!\cdot \!\sum_{m\geq 1} \frac{J(m)}{m} (-1)^m z^{-m} 
   \tau_1 ] \times \quad \quad \nn \\
\quad \times \exp [iT \!\cdot \!\!\!\sum_{m \leq -1} \!\!\frac{\hj_{10} (m\!+\!\frac{1}{2})}{m+\frac{1}{2}} (-1)^m z^{-(m+\frac{1}{2})} ] \exp [iT \!\cdot 
   \!\sum_{m \geq 0} \frac{\hj_{10} (m+\frac{1}{2})}{m+\frac{1}{2}} (-1)^m z^{-(m+\frac{1}{2})} ] \,.
\end{gather}
\end{subequations}
The integer-moded factors here comprise the untwisted vertex operator for NN open-string emission from an NN string at $\xi =\pi$, while the half-integer-moded
factors describe NN emission from a DN string -- and are closely related to well-known twisted vertex operators (see e.~g.~Refs.~\cite{Lep,Big'}) in  
orbifold theory. The following commutator with the Virasoro generators
\begin{gather}
[L_\s (m), \hat{\hg} (\st(T),\pi,z)]= (-1)^m z^m (z\pl_z +(4m+2) \Delta (T) ) \hat{\hg} (\st(T),\pi,z) \label{Eq4.21}
\end{gather}
is not difficult to verify by direct computation using standard properties of these vertex operators.

It is instructive to inquire into the compatibility of the $[L_\s (m), \hg_R]$ commutator in Eq.~\eqref{Eq4.18} with the $[L_\s (m), \hat{\hg}]$ commutator in 
Eq.~\eqref{Eq4.21}. In fact, the compatibility involves in an essential way the singularity near $\xi =\pi$ of the last term in the vertex operator 
equation \eqref{4.20b}. To see this, consider the $\sin (m\xi) \pl_\xi \hg_R$ term in the commutator \eqref{Eq4.18} and follow the steps
\begin{subequations}
\label{426}
\begin{gather}
\pl_\xi \hg_R \simeq \hg_R \frac{2\Delta(T)}{\xi -\pi} \,\,\,\text{near } \xi=\pi \label{4.26a} \\
\frac{\sin (m\xi)}{\xi -\pi} = (-1)^m m +\Ord (\xi-\pi) \\
[L_\s (m) ,\hg_R] = (-1)^m z^m \Big{(} z\pl_z +2\Delta(T) (m+1) +2\Delta(T) m \Big{)} \hg_R \,\,\,\text{near } \xi=\pi \label{4.26c} \\
[L_\s (m),\hat{\hg}(\pi,z)]= (-1)^m z^m \Big{(} z\pl_z + (4m+2) \Delta(T) \Big{)} \hat{\hg}(\pi,z) \label{4.26d}
\end{gather}
\end{subequations}
where \eqref{4.26a} shows the singularity, and the final result in Eq.~\eqref{4.26d} was obtained by dividing both sides of \eqref{4.26c} by the 
$(1+\cos (\xi))^{\Delta (T)}$ factor in $\hg_R$ and then taking the limit. This computation provides an important consistency check on the singularity 
near $\xi =\pi$ of the Minkowski-space prescription. 

Both the Euclidean-space and the Minkowski-space prescriptions give solutions of the vertex operator equations with the factorized form
\begin{gather}
\hg_R = \hat{\hg} f(\xi)
\end{gather}
where $\hat{\hg}$ is the vertex operator defined in Eq.~\eqref{AbOO-TVO}, and the factor $f(\xi)$ encodes the behavior of the charge and image charge 
as they approach each other at the strip boundary. As noted above, the last term of the vertex operator equation \eqref{4.20b} would not be singular at 
$\xi =\pi$ in the Euclidean-space prescription, and hence $f(\xi)$ is neither singular nor vanishing at $\xi =\pi$ in this prescription. Such behavior would 
disrupt the connection between Eqs.~\eqref{Eq4.18} and \eqref{Eq4.21} detailed in the previous paragraph. Then, since both of these equations hold independent 
of the prescription choice, the Euclidean-space singularity prescription of Ref.~\cite{Orient1} is simply {\it inconsistent}: The interaction of charge 
and image charge at $\xi =\pi$ must be singular, as dictated by the Minkowski-space prescription.

As a consequence, all the equations of Ref.~\cite{Orient1} should be read either with the Minkowski-space prescription of this paper, or with its Euclidean 
analogue:
\begin{subequations}
\label{Eucl-Rx} 
\begin{gather}
z =Re^{i\xi} ,\quad \bz =Re^{-i\xi} \\
\left( \frac{z_i}{z_j} \right)^{\!\bar{y}(r,u)} \!\!=\left( \frac{R_i}{R_j} \right)^{\!\bar{y}(r,u)} \!e^{i(\xi_i -\xi_j)\bar{y}(r,u)} ,\quad 
   \left( \frac{\bz_i}{\bz_j} \right)^{\!\bar{y}(r,u)} \!\!=\left( \frac{R_i}{R_j} \right)^{\!\bar{y}(r,u)} \!e^{-i(\xi_i -\xi_j)\bar{y}(r,u)} \label{451b} \\
\left( \frac{z_i}{\bz_j} \right)^{\!\bar{y}(r,u)} \!\!=\left( \frac{R_i}{R_j} \right)^{\!\bar{y}(r,u)} \!e^{i(\xi_i +\xi_j)\bar{y}(r,u)} ,\quad
   \left( \frac{\bz_i}{z_j} \right)^{\!\bar{y}(r,u)} \!\!=\left( \frac{R_i}{R_j} \right)^{\!\bar{y}(r,u)} \!e^{-i(\xi_i +\xi_j)\bar{y}(r,u)} \,. 
\end{gather}
\end{subequations}
The Euclidean prescriptions in Eq.~\eqref{451b} are not new, and except when $\bz \rightarrow z$ singularities are studied, the prescription \eqref{Eucl-Rx} 
does not change any of the conclusions of Ref.~\cite{Orient1}. In what follows, we use this prescription to detail the few instances in which 
the conclusions of Ref.~\cite{Orient1} are incorrect.

The first instance concerns the structures mentioned in Eq.~\eqref{Eq 4.11a} of this paper, which, although non-singular at $\xi =0$, are now seen to 
be singular at $\xi =\pi$:
\begin{subequations}
\label{Eq4.30}
\begin{gather}
\sum_{u=0}^1 (-1)^u \left( \frac{\bz}{z} \right)^{\!\!\bar{y}(r,u)} \!\frac{f (z,\bz ; \bar{y}(r,u))}{z -\bz} =\left( \frac{\bz}{z} 
   \right)^{\frac{\bar{n}(r)}{\r(\s)}} 
\left\{ \begin{array}{ll} \frac{1 -(\bz/z )^{\frac{1}{2}}}{z-\bz} & \text{when } \srac{\bar{n}(r)}{\r(\s)} < \srac{1}{2} \\ 
    \frac{1 -(z/\bz )^{\frac{1}{2}}}{z-\bz} & \text{when } \srac{\bar{n}(r)}{\r(\s)} \geq \srac{1}{2} \end{array} \right. \\
\bigspc \bigspc = \frac{2}{iR\sin \xi} e^{-2i\xi \bnrrs} \left\{ \begin{array}{ll} 1-e^{-i\xi} &\text{when } \srac{\bar{n}(r)}{\r(\s)} < \srac{1}{2} \\
   1-e^{i\xi} & \text{when } \srac{\bar{n}(r)}{\r(\s)} \geq \srac{1}{2} \end{array} \right. \\
\simeq \frac{4i e^{-\tp \bnrrs}}{R(\xi -\pi)} \,\,\,\text{near } \xi =\pi \,.
\end{gather}
\end{subequations}
This result should be read in place of the incorrect (non-singular) form in Eq.~($4.29$d) of Ref.~\cite{Orient1}.

Our second remark concerns the twisted Euclidean vertex operator equations in the general open-string orientation-orbifold sector
\begin{subequations}
\label{Eucl-TVOE}
\begin{align}
&\pl \tilde{\hg}(\st(T),\bz,z,\s) ={\cL}_{\sgb(\s)}^{\nrm;\mnrn} (\s) \sum_{u=0}^1 :\!\hj_\nrmu (z,\s) \tilde{\hg}(\st(T),\bz,z,\s) \st_{\mnrn ,-u}(T) \!:  \nn \\
&\quad \quad \quad ={\cL}_{\sgb(\s)}^{\nrm;\mnrn} (\s) \sum_{u=0}^1 \Big{\{} :\!\hj_\nrmu (z,\s) \tilde{\hg}(\st(T),\bz,z,\s) \!:_M  \nn \\
& \bigspc +\left( \frac{\bz}{z} \right)^{\!\bar{y}(r,u)} \frac{f(z,\bz;\bar{y}(r,u))}{z-\bz} \tilde{\hg}(\st(T),\bz,z,\s) \bar{\st}_\nrmu (\bT) \nn \\
&\bigspc -\frac{1}{z} \left( \bar{y}(r,u)-\theta (\bar{y}(r,u) \geq 1) \right) \tilde{\hg}(\st(T),\bz,z,\s) \st_\nrmu(T) \Big{\}} \st_{\mnrn ,-u}(T)
\end{align}\vspace{-0.12in}
\begin{align}
&\bpl \tilde{\hg}(\st(T),\bz,z,\s) ={\cL}_{\sgb(\s)}^{\nrm;\mnrn}(\s) \sum_{u=0}^1  :\!\hj_\nrmu (\bz,\s) \tilde{\hg}(\st(T),\bz,z,\s) \bar{\st}_{\mnrn ,-u} (\bT) \!: \nn \\
& \quad \quad \quad ={\cL}_{\sgb(\s)}^{\nrm;\mnrn}(\s) \sum_{u=0}^1 \Big{\{} :\!\hj_\nrmu (\bz,\s) \tilde{\hg}(\st(T),\bz,z,\s)\!:_M \nn \\
&\bigspc +\left( \frac{z}{\bz} \right)^{\!\bar{y}(r,u)} \frac{f(\bz,z;\bar{y}(r,u))}{\bz -z} \tilde{\hg}(\st(T),\bz,z,\s) \st_\nrmu (T) \nn \\
& \bigspc -\frac{1}{\bz} \left( \bar{y}(r,u)-\theta (\bar{y}(r,u) \geq 1) \right) \tilde{\hg}(\st(T),\bz,z,\s) \bar{\st}_\nrmu (\bT) \Big{\}} \bar{\st}_{\mnrn ,-u}(\bT)  \\
&\bigspc \st_\nrmu (T) =\st_\nrm (T) \tau_u ,\quad \bar{\st}_\nrmu (\bT) = \st_\nrm (\bT) (-1)^u \tau_u
\end{align}
\end{subequations}
which appeared in Eq.~($4.22$c,d) of Ref.~\cite{Orient1}. Here $:\,\cdot \,:$ is operator-product normal ordering and $:\,\cdot \,:_M$ is the mode normal 
ordering in Eq.~\eqref{Jg_M}. The equations \eqref{Eucl-TVOE} are term-by-term equivalent to the analytic continuation of our Minkowski-space vertex 
operator equations \eqref{OO-TVOE} when read with the new Euclidean prescription in Eq.~\eqref{Eucl-Rx}:
\begin{subequations}
\label{ERx-2}
\begin{gather}
\left( \frac{\bz}{z} \right)^{\!\bar{y}(r,u)} \frac{f(z,\bz;\bar{y}(r,u))}{z-\bz} = \Big{(} 1 +2i\sin \xi e^{i\xi} \theta (\bar{y}(r,u) \geq 1)\Big{)} 
   \frac{e^{-2i\xi \bar{y}(r,u)}}{2iR\sin \xi} \\
\left( \frac{z}{\bz} \right)^{\!\bar{y}(r,u)} \frac{f(\bz,z;\bar{y}(r,u))}{\bz-z} = -\Big{(} 1 -2i\sin \xi e^{-i\xi} \theta (\bar{y}(r,u) \geq 1) \Big{)}
   \frac{e^{2i\xi \bar{y}(r,u)}}{2iR\sin \xi} \,.
\end{gather}
\end{subequations}
The sums on $u$ of these terms then give structures of the form in Eq.~\eqref{Eq4.30}, which are singular at $\xi =\pi$ and regular at $\xi =0$. We remind the 
reader that regularity of the twisted affine primary fields at $\xi =0$ is a special characteristic of the open-string sectors of the WZW orientation 
orbifolds -- and is not expected generically in twisted open WZW strings. 

We finally remark on the normal-ordered products
\begin{gather}
:\! \hj(z) \hg(\st,\bw,w) \!: ,\quad :\! \hj(z) \hg(\st,\bw,w) \!:_M
\end{gather}
which are not singular as $z\rightarrow w$ and/or $z\rightarrow \bw$, but which may be singular as $w\rightarrow \bw$, because the twisted affine primary 
fields are themselves singular as $w \rightarrow \bw \in \Real^-$. For the same reason, normal-ordered products such as those which appear in the twisted
vertex operator equations \eqref{Eucl-TVOE} 
\begin{gather}
:\!\hj(z) \hg(\st,\bz,z) \!: ,\quad :\!\hj(z) \hg(\st,\bz,z) \!:_M
\end{gather}
may be singular as $z\rightarrow \bz \in \Real^-$. To investigate this issue, we recall the following relations between the two types of normal-ordered 
products \cite{Orient1}: 
\begin{subequations}
\begin{align}
&:\!\hj_\nrmu(z,\s) \hg(\st,\bz,z,\s)\!: \,=\,:\!\hj_\nrmu(z,\s) \hg(\st,\bz,z,\s)\!:_M \nn \\
& \bigspc -\frac{1}{z} (\bar{y}(r,u)- \theta(\bar{y}(r,u) \geq 1)) \hg(\st,\bz,z,\s) \st_\nrmu \nn \\
& \bigspc -\left\{\!\left( \frac{\bz}{z} \right)^{\!\bar{y}(r,u)} \!f(z,\bz;\bar{y}(r,u)) -1 \right\} \frac{1}{z-\bz} \tilde{\st}_\nrmu \hg(\st,\bz,z,\s) \\
&:\!\hj_\nrmu(\bz,\s) \hg(\st,\bz,z,\s)\!: \,=\, :\!\hj_\nrmu(\bz,\s) \hg(\st,\bz,z,\s)\!:_M \nn \\
& \bigspc +\left\{ \!\left( \frac{z}{\bz} \right)^{\!\bar{y}(r,u)} \!f(\bz,z;\bar{y}(r,u)) -1 \right\} \frac{1}{\bz-z} \hg(\st,\bz,z,\s) \st_\nrmu \nn \\
& \bigspc +\frac{1}{\bz} (\bar{y}(r,u) -\theta(\bar{y}(r,u) \geq 1)) \tilde{\st}_\nrmu \hg(\st,\bz,z,\s) \,.
\end{align}
\end{subequations}
Using the forms given in Eq.~\eqref{ERx-2}, we see that at least one of the normal-ordered products $:\,\cdot \,:$ or $:\,\cdot \,:_M$ is singular at 
$\xi =\pi$. This conclusion replaces that stated below Eq.~$(4.4)$ of Ref.~\cite{Orient1}.

\section{Boundary States of the General Twisted Open WZW String}

The discussion above developed the {\it open-string description} (in terms of a single set of current modes $\hj$) of the general twisted open WZW string,
including its twisted non-commutative geometry, branes and twisted open-string KZ equations. In this section we review and discuss the complementary
{\it closed-string} or {\it boundary state description} (in terms of $\hj$ and $\hjb$) of these twisted open strings.

Twisted WZW boundary states were first considered in Ref.~\cite{Big} and the general twisted boundary state equation was given in Ref.~\cite{TwGiu}:
\begin{subequations}
\label{Eq5.1}
\begin{gather}
\left( \hj_\nrm (\mnrrs) + \pom (n(r),\s)_\m{}^\n \hjb_{n(r),\n} (\mnrrs) \right) |B\rangle_\s =0 \label{Eq 5.1a} \\
\left( L_\s (m) - \bar{L}_\s (-m) \right) |B\rangle_\s =0  \label{Eq 5.1b} \\
\pom \in Aut (\gfrakh (\s)) ,\quad \bar{n}(r) \in \{ 0,\ldots ,\r(\s) -1 \} ,\quad \srange \,.
\end{gather}
\end{subequations}
Here $\hj$ and $L_\s$ are the same twisted left-mover current and Virasoro modes which appear above, while $\hjb$ and $\bar{L}_\s$ are the corresponding 
twisted right-mover modes -- all of which live in sector $\s$ of the general closed-string WZW orbifold $A_g(H)/H$. The quantity $\pom$ is again a T-duality 
automorphism of the twisted left- or right-mover current algebra, although (for a given twisted open string) this automorphism may not\footnote{For 
a given twisted open string, we expect that the $\pom$'s in the two descriptions should be closely related; for example, they could be identical, or inverses 
of each other or perhaps inner-automorphically equivalent. The precise relationship of the two $\pom$'s is beyond the scope of the present paper.} 
be exactly the same T-duality automorphism $\pom$ which appears above in the corresponding open-string description. 

The general twisted boundary state equation in Eq.~\eqref{Eq 5.1a} is consistent 
\begin{gather}
\left[ \Big{(} \hj +\pom \hjb \Big{)}_\nrm (\mnrrs) , \Big{(} \hj + \pom \hjb \Big{)}_\nsn (\nnsrs) \right] = \bigspc \bigspc \nn \\
    \bigspc = i \scf_{\nrm ;\nsn}{}^{n(r)+n(s) ,\de} (\s)  \Big{(}\hj +\pom \hjb \Big{)}_{n(r)+n(s),\de} (\mnnrnsrs)
\end{gather}
because the operator terms of the twisted left- and right-mover current algebras are the same, but the central terms have {\it opposite signs} \cite{Big}. 
(This circumstance also gives rise to the rectification problem \cite{Big,Big',Perm,so2n,TwGiu}, which need not concern us in the present context). The 
Virasoro boundary condition in Eq.~\eqref{Eq 5.1b} follows from Eq.~\eqref{Eq 5.1a} and the explicit form of the general twisted affine-Sugawara construction 
\cite{Big}.

In the special case of the untwisted sector $\s=0$ of any WZW orbifold $A_g(H)/H$ (including the trivial orbifold with $H =1$), the right mover modes
$\hjb (m)$ reduce to the untwisted right-mover modes $\bJ (-m)$ and $\pom$ reduces to an automorphism of Lie $g$. The general twisted boundary 
state equation \eqref{Eq 5.1a} then reduces to the standard {\it untwisted} boundary state equation 
\begin{gather}
\Big{(} J_a (m) +\omega_a{}^b \bJ_b (-m) \Big{)} |B\rangle =0 ,\quad a,b=1\ldots \text{dim }g ,\quad \omega \in Aut (g) \label{Eq5.2}
\end{gather}
which has been extensively studied (see for example Refs.~\cite{Ishi,FS}).

We emphasize that, beyond solutions of the untwisted case in Eq.~\eqref{Eq5.2}, no explicit twisted solutions have yet been obtained to the general 
boundary state equation \eqref{Eq 5.1a}. As a pedagogical guide for future work, however, we mention here some particularly simple classes of twisted boundary 
states (all of which are contained in the general equation).

\noindent $\bullet$ The Basic Class of Twisted Open WZW Strings

As noted in Ref.~\cite{TwGiu}, one expects that the basic class of twisted open WZW strings $A_g^{open}(H)/H$ (the open-string orbifolds) is described 
by the following basic class of twisted boundary states \cite{Big}
\begin{subequations}
\begin{gather}
\pom = \one :\quad \left( \hj_\nrm (\mnrrs) + \hjb_\nrm (\mnrrs) \right) |B \rangle_\s =0 \\
\bar{n}(r) \in \{ 0,\ldots ,\r(\s) -1 \} ,\quad \srange 
\end{gather}
\end{subequations}
in parallel with the choice $\pom =\thickone$ in the corresponding open-string picture of the basic class.

\noindent $\bullet$ The Permutation-Twisted Open WZW Strings

A large class of permutation-twisted open WZW strings was discussed in Subsec.~$3.4$. The corresponding boundary states of these twisted open strings are 
described by the following equation
\begin{subequations}
\begin{gather}
\pom (\srac{\hat{j}}{f_j(\s)} ,\s)_j{}^l = e^{\tp \frac{\n_j \hat{j}}{f_j(\s)}} (\pom_P )_j{}^l :\bigspc \bigspc \bigspc \bigspc \nn \\
\bigspc \Big{(} \hj_{\hat{j}aj} (m\!+\!\srac{\hat{j}}{f_j(\s)}) +e^{\tp \frac{\n_j \hat{j}}{f_j(\s)}} (\pom_P )_j{}^l \hjb_{\hat{j}al} 
   (m\!+\!\srac{\hat{j}}{f_j(\s)}) \Big{)} |B \rangle_\s =0 \\
a =1,\ldots ,\text{dim }\gfrak ,\quad \bar{\hat{j}} =0,\ldots ,f_j(\s) -1
\end{gather}
\end{subequations}
where we have again used the cycle notation of Refs.~[14-16]. We remind that $\pom_P$ is a permutation which can exchange cycles of the same length $f_j(\s)$ 
and $\n_j$ is a cycle-dependent integer. As noted for the corresponding open-string picture in Subsec.~$3.4$, this class of boundary states can be further 
generalized by including Lie automorphisms (which act on the Lie index $a$) in our $\pom$.

It is known that the WZW permutation orbifolds contain an extended Virasoro algebra \cite{Chr,DV2,Perm}
\begin{subequations}
\label{Perm-OVA}
\begin{gather}
\hat{T}_\s (\xi,t) = \sum_m L_\s (m) e^{-im (t+\xi)} ,\quad \hat{\bar{T}}_\s (\xi,t) = \sum_m \bar{L}_\s (m) e^{-im (t-\xi)} \\
L_\s (m) = \sum_j \hat{L}_{0,j} (m) ,\quad \bar{L}_\s (m) = \sum_j \hat{\bar{L}}_{0,j} (-m) ,\quad \hat{c} =\hat{\bar{c}} =K c_{\gfraks} 
\end{gather}
\begin{gather}
\hat{\Theta}_{\hat{j}j}(\xi,t) =\sum_m \hat{L}_{\hat{j}j} (m\!+\!\srac{\hat{j}}{f_j(\s)}) e^{-i (m+\frac{\hat{j}}{f_j(\s)}) (t+\xi)} \\
\hat{\bar{\Theta}}_{\hat{j}j}(\xi,t) =\sum_m \hat{\bar{L}}_{\hat{j}j} (m\!+\!\srac{\hat{j}}{f_j(\s)}) e^{i (m+\frac{\hat{j}}{f_j(\s)}) (t-\xi)} \\
[ \hat{L}_{\hat{j}j} (m\!+\!\srac{\hat{j}}{f_j(\s)}) , \hat{L}_{\hat{l}l} (n\!+\!\srac{\hat{l}}{f_l(\s)}) ] = \de_{jl} \Big{\{} (m\!-\!n\!+\!\srac{\hat{j} 
   -\hat{l}}{f_j(\s)}) \hat{L}_{\hat{j}+\hat{l} ,j} (m\!+\!n\!+\!\srac{\hat{j} +\hat{l}}{f_j(\s)}) +\nn \\
  \bigspc +\frac{c_\gfraks f_j(\s)}{12} (m\!+\!\srac{\hat{j}}{f_j(\s)}) ((m\!+\!\srac{\hat{j}}{f_j(\s)})^2 -1) \de_{m+n+\frac{\hat{j}+\hat{l}}{f_j(\s)} ,0} 
  \Big{\}} \\
[ \hat{\bar{L}}_{\hat{j}j} (m\!+\!\srac{\hat{j}}{f_j(\s)}) , \hat{\bar{L}}_{\hat{l}l} (n\!+\!\srac{\hat{l}}{f_l(\s)}) ] = 
   -\de_{jl} \Big{\{} (m\!-\!n\!+\!\srac{\hat{j}-\hat{l}}{f_j(\s)}) \hat{\bar{L}}_{\hat{j}+\hat{l} ,j} (m\!+\!n\!+\!\srac{\hat{j} +\hat{l}}{f_j(\s)}) +\nn \\
  \bigspc +\frac{c_\gfraks f_j(\s)}{12} (m\!+\!\srac{\hat{j}}{f_j(\s)}) ((m\!+\!\srac{\hat{j}}{f_j(\s)})^2 -1) \de_{m+n+\frac{\hat{j}+\hat{l}}{f_j(\s)} ,0} 
  \Big{\}} 
\end{gather}
\end{subequations}
where $K$ is the number of copies of simple $\gfrak$ in the underlying untwisted theory. It would be interesting to determine the action of this algebra 
on the boundary states, generalizing the Virasoro boundary condition in Eq.~\eqref{Eq 5.1b}.

The explicit forms [10,12-14,16,17] of the twisted current algebras in the various closed-string WZW orbifolds on simple $g$ can similarly 
be used to construct the twisted boundary state equations of the corresponding twisted open WZW strings.

\noindent $\bullet$ The Open-String Orientation-Orbifold Sectors

As a final class of examples, we consider the open-string sectors of the WZW orientation orbifolds $A_g(H_-)/H_-$. One expects that these twisted open 
strings are described by the following class of twisted boundary states
\begin{subequations}
\label{OO-BStEq}
\begin{gather}
\pom =(-1)^u :\quad \Big{(} \hj_\nrmu (m\!+\!\nrrs \!+\!\srac{u}{2}) + (-1)^u \hjb_\nrmu (m\!+\!\nrrs \!+\!\srac{u}{2}) \Big{)} |B \rangle_\s =0 \label{57a} \\
\bar{n}(r) \in \{ 0,\ldots ,\r(\s) -1 \} ,\quad \bar{u} \in \{ 0,1 \}
\end{gather}
\end{subequations}
in parallel with the open-string description of these sectors in Eq.~\eqref{48b}. In Eq.~\eqref{57a}, the left-mover modes $\hj$ satisfy the doubly-twisted
current algebra in Eq.~\eqref{Eq 4.1a}, and the right-mover modes $\hjb$ satisfy the same doubly-twisted current algebra, except with the usual sign-reversal
of the central term \cite{Big}. Following the characterization at the end of Subsec.~$4.1$, we know that both the left-mover and right-mover modes live in the
generalized $\Zint_2$ permutation orbifold $A_{g\oplus g}(H') /H'$ shown in Eq.~\eqref{Z2}. We remind the reader (see Subsec.~$4.2$) that these 
orientation-orbifold boundary states are T-dual to the boundary states of the generalized open-string permutation orbifold
\begin{gather}
\frac{A_{g\oplus g}^{open} (H')}{H'} ,\quad H' \subset \Zint_2 \times H ,\quad H\subset Aut(g)
\end{gather}
which presumably satisfy Eq.~\eqref{OO-BStEq} with $\pom =(-1)^u$ replaced by $\pom =1$.

In this class of examples, the general Virasoro boundary condition in Eq.~\eqref{Eq 5.1b} must have an extended form, including the half-integer moded Virasoro 
generators. To find this extension, we first give the explicit form of the left- and right-mover orbifold Virasoro generators in the general $\Zint_2$ 
permutation orbifold:
\begin{subequations}
\label{5.8}
\begin{gather}
\!\!\!\hat{L}_u (m\!+\!\srac{u}{2}) =\bigspc \bigspc \bigspc \bigspc \bigspc \bigspc \bigspc \quad \nn \\
  = {\cL}_{\sgb(\s)}^{\nrm;\mnrn}(\s) \sum_{v=0}^1 \Big{\{} \sum_{p\in \Zint} :\!\hj_{\nrm v} (p\!+\!\nrrs \!+ \!\srac{v}{2}) 
   \hj_{\mnrn ,u-v} (m\!-\!p\! -\nrrs \!+\!\srac{u-v}{2}) \!:_M \nn \\
   -i\scf_{\nrm;\mnrn}{}^{0,\de} (\s) \hj_{0,\de,u} (m\!+\!\srac{u}{2}) (\bar{y}(r,v) -\theta (\bar{y}(r,v) \geq 1)) \Big{\}} +\de_{m+\frac{u}{2},0} \gscfwt \\
\!\!\!\hat{\bar{L}}_u (m\!+\!\srac{u}{2}) =\bigspc \bigspc \bigspc \bigspc \bigspc \bigspc \bigspc \quad \nn \\
   = {\cL}_{\sgb(\s)}^{\nrm;\mnrn}(\s) \sum_{v=0}^1 \Big{\{} \sum_{p\in \Zint} :\!\hjb_{\nrm v} (p\!+\!\nrrs \!+ \!\srac{v}{2}) 
   \hjb_{\mnrn ,u-v} (m\!-\!p\! -\nrrs \!+\!\srac{u-v}{2}) \!:_{\bar{M}} \nn \\
   -i\scf_{\nrm;\mnrn}{}^{0,\de}(\s) \hj_{0,\de,u}(m\!+\!\srac{u}{2}) (\bar{y}_{\ast}(r,v) -\theta (\bar{y}_{\ast}(r,v) \geq 1)) \Big{\}} 
   +\de_{m+\frac{u}{2},0} \gscfwt \label{L-bar}
\end{gather}
\begin{gather}
[\hat{L}_u (m\!+\!\srac{u}{2}) ,\hat{L}_v (n\!+\!\srac{v}{2}) ] = (m\!-\!n \!+\! \srac{u-v}{2}) \hat{L}_{u+v}(m\!+\!n\!+\!\srac{u+v}{2}) +\bigspc \bigspc \nn \\
  \bigspc \bigspc + \de_{m+n+\frac{u+v}{2},0} \frac{2c_g}{12} (m\!+\!\srac{u}{2}) ((m\!+\!\srac{u}{2})^2 -1) \\
[\hat{\bar{L}}_u (m\!+\!\srac{u}{2}) ,\hat{\bar{L}}_v (n\!+\!\srac{v}{2}) ] = -(m\!-\!n \!+\! \srac{u-v}{2}) \hat{\bar{L}}_{u+v}(m\!+\!n\!+\!\srac{u+v}{2}) 
   \bigspc \bigspc \nn \\
   \bigspc \bigspc - \de_{m+n+\frac{u+v}{2},0} \frac{2c_g}{12} (m\!+\!\srac{u}{2}) ((m\!+\!\srac{u}{2})^2 -1) \\
[\hat{L}_u (m\!+\!\srac{u}{2}) ,\hat{\bar{L}}_v (n\!+\!\srac{v}{2}) ] = 0 ,\quad \hat{L}_0 (m) =L_\s (m) ,\quad \hat{\bar{L}}_0 (m) = \bar{L}_\s (-m) \\
\bar{y}(r,v) = \sbnrrs + \srac{\bar{u}}{2} ,\quad \bar{y}_{\ast} (r,v) \equiv \srac{\overline{-n(r)}}{\r(\s)} + \srac{\overline{-u}}{2} ,\quad \hat{\bar{c}} =\hat{c} =2c_g \,.
\end{gather}
\end{subequations}
Here $:\,\cdot \,:_{M,\bar{M}}$ denote the standard mode normal orderings introduced in Ref.~\cite{Big}. The left-mover generators in Eq.~\eqref{5.8} are 
identical to those of the orientation orbifolds in Ref.~\cite{Orient1}, and the form given here is equivalent to the operator-product normal-ordered form in 
Eq.~\eqref{43}. The conformal weight $\gscfwt$ of the scalar twist field is given in Eq.~$(3.38)$ of Ref.~\cite{Orient1}, and $c_g$ is the central charge of 
the left- or right-mover affine-Sugawara construction [42,50-53,37] on $g$. Finally, we determined the right-mover generators in 
Eq.~\eqref{L-bar} from the known form of the left-movers and the general current-current OPEs in Ref.~\cite{Big}.

Following steps analogous to those detailed in App.~A of Ref.~\cite{TwGiu}, it is then straightforward to verify the following {\it extended Virasoro boundary 
condition}
\begin{subequations}
\label{59}
\begin{gather}
\Big{(} \hat{L}_u (m\!+\!\srac{u}{2}) - (-1)^u \hat{\bar{L}}_u (m\!+\!\srac{u}{2}) \Big{)} |B\rangle =0   \\
[\hat{L}_u (m\!+\!\srac{u}{2}) -(-1)^u \hat{\bar{L}}_u (m\!+\!\srac{u}{2}) , \,\hat{L}_v (n\!+\!\srac{v}{2}) -(-1)^v \hat{\bar{L}}_v (n\!+\!\srac{v}{2})] = 
   \bigspc \bigspc \nn \\
  =(m\!-\!n \!+\! \srac{u-v}{2}) \Big{(} \hat{L}_{u+v}(m\!+\!n\!+\!\srac{u+v}{2}) -(-1)^{u+v} \hat{\bar{L}}_{u+v}(m\!+\!n\!+\!\srac{u+v}{2}) \Big{)}
\end{gather}
\end{subequations}
from the boundary state equation \eqref{OO-BStEq}. The ordinary Virasoro boundary condition \eqref{Eq 5.1b} is included in this result when $u=0$. 
The extension in Eq.~\eqref{59} may be considered as a prototype for the corresponding higher-order extensions in the permutation-twisted open WZW strings 
(see Eq.~\eqref{Perm-OVA}).

The orbifold program has successfully constructed the local theory of all closed-string WZW orbifolds [8-18] and this paper completes the description of 
all twisted open WZW strings [31,24-26]. Classification of the automorphisms of twisted current algebras will provide at least
a partial classification of the twisted open WZW strings. An important next step is to study open-closed string dualities among these twisted open and 
closed WZW strings.

\vspace{-.02in}
\bigskip

\noindent
{\bf Acknowledgements}

For helpful discussions, we thank M.~Aganagic, J.~de Boer, O.~Ganor, S.~Giusto, J.~Gomis, P.~Ho\v{r}ava and C.~Schweigert.

This work was supported in part by the Director, Office of Energy Research, Office of High Energy and Nuclear Physics, Division of High Energy Physics 
of the U.S. Department of Energy under Contract DE-AC03-76SF00098 and in part by the National Science Foundation under grant PHY00-98840.

\appendix

\section{Twisted Chiral Affine Primary Fields}

From Eq.~$(4.14)$ of Ref.~\cite{TwGiu} and the $\pom$-map in Eq.~\eqref{Eq3.0}, we find the {\it factorized form} \cite{W,H+O, Giusto,Big,Perm} of the general 
twisted affine primary field
\begin{subequations}
\label{EqA.1}
\begin{align}
\hg(\st,\xi,t) &\equiv \hg_- (\st,\xi,t) \hg_+ (\st,\xi,t) \\
[\hj_\nrm (\mnrrs) ,\hg_+ (\st,\xi,t)] &=\hg_+ (\st,\xi,t) \st_\nrm e^{i(\mnrrs )(t+\xi)} \\
\quad \quad [\hj_\nrm (\mnrrs) ,\hg_- (\st,\xi,t)] &=-\st_\nrm^\pom \hg_- (\st,\xi,t) e^{i(\mnrrs )(t-\xi)} \
\end{align}
\end{subequations}
in the general twisted open WZW string. Here $\st^\pom$ is the automorphic transform \eqref{st-pom-Defn} of the twisted representation matrix $\st$, 
$\hg_\pm$ are the {\it constituent} affine primary fields, and one also finds the dynamical relations for $\hg_{\pm}$
\begin{subequations}
\label{EqA.2}
\begin{align}
&\srac{1}{2} \pl_+ \hg_+ (\st,\xi,t)= 2i{\cL}_{\sgb(\s)}^{\nrm;\mnrn} (\s) :\!\hj_\nrm^{(+)} (\xi,t) \hg_+ (\st,\xi,t) \st_\mnrn \!:_M \nn \\
&\quad +i\hg_+ (\st,\xi,t) \Dg (\st) -2i{\cL}_{\sgb(\s)}^{\nrm;\mnrn}(\s) \sbnrrs \hg_+ (\st,\xi,t) \st_\nrm \st_\mnrn \\
&\srac{1}{2} \pl_- \hg_- (\st,\xi,t)= -2i{\cL}_{\sgb(\s)}^{\nrm;\mnrn} (\s) :\!\st_\nrm^\pom \hj_\mnrn^{(-)} (\xi,t) \hg_- (\st,\xi,t)\!:_M \nn \\
&\quad +i\Dg (\st^\pom) \hg_-(\st,\xi,t) -2i{\cL}_{\sgb(\s)}^{\nrm;\mnrn}(\s) \srac{\overline{-n(r)}}{\r(\s)} \st_\nrm^\pom \st_\mnrn^\pom \hg_- (\st,\xi,t) \\
&\bigspc \bigspc \quad \quad \pl_- \hg_+(\st,\xi,t) =\pl_+ \hg_-(\st,\xi,t) =0 \label{Chiral}
\end{align}
\end{subequations}
including their {\it chiralities} in Eq.~\eqref{Chiral}. The mode normal ordering $:\,\cdot \,:_M$ here is the same as that in Eq.~\eqref{Jg_M} with 
$\hg \rightarrow \hg_\pm$.

The chiral vertex operator equations in Eq.~\eqref{EqA.2} in fact provide the direct derivation of the full twisted vertex operator equations for $\hg$ in 
Eq.~\eqref{TwoS-TVOE} -- which were obtained in the text by using the $\pom$-map \eqref{Eq3.0} as a shortcut. As discussed in Refs.~\cite{Giusto,TwGiu}, this 
derivation requires a final {\it re-normal ordering} using Eq.~\eqref{EqA.1}. The re-normal ordering, which generates the singularities of $\hg$ at 
$\xi=0,\pi$, is necessary because the same set of twisted current modes act on both $\hg_+$ and $\hg_-$. The singularities themselves describe
the collision of non-abelian charge and image charge at the strip boundary. Such singularities of the affine primary fields are not found in closed string 
theory (including ordinary orbifolds) because the chiral constituents $\hg_\pm$ are independent in that case (or alternatively because there are no image 
charges in closed-string conformal field theories).

\vskip .5cm 
\addcontentsline{toc}{section}{References} 
 
\renewcommand{\baselinestretch}{.4}\rm 
{\footnotesize 
 
\providecommand{\href}[2]{#2}\begingroup\raggedright\endgroup

\pagebreak

\end{document}